\documentclass[aps,prd,a4paper,longbibliography,reprint,superscriptaddress,floatfix,nofootinbib,amsmath,amssymb,preprintnumbers]{revtex4-1}
\usepackage{graphicx}
\usepackage{array}
\usepackage[utf8]{inputenc}
\usepackage[bookmarks,
bookmarksopen = true,
bookmarksnumbered = true,
linktocpage,
colorlinks = true,
linkcolor = blue,
urlcolor  = blue,
citecolor = blue,
anchorcolor = green,
hyperindex = true,
hyperfigures]{hyperref}
\usepackage{cmap}
\newcommand{\be}{\begin{equation}}
\newcommand{\ee}{\end{equation}}
\newcommand{\nf}{{N_f}}
\newcommand{\mK}{M_K}
\newcommand{\mpi}{M_\pi}
\newcommand{\mev}{\,\textmd{MeV}}
\newcommand{\fm}{\,\textmd{fm}}
\newcommand{\gev}{\,\textmd{GeV}}
\newcommand{\deriv}{\mathrm{d}}
\begin{document}
\preprint{\begin{tabular}{r}WUB/16-03 \\ DAMTP-2016-50 \end{tabular}}
\title{Hadro-quarkonium from Lattice QCD}
\author{Maurizio Alberti}
\affiliation{Department of Physics, Bergische Universit\"at Wuppertal,
Gau\ss{}stra\ss{}e 20, 42119 Wuppertal, Germany}
\author{Gunnar~S.~Bali}
\affiliation{Institut f\"ur Theoretische Physik, Universit\"at Regensburg, Universit\"atsstra\ss{}e 31, 93053 Regensburg, Germany}
\affiliation{Department of Theoretical Physics, Tata Institute of Fundamental Research, Homi Bhabha Road, Mumbai 400005, India}
\author{Sara Collins}
\affiliation{Institut f\"ur Theoretische Physik, Universit\"at Regensburg, Universit\"atsstra\ss{}e 31, 93053 Regensburg, Germany}
\author{Francesco Knechtli}
\affiliation{Department of Physics, Bergische Universit\"at Wuppertal,
Gau\ss{}stra\ss{}e 20, 42119 Wuppertal, Germany}
\author{Graham Moir}
\affiliation{Department of Applied Mathematics and Theoretical Physics, Centre for Mathematical Sciences,
University of Cambridge, Wilberforce Road, Cambridge, CB3 0WA, UK}
\author{Wolfgang S\"oldner}
\affiliation{Institut f\"ur Theoretische Physik, Universit\"at Regensburg, Universit\"atsstra\ss{}e 31, 93053 Regensburg, Germany}
\begin{abstract}
The hadro-quarkonium picture [S.\ Dubinskiy and M.B.\ Voloshin, Phys.\ Lett.\ B
{\bf 666}, 344 (2008)] provides one
possible interpretation for the pentaquark candidates with
hidden charm, recently reported by the LHCb Collaboration, as well as for some
of the charmonium-like ``$X,Y,Z$" states.
In this picture, a heavy quarkonium core resides within a light hadron
giving rise to four- or five-quark/antiquark bound states.
We test this scenario in the heavy quark limit by investigating the modification
of the potential between a static quark-antiquark pair induced by the presence of a hadron.
Our lattice QCD simulations are performed on a Coordinated Lattice Simulations (CLS) ensemble with $\nf = 2+1$ flavours of non-perturbatively improved Wilson quarks at a pion mass of about $223\mev$ and a lattice spacing of about $a=0.0854\fm$.
We study the static potential in the presence of a variety of light mesons as well as of octet and decuplet baryons. In all these cases,
the resulting configurations are favoured energetically. The associated
binding energies between the quarkonium in the heavy quark limit and the
light hadron are found to be smaller than a few MeV, similar in
strength to deuterium binding. It needs to be seen if the small attraction
survives in the infinite volume limit and supports bound states or
resonances.
\end{abstract}
\keywords{lattice QCD, hadro-quarkonium, pentaquarks}
\maketitle

\section{Introduction}
Recently, the LHCb Collaboration found two structures in the
decay $\Lambda_b\rightarrow J/\psi pK$, which can be
interpreted as candidates for pentaquark states with hidden
charm, containing three light quarks, in addition to a
charm quark-antiquark pair~\cite{Aaij:2015tga,Aaij:2016phn}.
The most likely spin and parity assignments for these
candidates, labelled $P_c^+(4380)$
and $P_c^+(4450)$, are $J^P=3/2^-$ and $5/2^+$, respectively,
with $3/2^+$ and $5/2^-$ being another possibility.
While the nature of these (and of some other structures)
is still disputed~\cite{Szczepaniak:2015hya,Guo:2016bkl},
the number of established charmonium resonances certainly has
exploded during the past 15 years, see Ref.~\cite{Brambilla:2010cs}
and, e.g., Ref.~\cite{Chen:2016qju} for a more recent review.
Many of these are of an exotic nature and some clearly hint
at light quark-antiquark or --- in the case of the $P_c$ candidates ---
even at $qqq$ components, in addition to the charm quark and antiquark.

Many models can accommodate some, or if extended to include
states that contain five (anti-)quarks, even all of these resonances:
tetraquarks~\cite{Jaffe:1976ig,Weinstein:1982gc,Maiani:2007vr}
consisting of diquark-antidiquark pairs, including a recently
proposed ``dynamic'' picture~\cite{Brodsky:2014xia,Lebed:2016epe},
molecules of two
open charm mesons~\cite{Voloshin:1976ap,DeRujula:1976zlg,Novikov:1977dq,Tornqvist:1993ng,Close:2010wq},
hybrid states~\cite{Barnes:1977hg,Barnes:1982zs,Chanowitz:1982qj,Isgur:1985vy} containing a charm quark-antiquark
pair and additional valence gluons, hadro-charmonium
with a compact charmonium core bound
inside a light hadron~\cite{Dubynskiy:2008mq,Li:2013ssa},
and mixtures of the above.
Here we will specifically aim to establish if the last picture
(hadro-quarkonium) is supported in the heavy quark limit.

The standard way of addressing a strongly decaying
resonance and extracting the position of the associated
pole in the unphysical Riemann sheet
from simulations in Euclidean spacetime boxes was introduced
by L\"uscher~\cite{Luscher:1990ux}.
For applications of this and related methods to charmonium
spectroscopy, see, e.g., Ref.~\cite{Prelovsek:2015fra} and references
therein. In the case of charmonia, this is particularly challenging
since, in addition to ground states,
radial excitations need to be considered and the
number of different decay channels can be large, some with
more than two hadrons in the final state. Moreover,
while in principle resonance parameters can be computed,
at least below inelastic multi-particle thresholds,
these will not necessarily tell us much about the ``nature''
of the underlying state: how does the naive quark model need
to be modified to provide a guiding principle for the
existence or non-existence of an exotic resonance?

A direct computation of the scattering parameters of, e.g.,
a nucleon-charmonium resonance in a realistic setting
constitutes a serious computational challenge,
especially if one aims at conclusive results with meaningful
errors. Instead of directly approaching the problem at hand,
here we restrict ourselves to the heavy quark limit in which
the charm quarks can be considered as slowly moving in the background
of gluons, sea quarks and, possibly, light hadrons.

After integrating out the degrees of freedom
associated with the heavy quark mass $m_Q$, quarkonia
can be described in terms of an effective field theory:
non-relativistic QCD (NRQCD)~\cite{Caswell:1985ui}.
In the limit of small distances $r$, or equivalently,
large momentum transfers $m_Qv$, where $v$ is the interquark
velocity, the scale $m_Qv\sim 1/r$ can also be
integrated out, resulting in potential
NRQCD (pNRQCD)~\cite{Pineda:1997bj,Brambilla:1999xf}.
Then, to leading order in $r$ with respect to the pNRQCD multipole
expansion and to $v^2\sim\alpha_s$ in the NRQCD power counting,
quarkonium becomes
equivalent to a non-relativistic quantum mechanical system, where
the interaction potential is given by the static potential
$V_0(r)$ which can, e.g., be computed non-perturbatively from Wilson loop
expectation values $\langle W(r,t)\rangle$ in
Euclidean spacetime:
\be
\label{eq:potential}
V_0(r)=-\lim_{t\rightarrow\infty}\frac{\deriv}{\deriv t}\ln \langle
W(r,t)\rangle\,.
\ee
Here we investigate whether this potential becomes modified in the presence
of a light hadron. This would then lower or increase
quarkonium energy levels. If embedding the quarkonium
in the light hadron is energetically favourable, this would
suggest
a bound state, at least for sufficiently large quark masses.

This article is organized as follows.
In Sec.~\ref{sec_intro2} we briefly discuss previous studies
of nucleon-charmonium bound states and comment on the ordering
of scales that we consider.
In Sec.~\ref{sec_define} we define the observables that we
compute. Then, in Sec.~\ref{sec_technique} we describe details
of the simulation, before
numerical results are presented in Sec.~\ref{sec_numerics}.
Subsequently, in Sec.~\ref{sec_binding} we relate the modifications
of the static potential to quarkonium bound state energies,
before we summarize in Sec.~\ref{sec_summary}.

\section{Nucleon-charmonium bound states}
\label{sec_intro2}
Light meson exchanges between a single nucleon or nucleons bound in a
nucleus and quarkonium, which does not contain any light valence quarks,
are suppressed by the Zweig rule.
Therefore, such interactions should be dominated by
gluon exchanges.
In the heavy quark limit, quarkonium can
be considered essentially as a point particle of a heavy
quark and antiquark bound by the short-range perturbative
Coulomb potential. The first non-vanishing chromodynamical
multipole is then a dipole and quarkonium may interact with
the nuclear environment via colour dipole-dipole
van der Waals forces. For a recent discussion of the
relevant scales in the context of effective field theories,
see Ref.~\cite{Brambilla:2015rqa}.
Initially, using phenomenological interaction potentials,
nucleon-charmonium binding energies ranging from
$20\mev$~\cite{Brodsky:1989jd,Kaidalov:1992hd}
down to $10\mev$~\cite{Wasson:1991fb} were estimated
for nuclei consisting of
$A>3$~\cite{Brodsky:1989jd,Wasson:1991fb} and 
$A>10$~\cite{Kaidalov:1992hd} nucleons.
A first QCD based estimate~\cite{Luke:1992tm}
for the potential between quarkonium in the heavy quark
limit and a nucleus resulted in
$\Upsilon$ and $J/\psi$ binding energies of a few
$\mev$ and $10\mev$, respectively,
possibly with large relativistic and higher order multipole
corrections in the charmonium case. This discussion of light nuclei hosting
a quarkonium state may have contributed to the suggestion of
quarkonium states that are embedded within light hadrons,
hadro-quarkonia~\cite{Dubynskiy:2008mq}.

At present no (p)NRQCD lattice studies of baryon-charmonium states
exist. However, a few investigations employing relativistic charm quarks
have been carried out. In Ref.~\cite{Yokokawa:2006td},
the $\eta_c$ and $J/\psi$ charmonia were scattered with light pseudoscalar
and vector mesons as well as with the nucleon, in the quenched
approximation with rather large light quark mass values;
the ratio $\mpi/m_{\rho}$ ranged from 0.9
down to 0.68. Varying the lattice extent from $L=1.6\fm$
over $2.2\fm$ up to $3.2\fm$, in this pioneering work
scattering lengths were extracted,
indicating some attraction in all the channels investigated.
A similar study was performed in Ref.~\cite{Liu:2008rza},
combining staggered sea with domain wall light and Fermilab charm quarks,
however, unusually small scattering lengths were reported.
Finally, a pseudoscalar charm quark-antiquark pair was created
along with a nucleon and even with
light nuclei by the NPLQCD Collaboration~\cite{Beane:2014sda}. In this work
the binding energy reported for the nucleon case was about $20\mev$,
albeit at a rather large light quark mass
value, corresponding to $\mpi\approx
800\mev$, and for a coarse lattice spacing $a\approx 0.145\fm$.
This value of the binding energy is consistent with some of
the expectations for
charmonia in a nuclear environment discussed above.

Closest in spirit to the van der Waals interaction
picture, Kawanai and Sasaki~\cite{Kawanai:2010ev} in a quenched study,
again at rather large pion masses, $\mpi\geq 640\mev$,
computed a charmonium-nucleon Bethe--Salpeter wave function.
Plugging this into a Schr\"odinger equation, a potential
between the charmonium and the nucleon was extracted,
indicating very weak attractive forces.

Here we will not assume a non-relativistic light hadron of mass $m_H$,
whose dipole-dipole interaction with quarkonium can be described by a potential.
Instead, our light hadron is an extended relativistic object. We also go
beyond the point-dipole approximation in the heavy quark sector by
``pulling'' quark and antiquark apart by a distance $r$.
We then determine the modification of the interaction potential between the
heavy quark-antiquark pair, that we approximate as static sources, induced
by the presence of a light hadron. To be more precise, we will consider the
limit $m_Q\gg m_H$, $m_Q\gg\Lambda_{\mathrm{QCD}}$, where
$\Lambda_{\mathrm{QCD}}$ denotes a typical non-perturbative scale
of a few hundred $\mev$, and
$v^2\ll 1$. Since we determine the quark-antiquark
potential, i.e.\ the matching function between NRQCD and pNRQCD,
nonperturbatively, $m_Qv\sim 1/r$ does not need to
be much larger than $\Lambda_{\mathrm{QCD}}$. However,
we neglect colour octet contributions~\cite{Pineda:1997bj,Brambilla:1999xf},
which may become significant at distances
$r\gtrsim \Lambda_{\mathrm{QCD}}^{-1}$.

\section{Static potentials ``inside'' hadrons}\label{sec_define}
We denote an interpolator creating a static fundamental colour
charge $Q$ at a position
$\mathbf{z}+\mathbf{r}/2$ and destroying it at a position
$\mathbf{z}-\mathbf{r}/2$ as $\mathcal{Q}_{\mathbf{r}}^{\dagger}(\mathbf{z})$.
This will transform according to the fundamental $\mathbf{3}$ representation
of the gauge group at $\mathbf{z}+\mathbf{r}/2$ and according to
$\mathbf{3}^*$ at $\mathbf{z}-\mathbf{r}/2$ and hence it contains
a gauge covariant transporter connecting these
two points (usually a spatially smeared Schwinger line). The Wilson loop
can then be written as
\be
\label{eq:wilson}
\langle W(r,t)\rangle=
\langle 0|\mathcal{Q}_r \mathcal{T}^{t/a}\mathcal{Q}_r^{\dagger}|0\rangle\,,
\ee
where we assume rotational invariance is restored for
$r=|\mathbf{r}|\gg a$, and $\mathcal{T}=e^{-a\mathbb{H}}$
denotes the transfer matrix connecting
adjacent time slices.

Within the static approximation, there are different
strategies to investigate bound states containing a heavy quark-antiquark
pair and additional light quarks.
One method, which we are not going
to pursue here, amounts to creating a light hadron $H$ containing
either $\bar{q}q$ or $qqq$ along with the stringy $Q\overline{Q}$
state at equal Euclidean time. The interpolator for creating a zero momentum
projected tetra- or pentaquark state then has the form
\be
\label{eq:penta}
\overline{\mathcal{P}}_r=
\sum_{\mathbf{z}}\overline{\mathcal{H}}(\mathbf{z})\mathcal{Q}^{\dagger}_r(\mathbf{z})\,.
\ee
Note that the creation interpolator $\overline{\mathcal{H}}$ of a hadronic state
(as well as $\overline{\mathcal{P}}_r$) will carry a spinor
index, which we suppress.
The correlator of interest is now 
$\langle 0|\mathcal{P}_r\mathcal{T}^{t/a}\overline{\mathcal{P}}_r|0\rangle$.
Even without summing over positions $\mathbf{z}$ this
is automatically projected onto zero momentum at source and sink
as the light hadron
is tied in position space
to the static quarks, see Eq.~\eqref{eq:penta}. Numerous possibilities
exist for where to spatially place the light quarks relative to
the heavy sources within the interpolator $\mathcal{P}_r$
and how to transport and contract the colour such that
the interpolator respects the correct gauge transformation properties.
This freedom can be exploited to enhance the overlap of $\overline{\mathcal{P}}_r|0\rangle$
with the physical state and may also provide some insight into its
internal structure.

Subsequent to a pioneering lattice study~\cite{Richards:1990xf} of a light
$qq$ pair bound in the above way to two static anti-triplet sources,
quite a few simulations of a light $q\bar{q}$ pair bound to the string
state created by $\mathcal{Q}_r^{\dagger}$ have also been carried out.
Such results exist both for a light quark-antiquark pair with
isospin $I=1$~\cite{Pennanen:1999xi,Bali:2005fu,Bali:2010xa,Peters:2016wjm}
and $I=0$~\cite{Bali:2005fu,Bali:2011gq}.
In contrast, a static quark-antiquark pair accompanied by three
light quarks has not been investigated on the lattice so far.

Instead of creating tetra- or pentaquark states containing a
heavy or static quark and the corresponding antiquark, here we
wish to ``directly'' address a particular picture of such bound states,
hadro-quarkonium~\cite{Dubynskiy:2008mq,Li:2013ssa}.
This will be achieved by computing
differences between the static potential in the presence of
a light hadron, relative to the static potential in the vacuum.
The former can be obtained from the large Euclidean time decay of
\be
\label{eq:expe}
\langle H|\mathcal{Q}_r \mathcal{T}^{t/a}\mathcal{Q}_r^{\dagger}|H\rangle\,,
\ee
where $|H\rangle$ is the ground state that is destroyed by the
zero momentum interpolator
\be
\mathcal{H}\equiv \sum_{\mathbf{x}}\mathcal{H}(\mathbf{x})\,.
\ee

\begin{figure}
\includegraphics[width=0.48\textwidth]{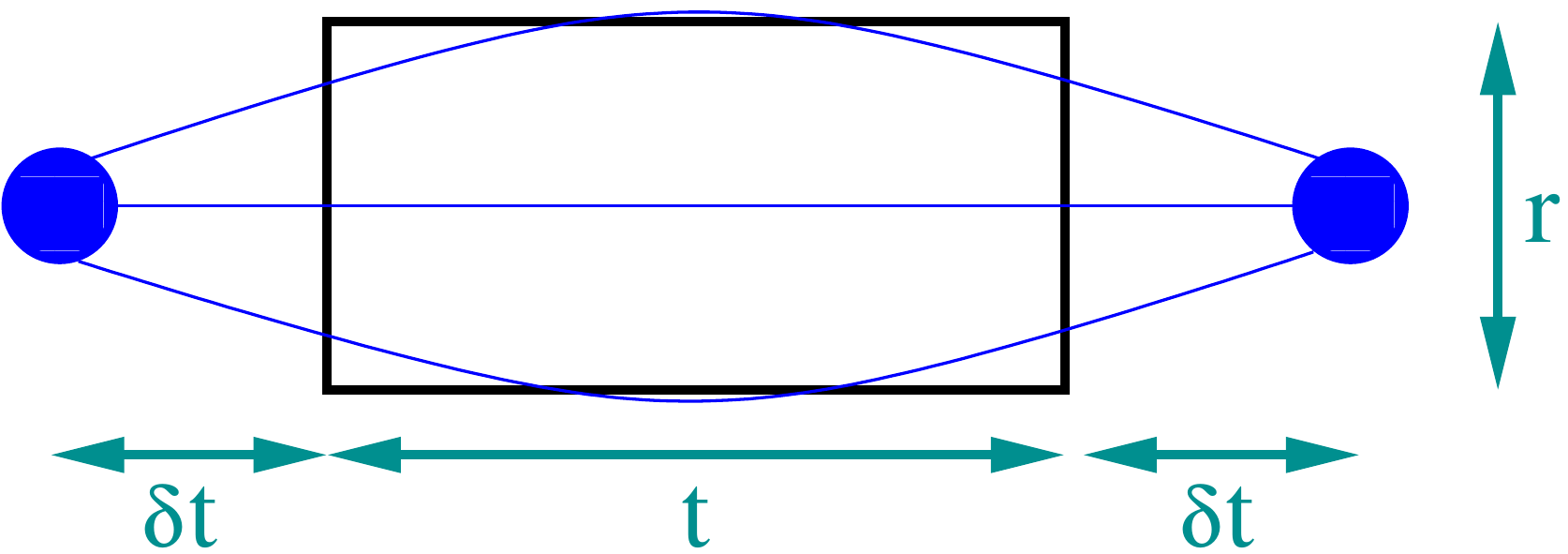}
\caption{Graphical representation of the four-point correlation
function in the numerator of
Eq.~\eqref{eq:hadroloop} for the example of a static quark-antiquark
pair at a distance $r$ embedded in a baryon. Thin blue lines
correspond to light quark
propagators and the black rectangle to the Wilson loop.}
\label{fig:correlator}
\end{figure}

In order to evaluate the expectation value Eq.~\eqref{eq:expe}
we create a
hadronic state at time $0$. We then let it propagate to
$\delta t$ to achieve ground state dominance.
At this time we create an additional
quark-antiquark string by inserting a
(smeared) Wilson loop of time extent $t$, which terminates
at $t+\delta t$.
Finally, we destroy the light hadron at the time $t+2\delta t$.
Then
\begin{align}\label{eq:hadroloop}
&\langle H|\mathcal{Q}_r \mathcal{T}^{t/a}\mathcal{Q}_r^{\dagger}|H\rangle\nonumber\\
&\propto \lim_{\delta t\rightarrow\infty}
\frac{\langle 0|\mathcal{H}\mathcal{T}^{\delta t/a}\mathcal{Q}_r \mathcal{T}^{t/a}
\mathcal{Q}_r^{\dagger}\mathcal{T}^{\delta t/a}\overline{\mathcal{H}}|0\rangle}{\langle 0|
\mathcal{H}\mathcal{T}^{(t+2\delta t)/a}\overline{\mathcal{H}}|0\rangle}\,,
\end{align}
where we average over all spatial Wilson loop positions $\mathbf{z}$ and
light hadronic sink positions $\mathbf{x}$. Zero momentum projection
at the light hadronic source can be avoided, due to the translational
invariance
of expectation values. The correlator of interest is depicted in
Fig.~\ref{fig:correlator}.

We can now define the potential in the background of the hadron as
\be
V_H(r)=-\lim_{t\rightarrow\infty}\frac{\deriv}{\deriv t}\ln
\langle H|\mathcal{Q}_r \mathcal{T}^{t/a}\mathcal{Q}_r^{\dagger}|H\rangle\,,
\ee
in analogy  to Eqs.~\eqref{eq:potential} and \eqref{eq:wilson}.
In the end we will compute differences
\begin{align}
\Delta V_H(r)&=V_H(r)-V_0(r) \nonumber\\
&=-\lim_{t\rightarrow\infty}\frac{\deriv}{\deriv t}
\ln
\frac{\langle H|\mathcal{Q}_r \mathcal{T}^{t/a}\mathcal{Q}_r^{\dagger}|H\rangle}{\langle 0|\mathcal{Q}_r \mathcal{T}^{t/a}\mathcal{Q}_r^{\dagger}|0\rangle}\nonumber
\\
&=-\lim_{t\rightarrow\infty}\frac{\deriv}{\deriv t}
\ln
\frac{\langle W(r,t)C_{H,\mathrm{2pt}}(t+2\delta t)\rangle}{\langle W(r,t)\rangle
\langle C_{H,\mathrm{2pt}}(t+2\delta t)\rangle}
\label{eq:energy_shift}
\,,
\end{align}
where the argument of the logarithm is simply the correlator
of a light hadronic two-point function with the Wilson loop inserted,
divided by the Wilson loop expectation value
times the hadronic two-point function
$\langle C_{H,\mathrm{2pt}}(t+2\delta t)\rangle=\langle 0|
\mathcal{H}\mathcal{T}^{(t+2\delta t)/a}\overline{\mathcal{H}}|0\rangle$,
see the denominator of Eq.~\eqref{eq:hadroloop}.

We are now in the position to address the question
within what hadronic channels
$\Delta V_H(r)$ will be attractive and in what cases repulsive.
This may serve as an indicator for the stability of related
hadro-quarkonia. In view of the recent LHCb
result~\cite{Aaij:2015tga,Aaij:2016phn} baryonic states $|H\rangle$ are
particularly interesting. For instance, adding the
mass of the $\Delta$ to that of the $J/\psi$
gives $4329\mev$~\cite{Agashe:2014kda}, which is not far away from
the mass of the $P_c(4380)$. Furthermore,
$J^P=3/2^+$ can couple to $1^-$ to give $3/2^-$. Another example is the sum of
the nucleon ($N$) and $\chi_{c2}$ masses, $4496\mev$,
which is close to the mass of the $P_c^+(4450)$. Again,
$1/2^+$ and $2^+$ can couple to $J^P=5/2^+$.

\section{Implementation and technical details}\label{sec_technique}
We analyse the $\nf=2+1$ ensemble ``C101'', which has a volume of
$96\times 48^3$ sites and was generated by the Coordinated
Lattice Simulations (CLS) effort~\cite{Bruno:2014jqa} using the
{\sc openQCD} simulation program~\cite{ddopenqcd,Luscher:2012av}.
Open boundary conditions in time and non-perturbatively
order-$a$ improved Wilson Fermions on top of the tree level Symanzik
improved Wilson gauge action are employed,
see Ref.~\cite{Bruno:2014jqa} for details on the simulation.
To determine the lattice spacing we extrapolate the
scale parameter $t_0$~\cite{Luscher:2010iy} to the physical point,
where we obtain $\sqrt{8t_0}/a=4.852(7)$~\cite{Bali:2016umi}.
Using the continuum limit result
$\sqrt{8t_0}=0.4144(59)(37)\fm$~\cite{Borsanyi:2012zs},
gives $a=0.0854(15)\fm$.
The pion and kaon masses on this ensemble are $\mpi\approx
223\mev$ and $\mK\approx 476\mev$,
respectively. Note that while the pion is heavier than
in nature the kaon is somewhat lighter
since the sum of quark masses $2m_{\ell} +m_s$ ($m_{\ell}=m_u=m_d$)
was adjusted to a value close to the physical one and kept constant
within the main set of CLS simulations~\cite{Bruno:2014jqa}.
The spatial lattice extent reads
$L\approx 4.6/\mpi\approx 4.1\fm$. For details
see Ref.~\cite{Bali:2016umi}.

We analyse 1552 configurations, separated by
four molecular dynamic units. On each of these
configurations we place
hadronic sources on 12 different time slices
($30, 43, 44,\ldots,52,53,65$)  at random spatial
positions to reduce autocorrelations. Due to the
use of open boundary conditions, we have to discard
the boundary regions from our analysis.
After carefully checking for translational
invariance in time, we use forward and backward
propagating hadronic two-point functions for the 11\footnote{
On time slice 47 two different spatial source positions were used.}
sources placed in the central region of the lattice
but propagate only forward from $t_0/a=30$ and backward from
$(t-t_0)/a=65$. This gives a total of
$24\times 1552$ two-point functions for each light hadron and
spin polarization considered.
Since $\delta t$ needs to be kept
small to obtain statistically meaningful results,
the quark propagators entering these two-point functions are Wuppertal smeared
at source and sink, using spatially smeared gauge transporters,
to improve the overlap with the physical ground states.

We measure the Wilson loops using the publicly available
{\sc wloop} package~\cite{wloop},
following the method described in Ref.~\cite{Donnellan:2010mx}.
In a first step,
all gauge links are smeared using a single iteration of hypercubic 
(HYP) blocking~\cite{Hasenfratz:2001hp}.
Smearing the temporal links corresponds to a particular discretization
choice of the static action and results in an exponential improvement of the
signal-to-noise ratio of correlators involving static
quarks~\cite{DellaMorte:2003mn}; HYP links
reduce the coefficient of the divergent contribution to the self-energy
of a static quark
\cite{Hasenfratz:2001tw,DellaMorte:2005yc,Bali:2005fu,Grimbach:2008uy}.
In a second step we construct a variational basis of Wilson loops using
four different levels ($0,5,7,12$)
of HYP  smearing restricted to the three space dimensions.

To enable the construction of the correlators Eq.~\eqref{eq:energy_shift},
we separately average the Wilson loops for each direction
of $\mathbf{r}$, pointing along one of the three spatial lattice axes,
and for different temporal positions.
As detailed above, due to the use of open
boundary conditions, our hadronic two-point functions $C_{H,\mathrm{2pt}}(t)$
are confined to the central time region of the lattice. We checked
that ratios of Wilson loop expectation values, averaged
over different temporal domains, centred about the middle
of the lattice, were statistically consistent
with 1. Furthermore, Eq.~\eqref{eq:energy_shift} was evaluated
in two ways, restricting the Wilson loop average in the denominator
to the same time slices as the averaging performed within the numerator
as well as averaging the Wilson loop expectation value in the
denominator within the whole region where boundary effects were negligible,
from time slice $24$ to $72$. The two results obtained for
each quantity were statistically compatible with each other and below we
will make use of the larger averaging region as this resulted
in slightly smaller statistical errors.

For the error analysis, we apply the standard method of Ref.~\cite{Wolff:2003sm}.
We include the reweighting factors due to twisted-mass reweighting and the
rational approximation for the strange quark, see Ref.~\cite{Luscher:2012av}.
We checked that carrying out a more conservative analysis, estimating
the effect of slow modes~\cite{algo:csd}, only affects the errors
in very few cases and never by more than 30\%.

The distance $\mathbf{r}$ between the static sources breaks
the continuum $\textmd{O}(3)$ symmetry down to the cylindrical subgroup
$\textmd{O}(2)\otimes \mathbb{Z}_2=\textmd{D}_{\infty h}$. Regarding Fermionic
representations, i.e.\ for baryons, the double cover is reduced accordingly.
In our implementation the static source-antisource distance
$\mathbf{r}$ is kept parallel to lattice axes.
This means that the 48 element octahedral crystallographic group
with reflections $\textmd{O}_h$ is reduced to its 16 element
subgroup $\textmd{D}_{4h}$ (and its double cover $\textmd{O}_h'$ to
$\textmd{Dih}_4\otimes \textmd{Dih}_1$). Therefore, when
correlating hadrons with a continuum spin assignment $J\geq 1$
with the string state in the $\Sigma_g^+$ irreducible
representation (irrep) of $D_{\infty h}$ ($A_{1g}$ of $D_{4h}$ on the
lattice), care has to be taken to construct the adequate
irrep of the cylindrical group. Below we address the continuum
situation but we have checked that the same arguments hold
regarding the lattice irreps that we use.
In the case of vector mesons, for example the $\phi$ meson,
the $1^-$ $\textmd{O}(3)$ irrep will split into
$\Pi_u$ and $\Sigma_u^+$, the latter also appearing in
the pseudoscalar channel. To block out this undesired contribution,
we need to correlate a Wilson loop with
$\mathbf{r}$ pointing in the $z$ direction
with the vector
state destroyed by a polarized interpolator $(\phi_x+i\phi_y)/\sqrt{2}$.
We average over cyclic permutations of $x$, $y$ and $z$.
The decuplet baryon interpolator we use,
for example for the $\Delta$ baryon, gives a state maximally polarized in
the $z$ direction. This then has to be correlated
with a Wilson loop pointing in the $z$ direction too, to guarantee
$\Lambda=|J_z|=3/2$ and to avoid mixing with spin $1/2$ baryonic states.
In this case we only used one polarization and therefore we cannot
exploit averaging over different directions.

\section{Numerical results}\label{sec_numerics}
Our strategy for testing the hadro-quarkonium picture is to determine the
potential between two static quarks in the vacuum and to compare this with
its counterpart in the presence of a hadron. An energetically favourable
difference may signal a tendency of the system to bind. In Sec.~\ref{light_hadrons}
we discuss the quality of our light hadronic effective masses and in
Sec.~\ref{subsec_stat_pot} we determine the potential in the vacuum, before
moving on to Sec.~\ref{subsec_energy_diffs}, where we investigate the modifications
induced by the presence of hadrons. We delay the discussion of the phenomenological consequences
to Sec.~\ref{sec_binding}.

\subsection{Light hadronic effective masses}\label{light_hadrons}
\begin{figure}
\includegraphics[width=0.48\textwidth]{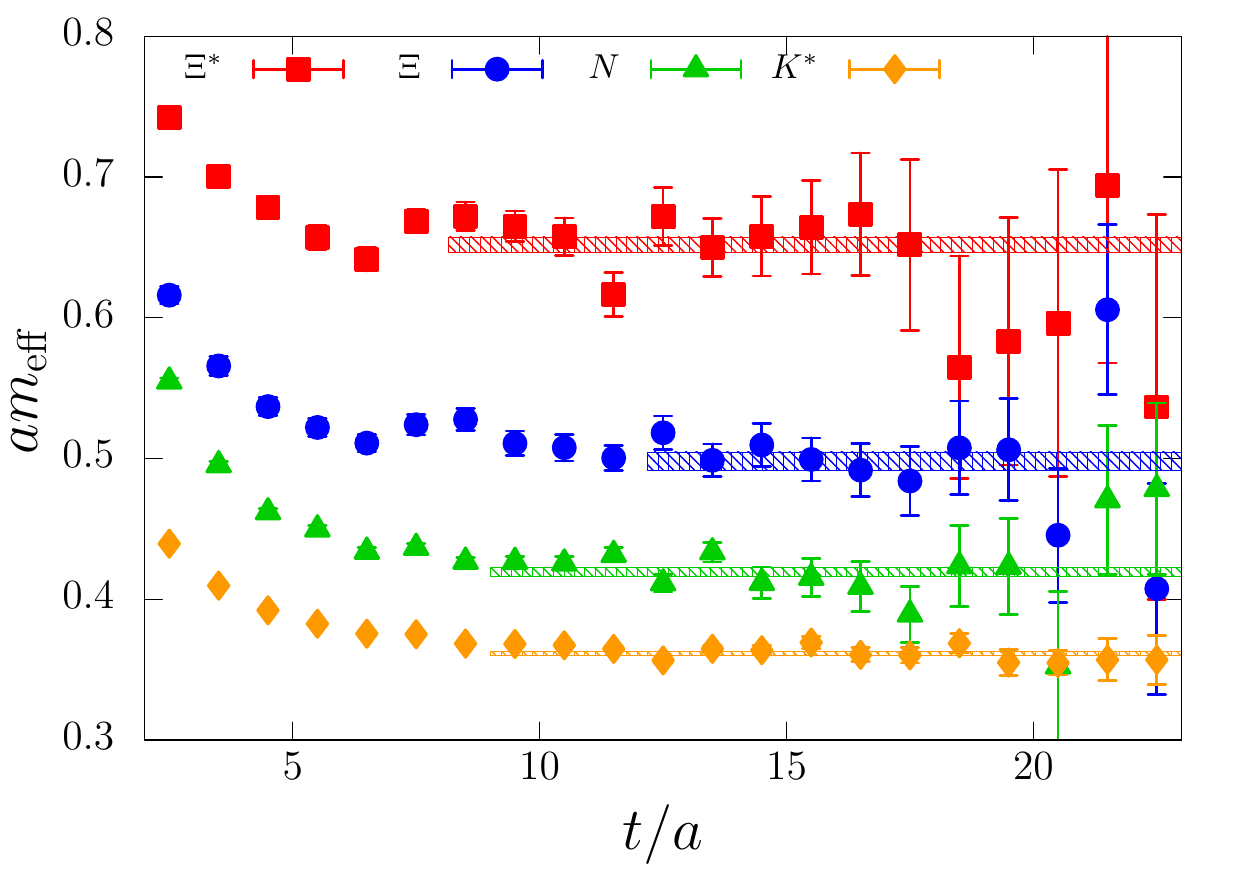}
\caption{Effective masses Eq.~\eqref{eq:effmassh}, extracted from
various hadronic two-point functions, together with results from
one-exponential fits (shaded regions).}
\label{fig:effmassh}
\end{figure}
In the determination of $\Delta V_H(r)$ below we will
quote the $\delta t = \delta t_{\mathrm{opt}}=
5a\approx 0.43\fm$ estimates as our final
results. With this $\delta t$ value, the fit in $t$ to
the right hand side of Eq.~\eqref{eq:energy_shift} is dominated
by data with $t\leq t_{\max}=10a$. Therefore, the hadronic effective
masses
\be
\label{eq:effmassh}
m_{H,\mathrm{eff}}(t+a/2)\equiv a^{-1}\ln\frac{C_{H,\mathrm{2pt}}(t)}{C_{H,\mathrm{2pt}}(t+a)}
\ee
should ideally exhibit plateaus for $t\ll t_{\max}+2\delta t_{\mathrm{opt}}=
20a\approx 1.7\fm$. We wish to check whether this is the case
within the given statistics and for the quark smearing that we employ.

In Fig.~\ref{fig:effmassh} we display effective
masses for some representative hadrons, namely the $K^*$, the nucleon $N$,
and the cascades $\Xi$ and $\Xi^*$, together with one-exponential
fits to the plateau region. This region
was determined from the requirement
that the contribution of the second exponent of a two-exponential fit
to data starting at $t=3a$
amounted to less than 25\% of the error of the correlation
function. Using this criterion, indeed, in almost all the cases
the plateau starts at $t<10a=2\delta t_{\mathrm{opt}}=(t_{\max}+2\delta t_{\mathrm{opt}})/2$.
One of the few exceptions, that may very well be due to
a statistical fluctuation, is the $\Xi$ shown in the figure.
We conclude that the ground state overlap achieved for
the light hadrons is sufficient for our purposes.

\subsection{The static potential in the vacuum}\label{subsec_stat_pot}
\begin{figure}
\includegraphics[width=0.48\textwidth]{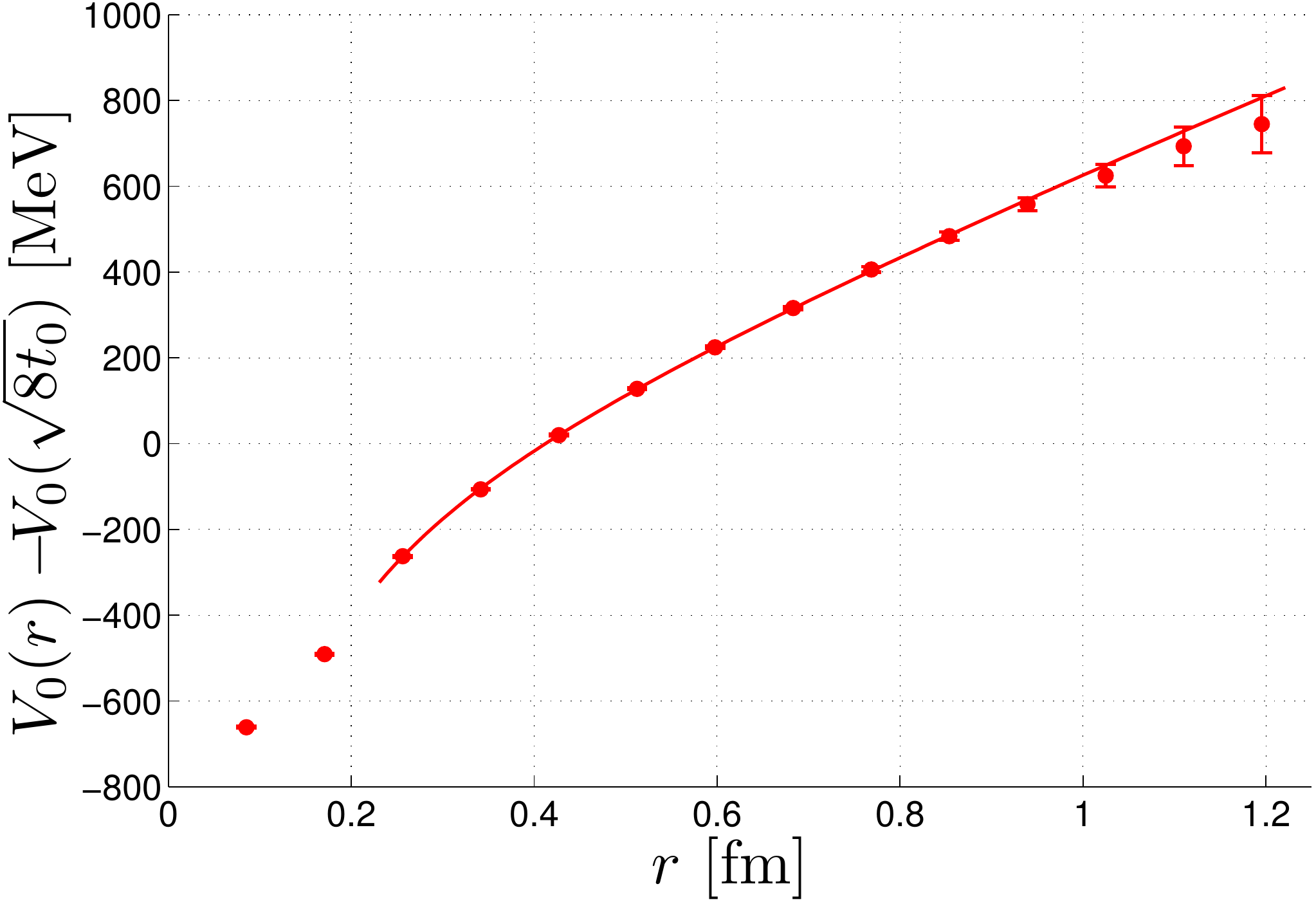}
\caption{The quantity $V_0(r)-V_0(\sqrt{8t_0})$, where $V_0(r)$ denotes the
static quark-antiquark potential in the vacuum, together with the
Cornell fit Eq.~\eqref{eq:cornell}.}
\label{fig:statpot}
\end{figure}

As described in Sec.~\ref{sec_technique}, we determine the static potential, $V_{0}(r)$, from a variational
procedure applied to a matrix of correlation functions consisting of spatially smeared Wilson loops. In
Fig.~\ref{fig:statpot}, we show the physical quantity, $V_0(r)-V_0(\sqrt{8t_0})$,
where the subtraction ensures that the self-energies of the static quarks are removed. The value of $V_0(r)$ at $r = \sqrt{8t_0}$ was obtained
from a local interpolation, cf.~Ref.~\cite{Necco:2001xg}.
For later use we also performed
a fit to the Cornell parametrization~\cite{Eichten:1974af}
\be
\label{eq:cornell}
V_0(r) = \mu - \frac{c}{r} + \sigma r \,,
\ee
where $\mu$ denotes a constant off-set (that diverges
in the continuum limit), $\sigma$ is the string tension
and the Coulomb coefficient reads $c=4\alpha_s/3$ at tree level.
The fit with the parameter values,
\begin{align}
\mu&=0.721(14)\gev\,,\quad
c=0.468(14)\,,\nonumber\\
\label{eq:fitparamspot}
\sigma&=0.906(16)\gev/\fm\,,
\end{align}
where we used $a=0.0854\fm$, is also shown in the figure.

To ensure that our results are not tainted by the breaking of the ``string" between the static quarks,
we only consider the static potential up to $\sim 1.2\fm\approx 14a$, the distance for which string breaking is expected
to occur~\cite{Bali:2005fu,Koch:2015qxr}. At larger distances, the
phenomenological parametrization Eq.~\eqref{eq:cornell}
is no longer valid and additional interpolating operators would be required to extract the true ground state.
From the static potential, we compute the static force $F=V'(r)$ and determine the Sommer scale~\cite{Sommer:1993ce},
$r_{0}\approx 0.5\fm$, from the equation $r^2F(r)|_{r=r_0}=1.65$,
obtaining $r_0/a =5.890 (41)$.
We determine $r_{0}$ from a local interpolation of the static force as it is
explained in \cite{Donnellan:2010mx}.
Indeed, at our lattice spacing and quark mass values, $r_0\approx 5.89 a\approx 5.89\times 0.0854\fm\approx 0.50\fm$.

\subsection{The static potential within a hadron}\label{subsec_energy_diffs}
\begin{figure}
\includegraphics[width=0.48\textwidth]{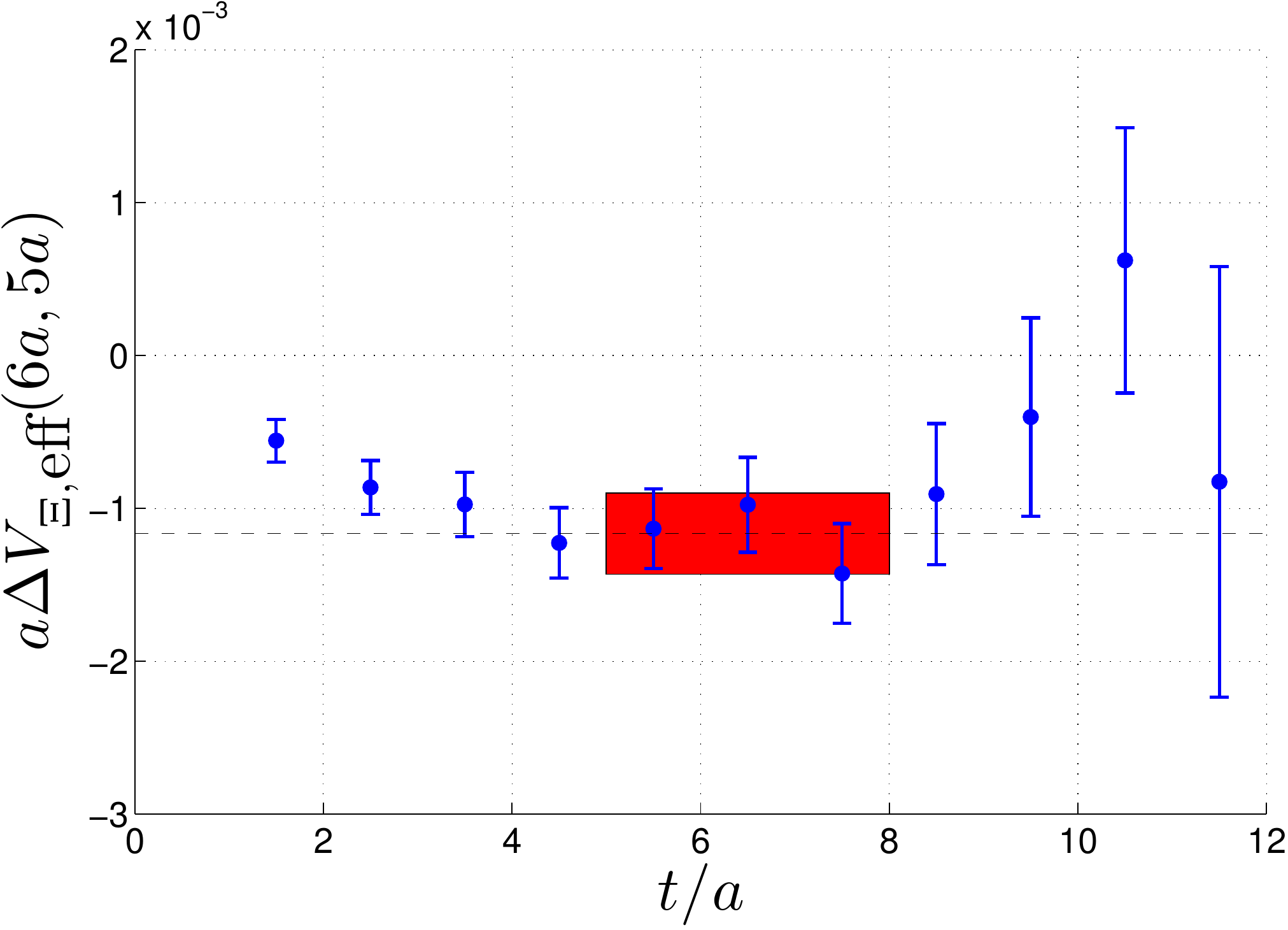}
\caption{Effective energy for $\Delta V_{\Xi}(r=6a,\delta t=5a)$,
defined in Eqs.~\eqref{eq:correlator} and \eqref{eq:energy_shift_dt},
as a function of $t$.
For the definition of effective energies, see Eq.~\eqref{eq:effmassh}.
The error band shows our estimate for $\Delta V(r,\delta t)$, obtained
from a linear fit to $\ln C_H(r, \delta t, t)$.}
\label{fig:effmass1}
\includegraphics[width=0.48\textwidth]{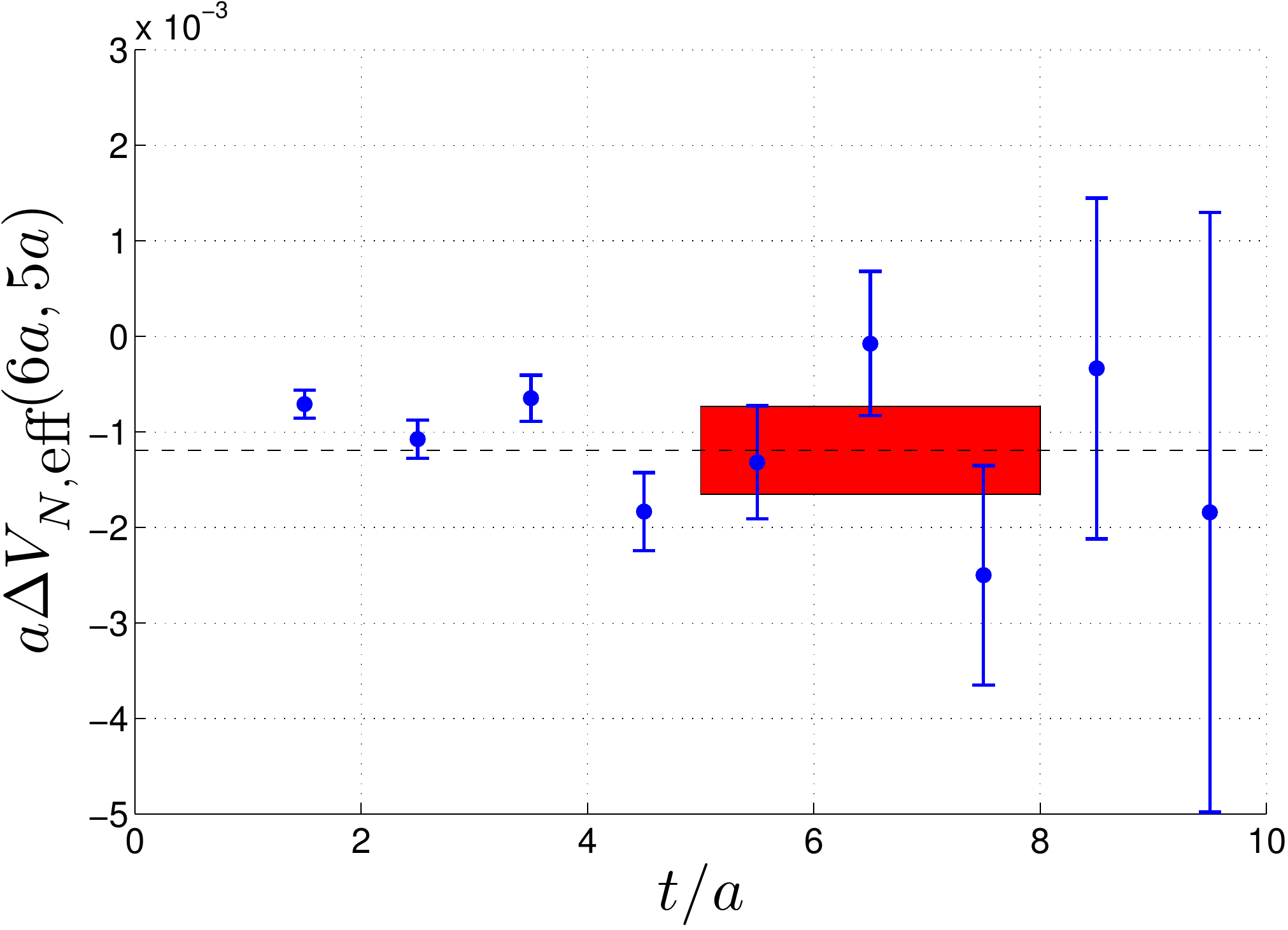}
\caption{The same as in Fig.~\ref{fig:effmass1} for the nucleon.}
\label{fig:effmass2}
\end{figure}

We now determine how the presence of a hadron alters the static potential. As discussed in Sec.~\ref{sec_define},
we compute correlation functions
\be\label{eq:correlator}
C_H(r, \delta t, t) =
\frac{\langle W(r,t)C_{H,\mathrm{2pt}}(t+2\delta t)\rangle}{\langle W(r,t)\rangle
\langle C_{H,\mathrm{2pt}}(t+2\delta t)\rangle}\,,
\ee
where we average over the spatial Wilson loop and hadronic sink positions, for different hadrons $H$. For sufficiently
large values of $t$ and for fixed values of $r$ and $\delta t$, we
can extract the difference between the
static potential in the presence of the hadron,
$V_H(r,\delta t)\stackrel{\delta t\rightarrow\infty}{\longrightarrow} V_H(r)$,
and the vacuum static potential, $V_0(r)$, from the exponential
decay of this function in Euclidean time:
\begin{align}
\nonumber
\Delta V_H(r,\delta t)&\equiv V_H(r,\delta t)-V_0(r)\\
&=-\lim_{t\rightarrow\infty}\frac{\deriv}{\deriv t}\ln [C_H(r, \delta t, t)]\,.
\label{eq:energy_shift_dt}
\end{align}
As the clover term that appears within the Fermionic action
extends one unit in time and we have also
applied one level of four-dimensional HYP smearing to the
Wilson loop, we only consider $\delta t\geq 2a$.
In practice, we obtain statistically meaningful results
for $\delta t\lesssim 8a$, and in some channels even larger
values are possible. Note that within Eq.~\eqref{eq:correlator}
no variational optimization is performed but we restrict
ourselves to our highest
level of twelve spatial HYP smearing iterations for the
Wilson loops.

For a given hadron and for each combination of $r$ and $\delta t$,
we perform linear fits in $t$ to $\ln [C(t, \delta t, t)]$
within the effective energy plateau range. For examples
see Figs.~\ref{fig:effmass1} and \ref{fig:effmass2}, where
we display effective energies for the cascade and the nucleon,
respectively, for $r=6a\approx 0.51\fm$ and $\delta t =5a$,
together with the results of the corresponding fits.
The errors are determined following Ref.~\cite{Wolff:2003sm}.
Below we will assign an additional systematic error to our
results from varying the fit range.

We will approximate $\Delta V_H(r)$ by
$\Delta V_H(r,\delta t =5a)$. The functional form is well described
by the Cornell parametrization
\be
\label{eq:cornell2}
\Delta V_H(r) = \Delta\mu_H - \frac{\Delta c_H}{r}+ \Delta\sigma_H r\,.
\ee
The errors on the fit parameters $\Delta\mu_H$, $\Delta c_H$ and
$\Delta\sigma_H$ which we will quote below will be indicative, since
they only take into account the statistical errors of $\Delta V_H$ and 
neglect their correlations.
Below we summarize our results for the hadron $H$ being
a pseudoscalar or vector meson, a positive parity octet
or decuplet baryon and a negative
parity baryon, respectively.

Note that the $\rho$ and $K^*$ mesons as well as the
negative parity baryons are not stable for our light quark mass value
and lattice volume. However, using only quark-antiquark and three-quark
interpolators, we are unable to detect their decays into pairs of
$p$-wave pions, pion plus kaon and $s$-wave pion plus positive parity baryon,
respectively. As we see effective energy plateaus, we also quote
results for these channels. Clearly, this needs
to be digested with some caution. We also note that
the disconnected quark line contribution was neglected
for the $\phi$ meson.

\begin{table*}
\caption{Values of the difference in the static potential for the mesons,
measured at $\delta t = 5a$. Errors are statistical and systematic, respectively.}
\label{t:shift_mesons}
\begin{ruledtabular}
\begin{tabular}{cccccc}
$r/a$&$\Delta V_{\pi}\,[\mev]$&$\Delta V_{K}\,[\mev]$&$\Delta V_{\rho}\,[\mev]$& $\Delta V_{K^{\star}}\,[\mev]$&$\Delta V_{\phi}\,[\mev]$\\
\hline
1&-0.16(3)(1)&-0.10(3)(1)&-0.07(6)(5)&-0.11(3)(3)&-0.08(2)(3)\\
2&-0.40(8)(4)&-0.24(8)(3)&-0.17(17)(20)&-0.27(8)(7)&-0.22(7)(6)\\
3&-0.80(16)(19)&-0.53(14)(09)&-0.29(33)(56)&-0.50(17)(08)&-0.49(16)(9)\\
4&-1.21(26)(30)&-0.91(24)(18)&-0.46(52)(1.03)&-0.78(28)(21)&-0.85(26)(22)\\
5&-1.71(40)(56)&-1.43(37)(27)&-0.67(73)(1.24)&-1.22(41)(45)&-1.39(38)(49)\\
6&-2.24(61)(71)&-2.02(51)(45)&-1.33(96)(2.09)&-1.91(55)(83)&-2.09(52)(80)\\
7&-2.73(80)(86)&-2.66(68)(71)&-2.03(1.20)(3.19)&-2.48(67)(1.36)&-2.78(66)(1.38)\\
8&-3.27(1.06)(63)&-3.40(89)(1.02)&-2.77(1.46)(4.75)&-3.15(84)(2.24)&-3.43(84)(2.10)
\end{tabular}
\end{ruledtabular}
\end{table*}
\subsubsection{Mesons}\label{mesons}
\begin{figure}
\includegraphics[width=0.48\textwidth]{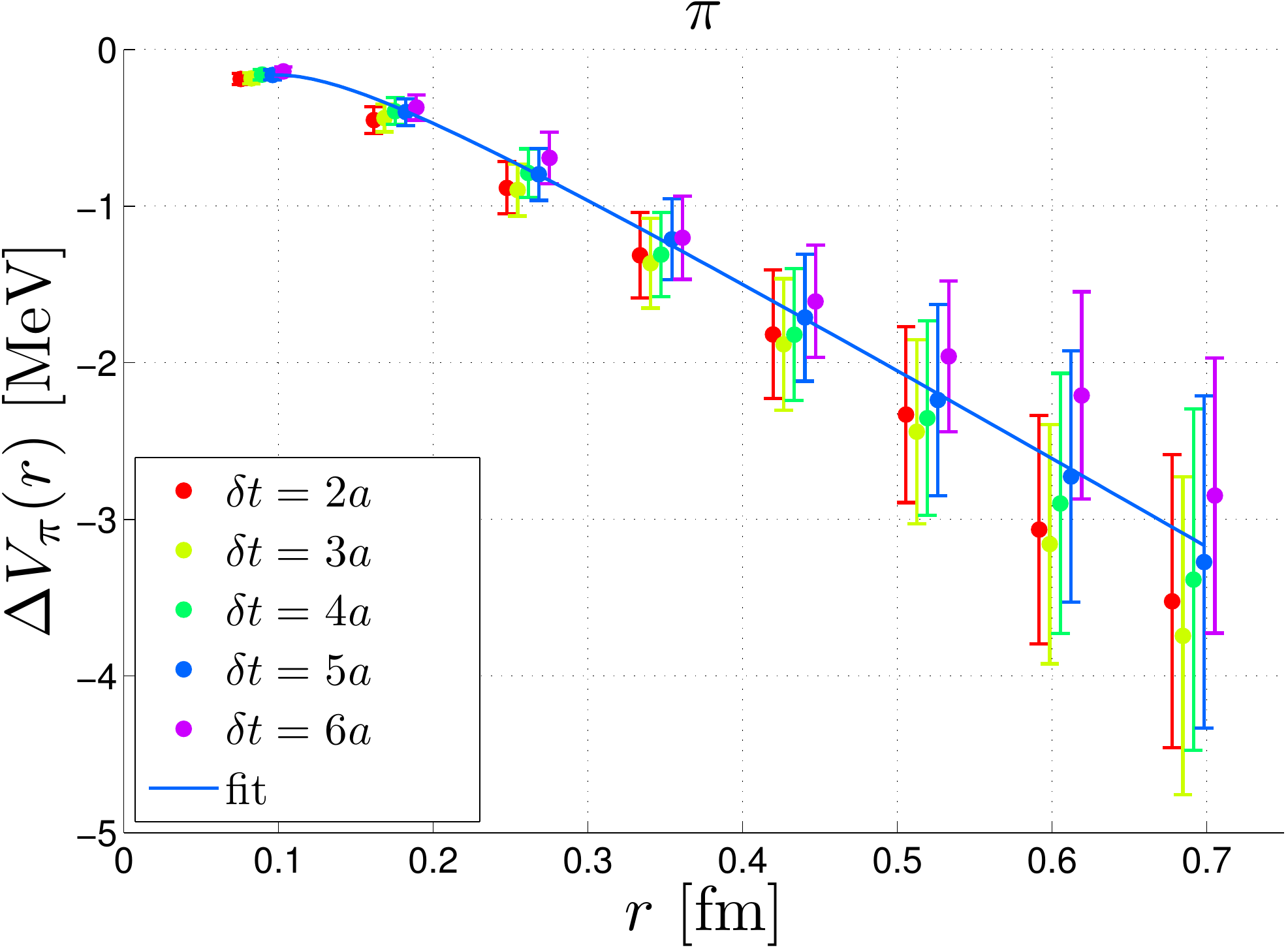}
\caption{The modification to the static potential in the presence of a pion, $\Delta V_{\pi}(r,\delta t)$.
The colour coding corresponds to different values of $\delta t$ as indicated in the legend, where the left-most point within a group corresponds to
$\delta t =2a$. The curve shown is the result of a fit of the $\delta t=5a$ data to Eq.~\eqref{eq:cornell2}.}
\label{fig:shift_pion}
\end{figure}
\begin{figure}
\includegraphics[width=0.48\textwidth]{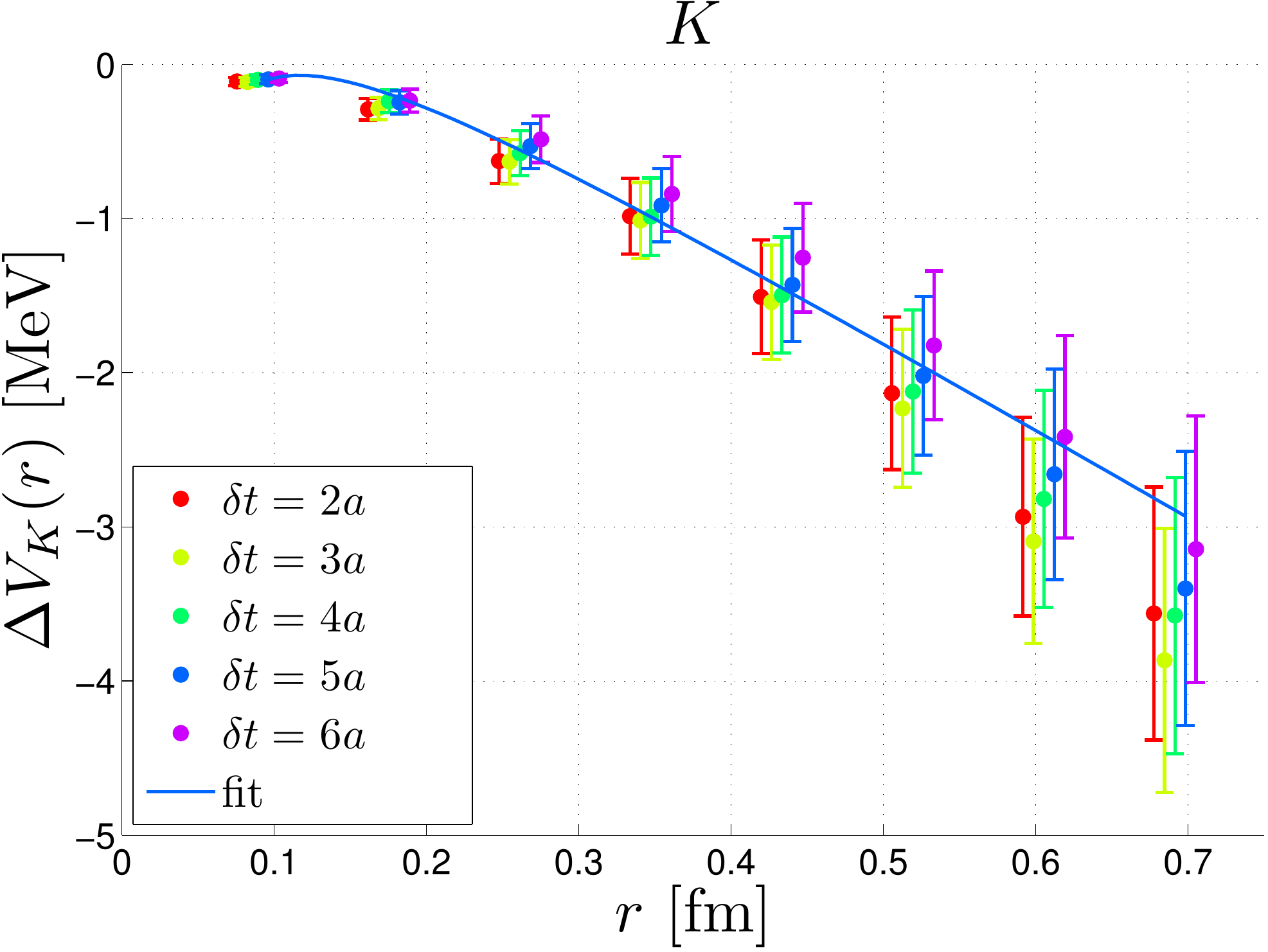}
\caption{The same as in Fig.~\ref{fig:shift_pion} for the kaon.}
\label{fig:shift_kaon}
\end{figure}
\begin{figure}
\includegraphics[width=0.48\textwidth]{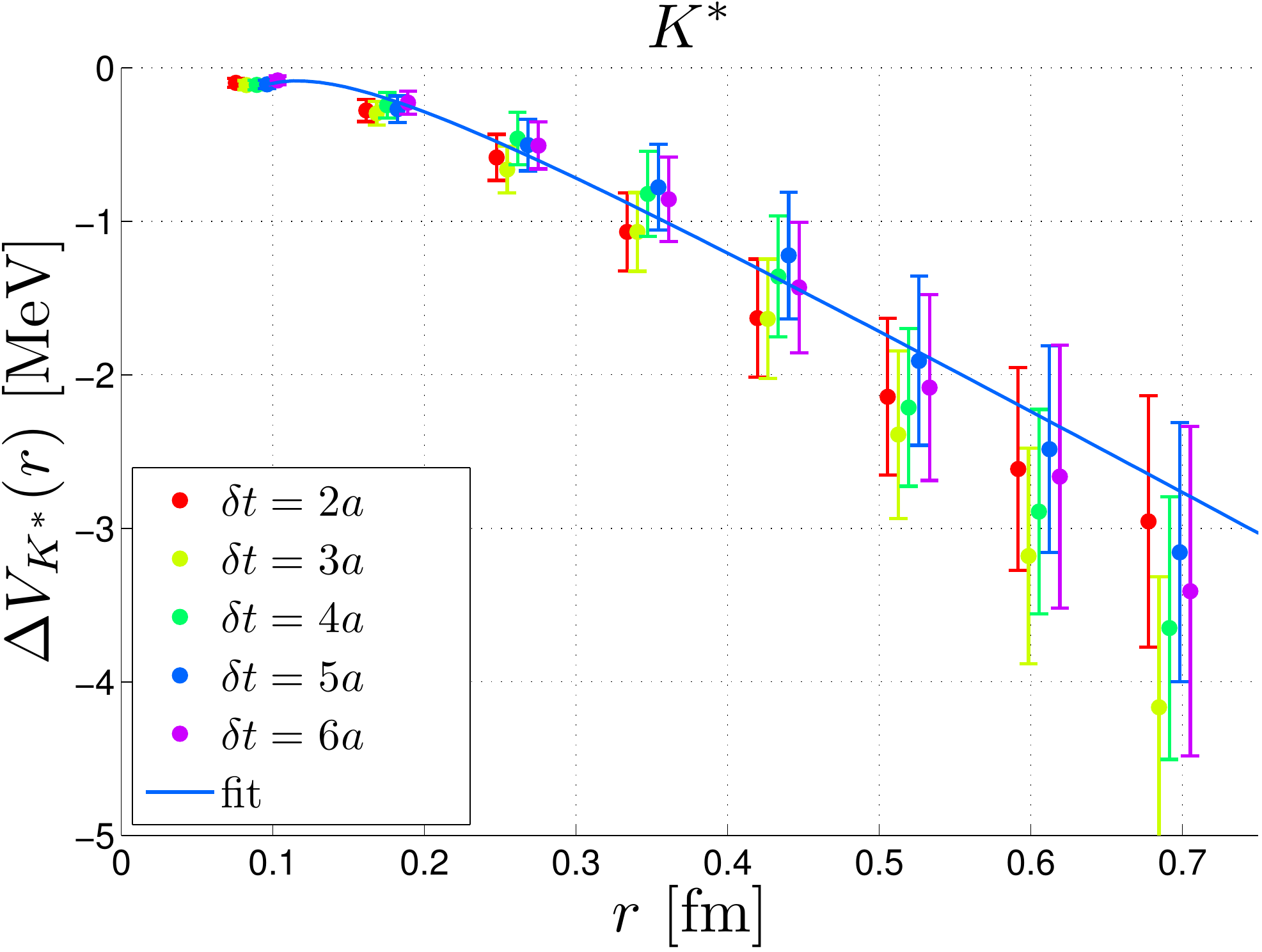}
\caption{The same as in Fig.~\ref{fig:shift_pion} for the $K^{\star}$ meson.}
\label{fig:shift_kstar}
\end{figure}
\begin{figure}
\includegraphics[width=0.48\textwidth]{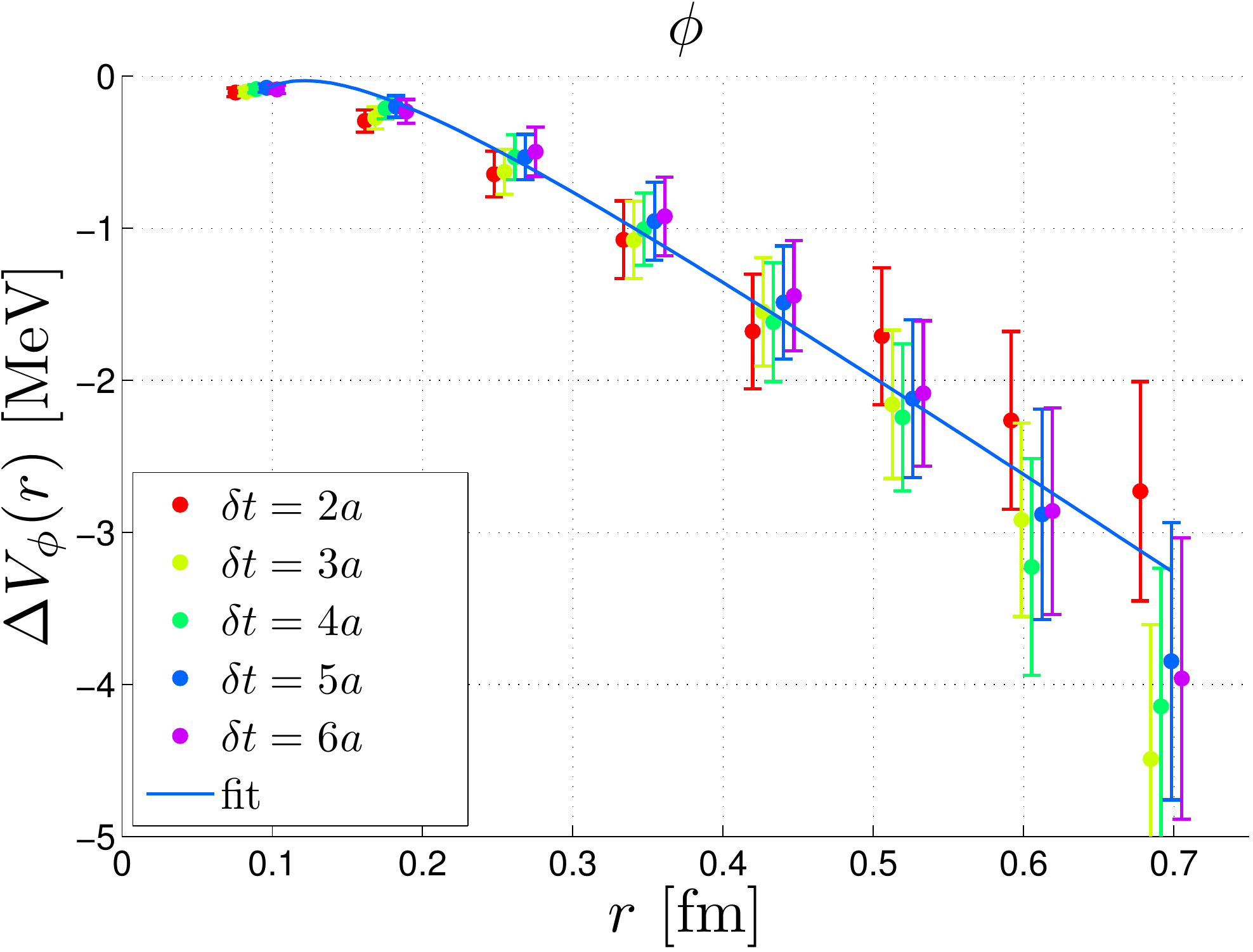}
\caption{The same as in Fig.~\ref{fig:shift_pion} for the $\phi$ meson.}
\label{fig:shift_phi}
\end{figure}
\begin{table}
\caption{Fit parameters for the difference of the potential for the mesons,
see Eq.~\eqref{eq:cornell2}.}
\label{t:shift_fit_mesons}
\begin{ruledtabular}
\begin{tabular}{cccc}
Meson $H$&$\Delta\mu_H\,[\mev]$&$\Delta c_H\,[10^{-4}]$&$\Delta\sigma_H\,[\mev/\fm]$\\\hline
$\pi$&0.858(39)&2.30(13)&-5.75(11)\\
$K$&1.167(15)&3.34(52)&-5.82(42)\\
$\rho$&2.28(38)&6.62(1.31)&-10.19(1.02)\\
$K^{\star}$&1.38(16)&4.10(59)&-6.47(46)\\
$\phi$&1.45(12)&4.18(42)&-6.67(32)
\end{tabular}
\end{ruledtabular}
\end{table}
Several hidden charm resonances such as the $Y(4260)$ have been
interpreted as tightly bound quarkonium states,
embedded within light mesonic matter
\cite{Dubynskiy:2008mq,Li:2013ssa}. Here we follow the procedure described
above to calculate the modification of the
static potential, $\Delta V_H(r,\delta t)$, for several light mesons.

In Figs.~\ref{fig:shift_pion}, \ref{fig:shift_kaon}, \ref{fig:shift_kstar} and \ref{fig:shift_phi},
we show our determinations for the $\pi$, the $K$, the $K^{\star}$ and
the $\phi$ mesons, respectively,
where the colour coding corresponds to different values of $\delta t$
which are displaced horizontally in the plots for clarity.

In all the cases we find $\Delta V_H(r,\delta t) < 0$.
When considering the dependence on the spatial distance between
the static sources, we observe a similar
pattern for all the mesons; the modification of the static potential becomes
more pronounced toward large distances $r$. For distances up to
about $0.3\fm$, we generally find $|\Delta V_H(r,\delta t)| \lesssim 1\mev$,
and at our largest shown
distance of about $0.7\fm$, we always find
$|\Delta V_H(r,\delta t)| \lesssim 4\mev$. The values of
$\Delta V_H(r)$ should be determined from the extrapolation
$\delta t\rightarrow\infty$.
In practice we find that all results for $\delta t \gtrsim 3a$
agree. The numbers obtained for $\delta t =5a$ represent a
compromise between a value of $\delta t$ as large as possible and a
reasonable signal-to-noise ratio.
These should be
considered as our final results and are displayed in
Table~\ref{t:shift_mesons}.

Our data are well described by the parametrization given in
Eq.~\eqref{eq:cornell2}. The resulting fit parameters for the
different mesons are displayed in Table~\ref{t:shift_fit_mesons} and the corresponding
curves are also shown in the figures. Note that, although the
fit parameters appear to indicate a somewhat different behaviour
for the $\rho$ meson, the data points alone, that are displayed in
Table~\ref{t:shift_mesons}, do not show any statistically significant deviation.

We will take the analysis one step further in Sec.~\ref{sec_binding}.
However, taking the above results
at face value, we can already make two interesting observations.
The first one is that, for identical
valence quark content, there is no difference between the tendency
of light pseudoscalars, such as
the pion or the kaon, and vector mesons, such as the $\rho$ or the
$K^{\star}$, to bind with quarkonium.
The second observation is
that there appears to be little or no difference increasing or
decreasing the strangeness of the light mesonic matter.

\begin{figure}
\includegraphics[width=0.48\textwidth]{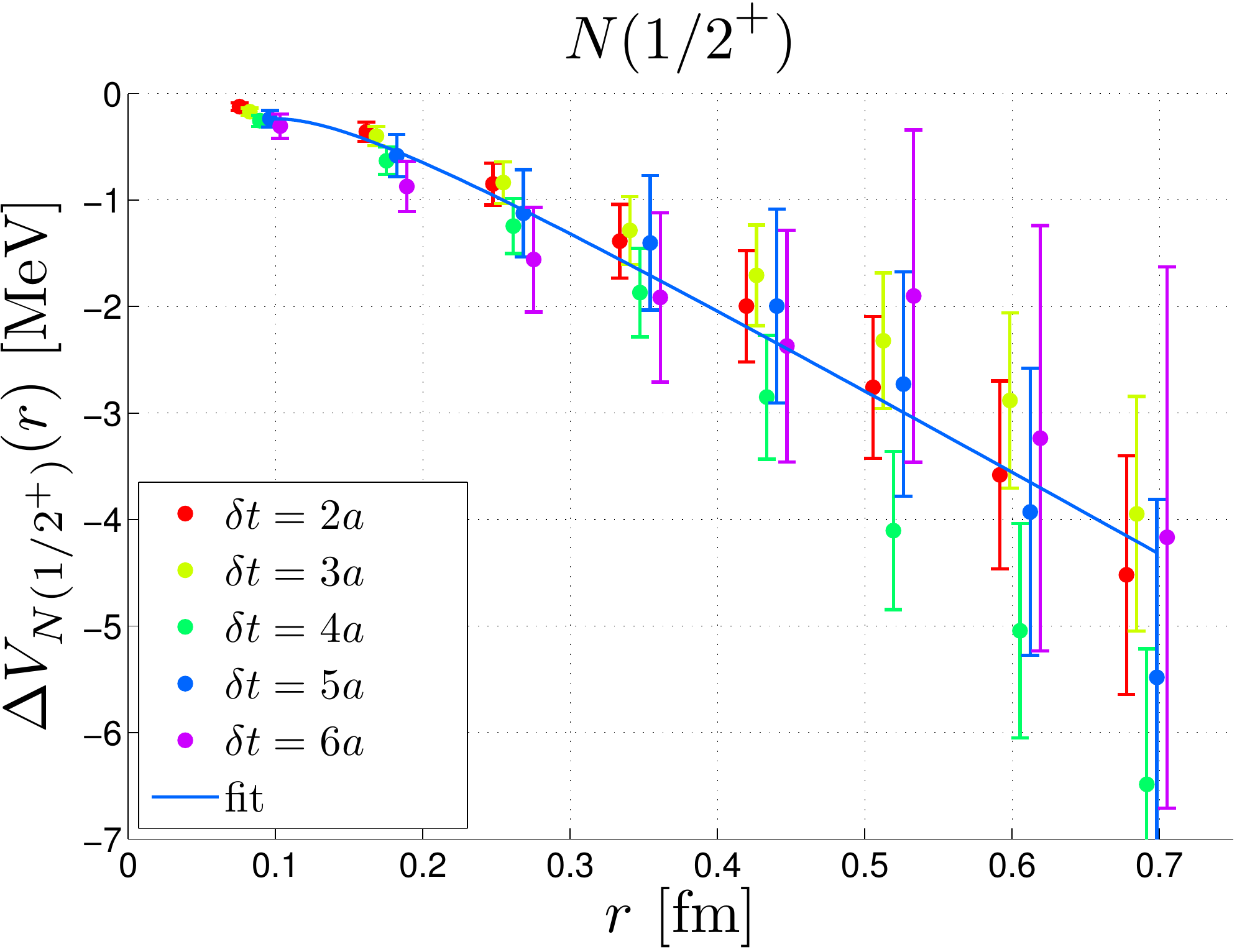}
\caption{The same as in Fig.~\ref{fig:shift_pion} for the positive parity nucleon.}
\label{fig:shift_nucleon_p}
\end{figure}
\begin{figure}
\includegraphics[width=0.48\textwidth]{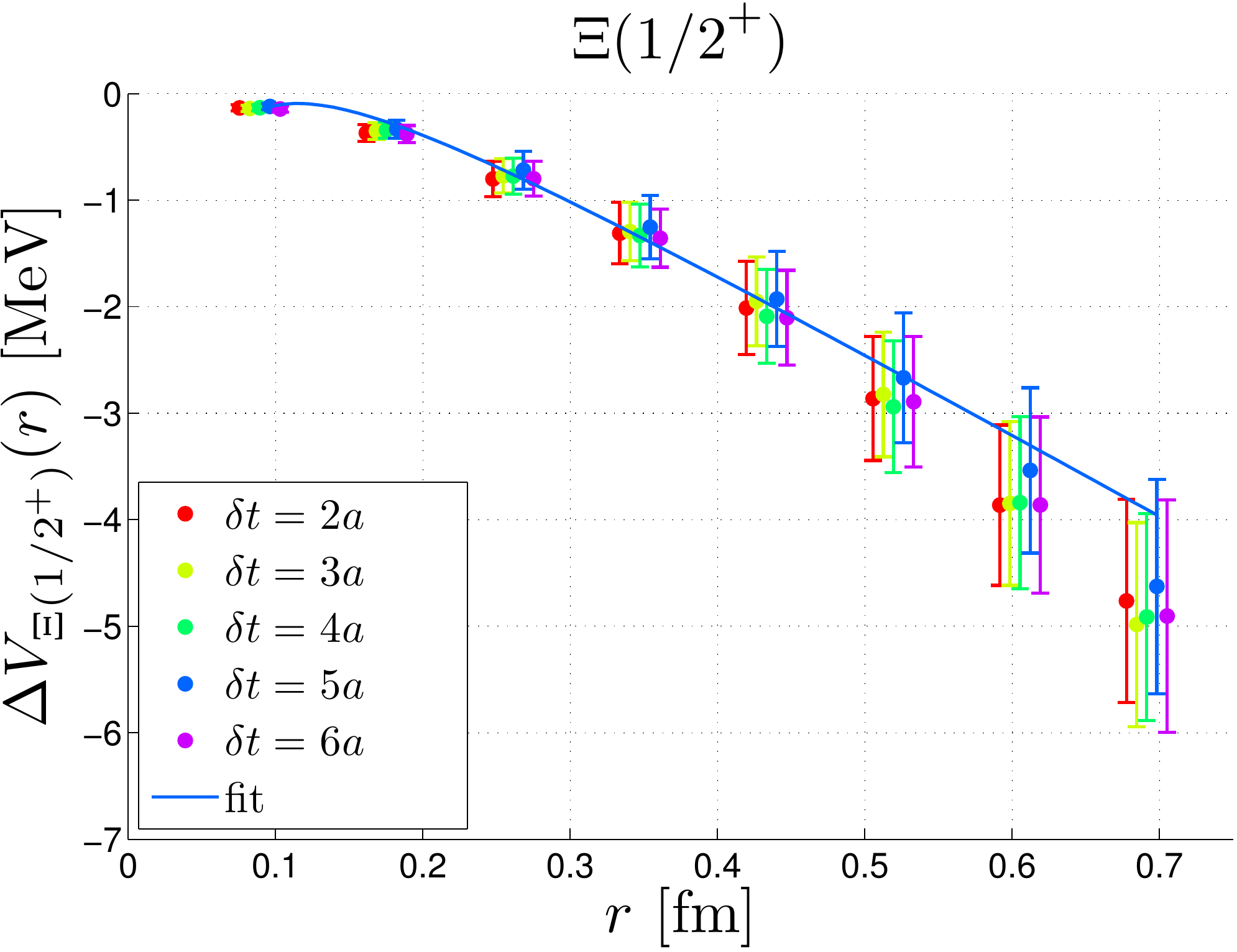}
\caption{The same as in Fig.~\ref{fig:shift_pion} for the positive parity cascade.}
\label{fig:shift_xi_p}
\end{figure}
\begin{figure}
\includegraphics[width=0.48\textwidth]{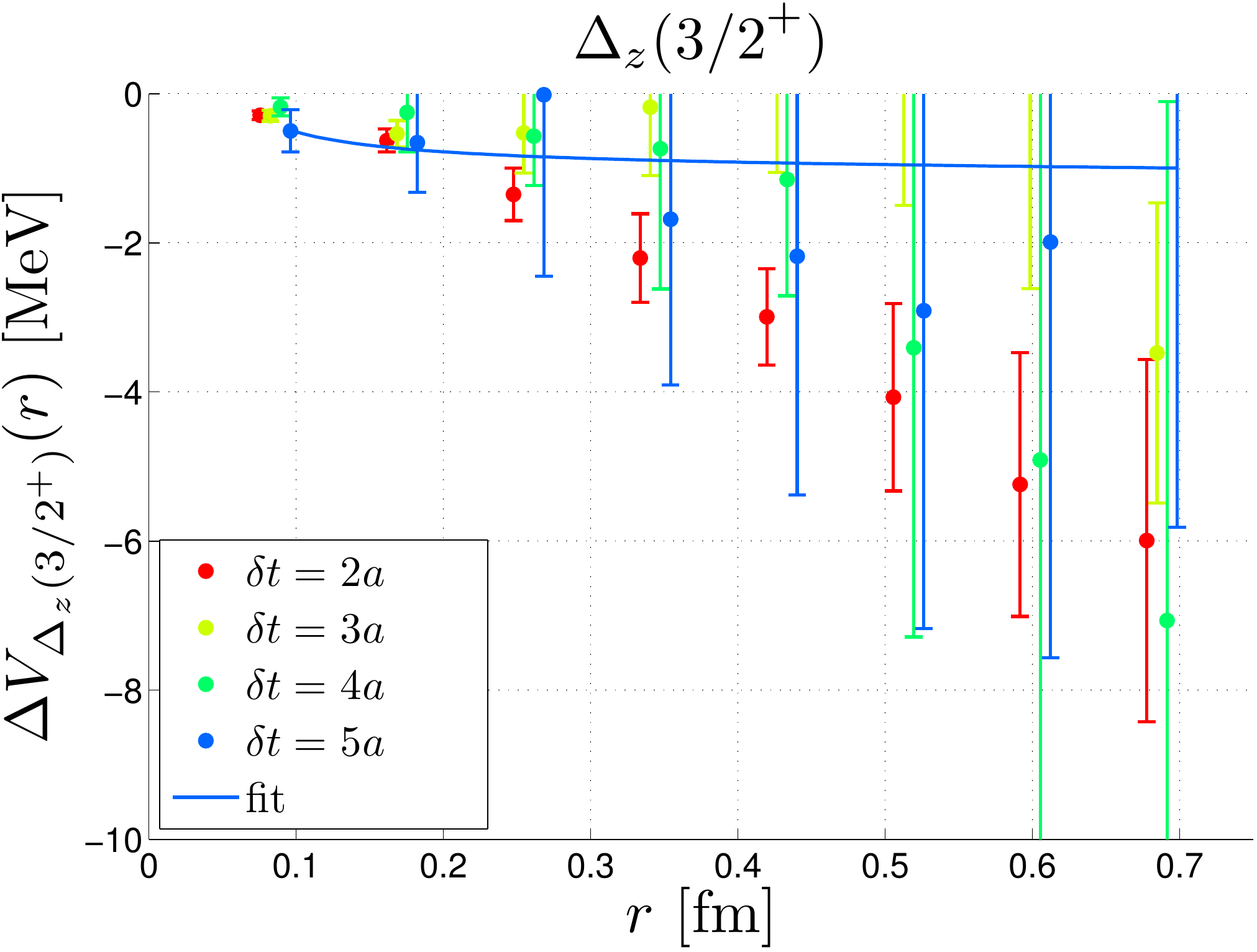}
\caption{The same as in Fig.~\ref{fig:shift_pion} for the positive parity
$\Delta$ baryon. The subscript $z$ of the
baryon label refers to the projection along the $z$ axis $J_z=3/2$,
ensuring no mixing with $J^{P} = 1/2^{+}$ states takes place.}
\label{fig:shift_delta_p}
\end{figure}
\begin{figure}
\includegraphics[width=0.48\textwidth]{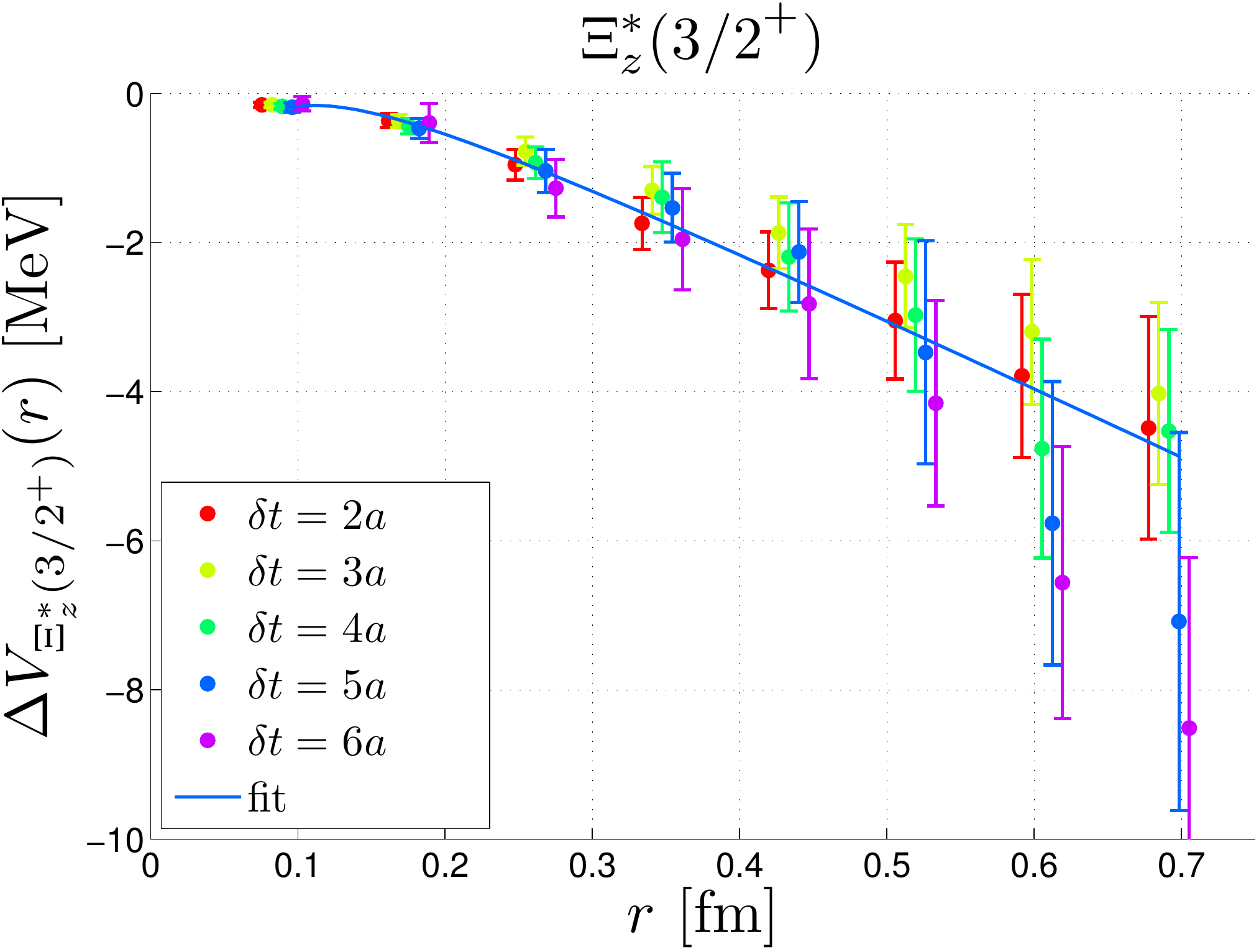}
\caption{The same as in Fig.~\ref{fig:shift_delta_p} for the positive parity $\Xi^{\star}$ baryon.}
\label{fig:shift_xistar_p}
\end{figure}
\begin{table*}
\caption{Values of the difference in the static potential for the positive parity baryons, measured at $\delta t = 5a$.}
\label{t:shift_octet_p}
\begin{ruledtabular}
\begin{tabular}{ccccc}
$r/a$&$\Delta V_{N\left(1/2^+\right)}\,[\mev]$&$\Delta V_{\Sigma\left(1/2^+\right)}\,[\mev]$&$\Delta V_{\Lambda\left(1/2^+\right)}\,[\mev]$&$\Delta V_{\Xi\left(1/2^+\right)}\,[\mev]$\\\hline
1&-0.24(8)(13)&-0.12(3)(10)&-0.24(4)(5)&-0.12(3)(5)\\
2&-0.58(19)(33)&-0.32(9)(20)&-0.60(10)(15)&-0.33(8)(9)\\
3&-1.12(41)(68)&-0.67(20)(38)&-1.12(22)(29)&-0.72(18)(15)\\
4&-1.40(63)(72)&-1.22(32)(33)&-1.64(35)(28)&-1.25(30)(22)\\
5&-1.99(91)(67)&-2.03(49)(54)&-2.49(60)(46)&-1.93(44)(40)\\
6&-2.73(1.05)(1.08)&-2.87(68)(91)&-3.21(80)(59)&-2.67(61)(51)\\
7&-3.93(1.35)(1.53)&-3.62(90)(1.08)&-4.19(1.00)(1.30)&-3.54(78)(1.00)\\
8&-5.48(1.67)(2.28)&-4.40(1.16)(1.47)&-5.34(1.23)(1.84)&-4.63(1.01)(1.80)\\\hline

&$\Delta V_{\Delta\left(3/2^+\right)}\,[\mev]$&$\Delta V_{\Sigma^{\star}\left(3/2^+\right)}\,[\mev]$&$\Delta V_{\Xi^{\star}\left(3/2^+\right)}\,[\mev]$&$\Delta V_{\Omega\left(3/2^+\right)}\,[\mev]$\\
\hline
1&-0.50(28)(46)&-0.23(9)(7)&-0.18(6)(5)&-0.15(4)(5)\\
2&-0.65(66)(58)&-0.49(24)(13)&-0.47(14)(20)&-0.40(10)(16)\\
3&-0.01(2.43)(1.29)&-1.27(52)(31)&-1.04(29)(48)&-0.91(22)(40)\\
4&-1.68(2.22)(1.22)&-1.96(84)(45)&-1.53(46)(72)&-1.45(37)(70)\\
5&-2.18(3.20)(3.04)&-3.27(1.18)(65)&-2.12(67)(99)&-2.07(53)(1.22)\\
6&-2.91(4.26)(3.64)&-5.33(2.81)(1.65)&-3.47(1.50)(1.04)&-3.31(1.00)(1.41)\\
7&-1.99(5.75)(2.12)&-5.41(2.09)(1.86)&-5.76(1.90)(1.87)&-5.70(1.23)(2.24)\\
8&9.5(15.3)(11.4)&-6.14(2.79)(2.02)&-7.08(2.53)(3.02)&-7.35(1.66)(3.37)
\end{tabular}
\end{ruledtabular}
\end{table*}
\begin{table}
\caption{Fit parameters for the difference of the potential for the positive parity
baryons, see Eq.~\eqref{eq:cornell2}.}
\label{t:shift_fit_octet_p}
\begin{ruledtabular}
\begin{tabular}{cccc}
Baryon $H$&$\Delta\mu_H\,[\mev]$&$\Delta c_H\,[10^{-4}]$&$\Delta\sigma_H\,[\mev/\fm]$\\\hline
$N\left(1/2^+\right)$&1.17(37)&3.21(1.30)&-7.83(97)\\
$\Sigma\left(1/2^+\right)$&1.62(21)&4.63(73)&-7.99(60)\\
$\Lambda\left(1/2^+\right)$&1.28(20)&3.46(69)&-8.49(57)\\
$\Xi\left(1/2^+\right)$&1.54(19)&4.32(75)&-7.81(55)\\
$\Delta\left(3/2^+\right)$&-0.99(1.75)&-2.22(6.16)&-0.10(4.77)\\
$\Sigma^{\star}\left(3/2^+\right)$&2.15(37)&6.14(1.30)&-11.38(1.01)\\
$\Xi^{\star}\left(3/2^+\right)$&1.74(36)&4.90(1.41)&-9.40(1.03)\\
$\Omega\left(3/2^+\right)$&2.34(49)&6.77(1.68)&-11.02(1.41)\\
\end{tabular}
\end{ruledtabular}
\end{table}
\begin{table*}
\caption{Values of the difference in the static potential for the negative parity baryons,
measured at $\delta t = 5a$.}
\label{t:shift_octet_n}
\begin{ruledtabular}
\begin{tabular}{ccccc}
$r/a$&  $\Delta V_{N\left(1/2^-\right)}\,[\mev]$  & $\Delta V_{\Sigma\left(1/2^-\right)}\,[\mev]$  & $\Delta V_{\Lambda\left(1/2^-\right)}\,[\mev]$ &  $\Delta V_{\Xi\left(1/2^-\right)}\,[\mev]$\\\hline
1& -1.73(91)(43)&-0.68(50)(77)&-1.09(48)(95)&-0.31(26)(29)\\
2&-4.19(2.30)(1.74)&-1.46(1.27)(1.61)&-2.59(1.24)(1.92)&-0.60(69)(51)\\
3&-8.43(5.48)(4.19)&-2.93(2.59)(3.16)&-5.31(2.69)(3.68)&-1.13(1.41)(58)\\
4&-12.27(7.53)(5.25)&-5.12(4.06)(4.76)&-8.08(4.12)(6.48)&-3.29(2.28)(2.08)\\
5&-12.7(11.4)(9.7)&-6.21(6.01)(6.09)&-12.1(6.1)(10.1)&-5.28(3.32)(3.80)\\
6&1.1(12.7)(22.9)&-5.80(8.06)(5.20)&-14.2(6.4)(9.6)&-8.93(4.61)(6.57)\\
7&4.6(18.7)(31.8)&-7.3(10.2)(5.9)&-7.36(6.80)(7.91)&-9.16(5.75)(6.29)\\
8&0.9(24.8)(35.4)&-9.1(12.9)(7.1)&-3.9(8.8)(19.5)&-6.65(7.10)(5.83)\\\hline
&$\Delta V_{\Delta\left(3/2^-\right)}\,[\mev]$  &  $\Delta V_{\Sigma^{\star}\left(3/2^-\right)}\,[\mev]$ & $\Delta V_{\Xi^{\star}\left(3/2^-\right)}\,[\mev]$ & $\Delta V_{\Omega\left(3/2^-\right)}\,[\mev]$\\
\hline
1&0.98(7.20)(1.66)&-0.70(1.22)(0.28)& 0.01(28)(25)&-0.11(20)(10)\\
2&0.1(16.7)(7.5)&-3.74(4.87)(1.49)&-1.72(1.52)(1.02)&-0.46(47)(13)\\
3&1.0(29.2)(16.1)&-8.06(5.46)(3.29)&-2.47(2.17)(1.57)&-0.64(96)(55)\\
4&49.3(64.4)(7.6)&-13.8(8.5)(3.9)&-2.47(1.79)(0.83)&-0.48(1.50)(1.30)\\
5&35.8(95.8)(12.5)&-12.6(6.8)(12.6)&-3.26(2.48)(3.05)& 0.01(2.11)(2.11)\\
6&56.9(96.8)(17.2)&-11.0(32.4)(4.3)&-4.35(3.19)(5.40)&-0.46(2.87)(2.76)\\
7&1.4(132.0)(34.6)&-8.1(42.7)(9.7)&-6.06(3.99)(5.70)&-1.62(3.73)(2.77)\\
8&-25.6(160.0)(49.7)&-21.5(13.2)(11.9)&-8.09(4.90)(6.33)&-3.33(4.77)(3.23)
\end{tabular}
\end{ruledtabular}
\end{table*}
\begin{figure}
\includegraphics[width=0.48\textwidth]{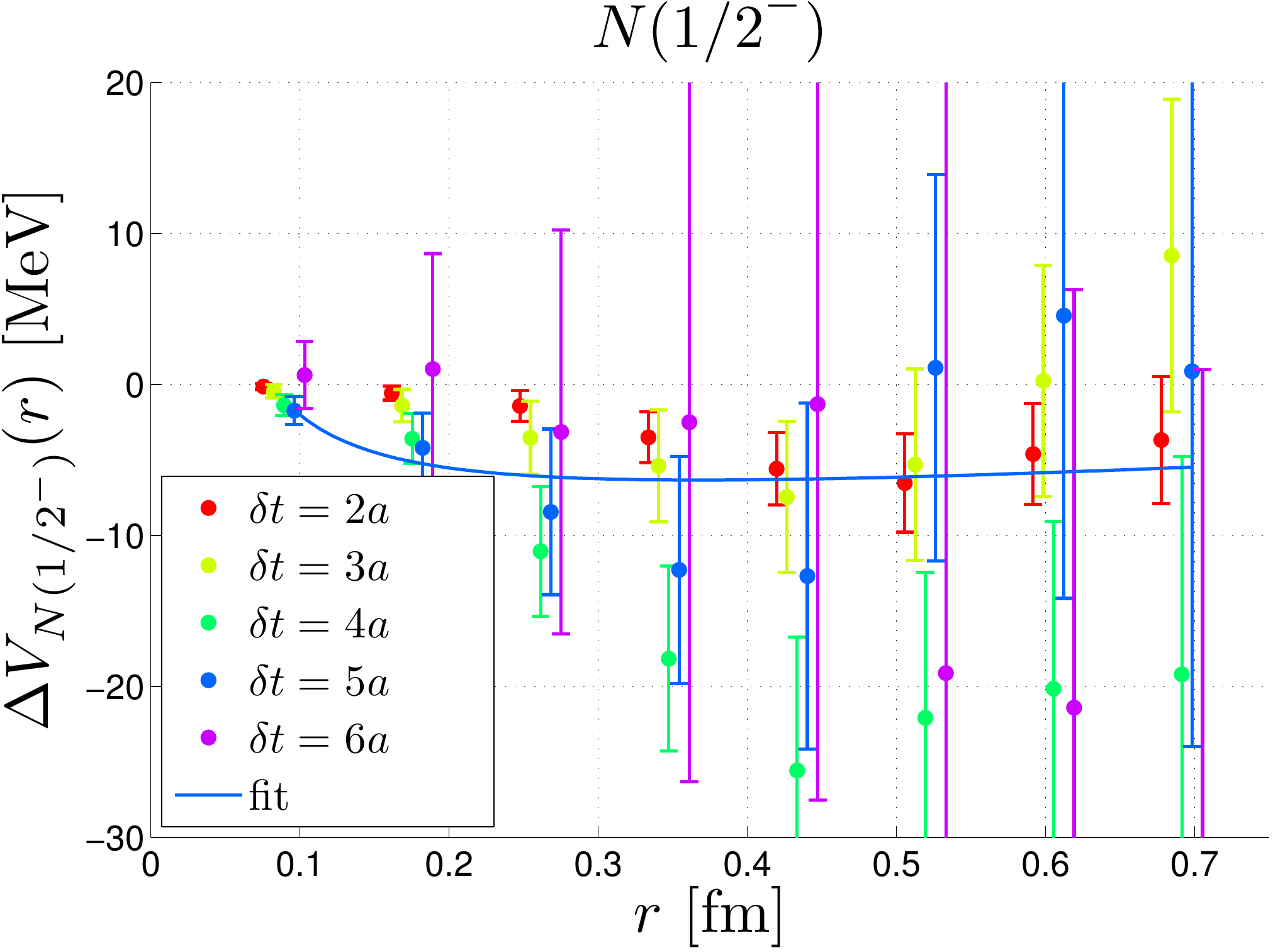}
\caption{The same as in Fig.~\ref{fig:shift_pion} for the negative parity nucleon.}
\label{fig:shift_nucleon_n}
\end{figure}
\begin{figure}
\includegraphics[width=0.48\textwidth]{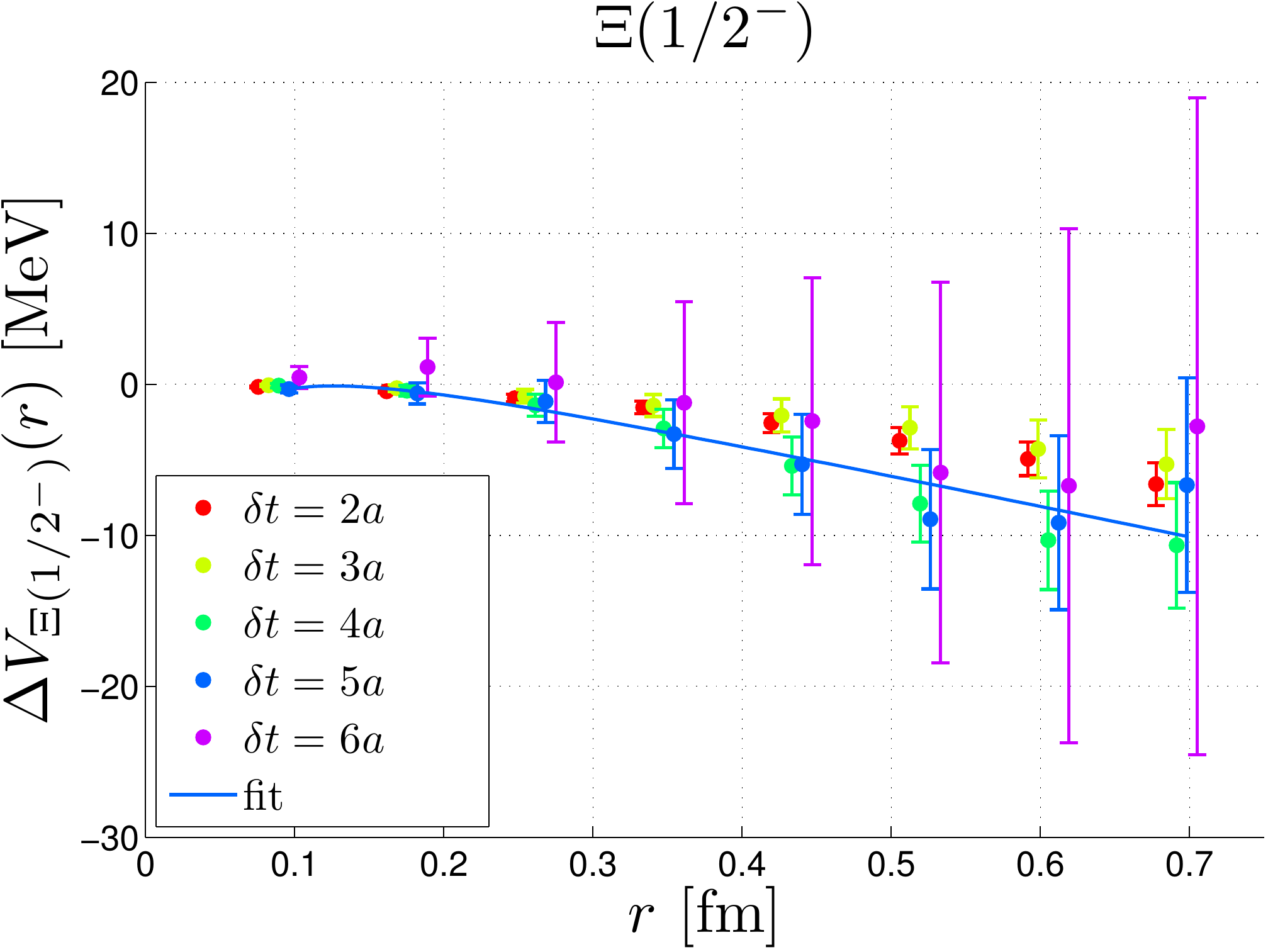}
\caption{The same as in Fig.~\ref{fig:shift_pion} for the negative parity cascade.}
\label{fig:shift_xi_n}
\end{figure}
\begin{figure}
\includegraphics[width=0.48\textwidth]{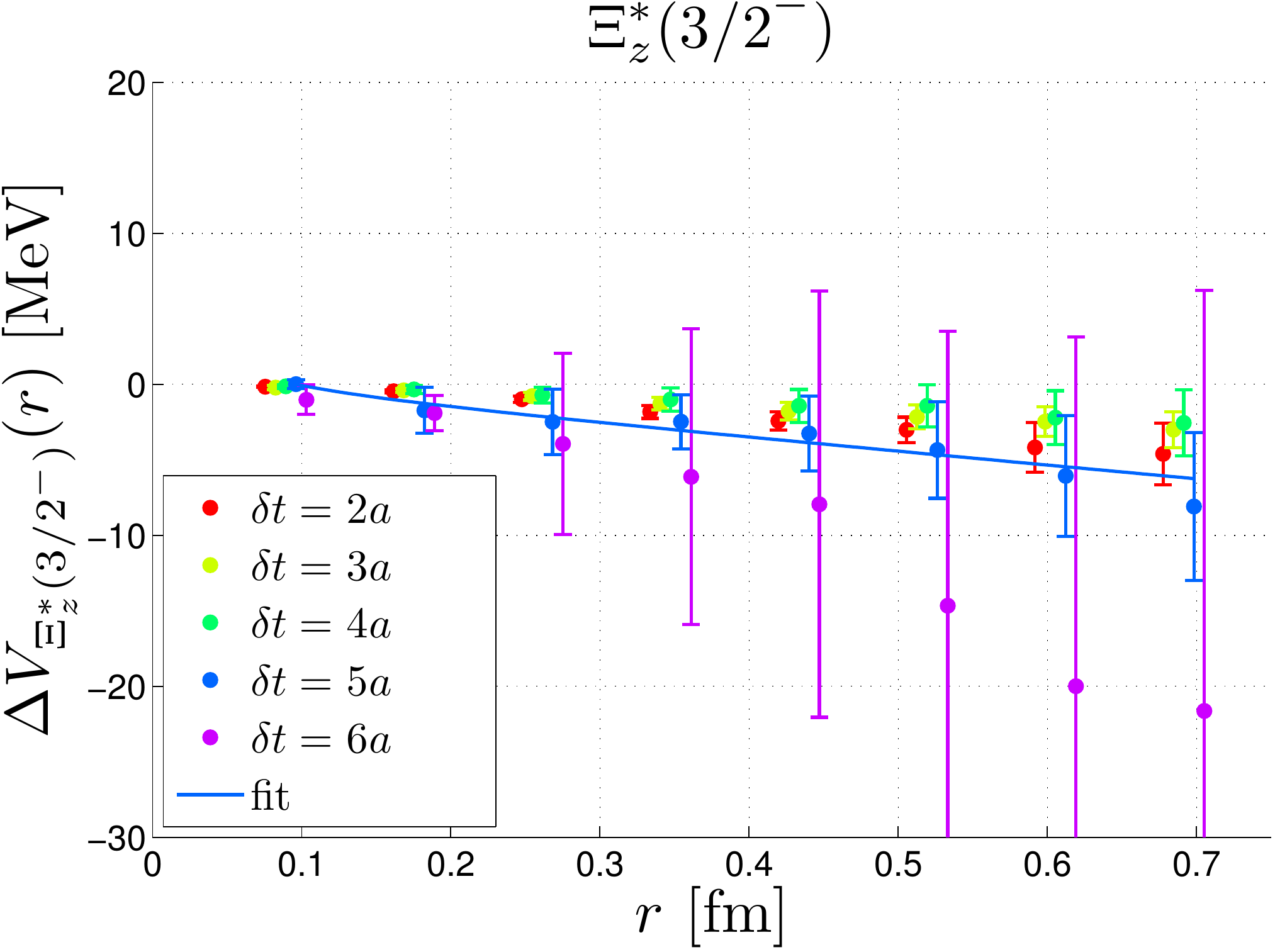}
\caption{The same as in Fig.~\ref{fig:shift_delta_p} for the negative parity $\Xi^{\star}$ baryon.}
\label{fig:shift_xistar_n}
\end{figure}
\begin{table}
\caption{Fit parameters for the difference of the potential for the negative parity
baryons, see Eq.~\eqref{eq:cornell2}.}
\label{t:shift_fit_octet_n}
\begin{ruledtabular}
\begin{tabular}{cccc}
Baryon $H$&$\Delta\mu_H\,[\mev]$&$\Delta c_H\,[10^{-4}]$&$\Delta\sigma_H\,[\mev/\fm]$\\\hline
$N\left(1/2^-\right)$&-10.18(6.43)&-3.50(22.58)&5.39(17.46)\\
$\Sigma\left(1/2^-\right)$&1.88(83)&4.84(2.91)&-16.89(2.26)\\
$\Lambda\left(1/2^-\right)$&-0.77(3.51)&-5.03(12.61)&-16.92(8.93)\\
$\Xi\left(1/2^-\right)$&4.74(1.18)&14.21(4.11)&-20.93(3.25)\\
$\Delta\left(3/2^-\right)$&-23.1(27.6)&-69.34(96.85)&94.8(76.2)\\
$\Sigma^{\star}\left(3/2^-\right)$&-0.853(3.26)&-12.11(11.90)&-30.3(7.22)\\
$\Xi^{\star}\left(3/2^-\right)$&-0.23(1.25)&-4.37(4.59)&-8.92(2.61)\\
$\Omega\left(3/2^-\right)$&-0.47(62)&-1.92(2.18)&-0.99(1.67)\\
\end{tabular}
\end{ruledtabular}
\end{table}
\subsubsection{Positive parity baryons}\label{positive}
We now turn our attention to modifications of the static potential in the presence of a
positive parity octet ($J^P=1/2^+$) or decuplet ($J^P=3/2^+$) baryon. As explained
at the end of Sec.~\ref{sec_technique}, in the latter case we are restricted to
employing a particular polarization to avoid mixing with $J=1/2$ states. In our case
we project onto $J_z=3/2$ with respect to the $z$ axis. We remark that embedding charmonium states within baryons
of vanishing strangeness could be an interpretation of the ``pentaquark'' structures that
were recently reported by the LHCb Collaboration~\cite{Aaij:2015tga,Aaij:2016phn}, for examples see the last paragraph of Sec.~\ref{sec_define}.

In Figs.~\ref{fig:shift_nucleon_p}, \ref{fig:shift_xi_p}, \ref{fig:shift_delta_p} and \ref{fig:shift_xistar_p} we show
$\Delta V_H(r,\delta t)$ for the nucleon, the cascade $\Xi$, the $\Delta$
and the decuplet cascade $\Xi^*$, respectively.
Again, in all the cases we observe $\Delta V_H(r,\delta t) < 0$.
The results for the positive parity baryons are collected in Table~\ref{t:shift_octet_p}
and are very similar to the values discussed above for the pseudoscalar and vector mesons.
Note however, that the errors of $\Delta V$ in the presence of decuplet baryons become
rather large. In particular, this is so for the $\Delta$, which is why in this
case we only show the data up to $\delta t =5a$.
The Cornell fit parameters Eq.~\eqref{eq:cornell2} are displayed in Table~\ref{t:shift_fit_octet_p}.

\subsubsection{Negative parity baryons}\label{decuplet}
The modification of the potential in the presence of negative parity baryons appears statistically consistent to
the positive parity case, however, due to the much larger statistical errors, we cannot exclude
a more rapid decrease of $\Delta V_H(r)$ as a function of $r$.
As examples we show in Figs.~\ref{fig:shift_nucleon_n}, \ref{fig:shift_xi_n} and \ref{fig:shift_xistar_n}
the results for the negative parity partners of the nucleon, the cascade and the decuplet cascade, respectively.
The corresponding numerical values for $\delta t=5a$ are displayed in Table~\ref{t:shift_octet_n} and
the Cornell fit parameters in Table~\ref{t:shift_fit_octet_n}.

\subsubsection{Summary}
\begin{figure}
\includegraphics[width=0.48\textwidth]{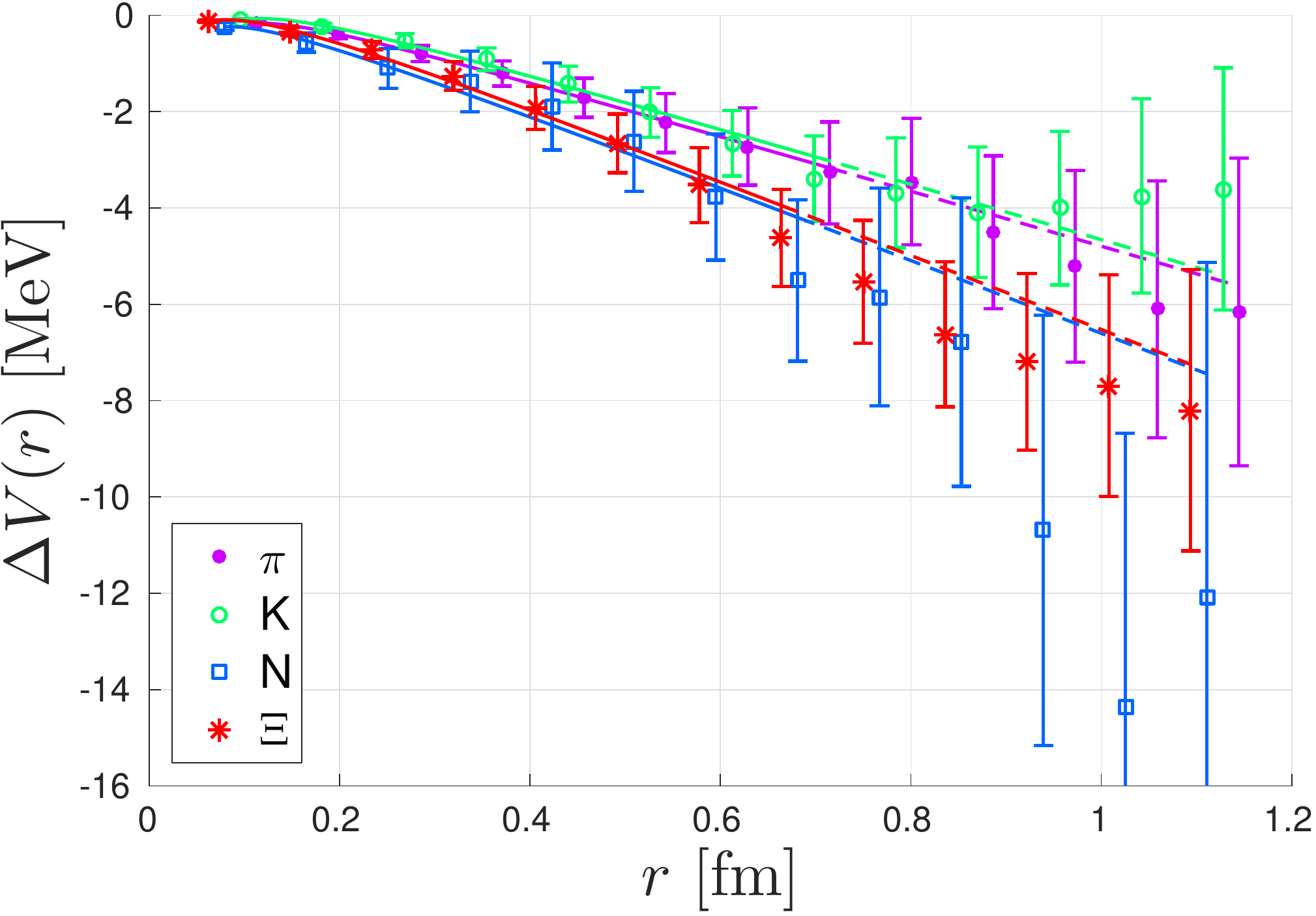}
\caption{The difference in the static potential for the pion, the kaon
and the positive parity nucleon and cascade,
measured at $\delta t = 5a$, up to a distance of $1.2\,\fm$.
The curves correspond to the parametrization Eq.~\eqref{eq:cornell2}
with the parameters (obtained by fitting the $r<0.7\,\fm$ data points)
displayed in Tables
\ref{t:shift_fit_mesons} and \ref{t:shift_fit_octet_p}.}
\label{fig:largedist}
\end{figure}

Regardless of meson or baryon, spin, strangeness or parity, the modifications
of the static potential are well described by the parametrization
Eq.~\eqref{eq:cornell2},
with the main effects being a reduction of the linear slope and increases
of the Coulomb coefficient $c$ and of the off-set $\mu$. All data are consistent
with a decrease of the static potential at the distance $r=0.5\fm$ by about
$2$--$3\mev$.

For $r>0.7\,\fm$ the statistical errors grow substantially as a result
of the deteriorating signal-to-noise ratio. Fortunately, larger distances
exceed the size both of charmonium and of the hosting hadron and will
not be relevant for the discussion of Sec.~\ref{sec_binding} below.
However, one may wonder if the reduction persists. In
Fig.~\ref{fig:largedist} we show the data for the pion, the kaon,
the nucleon and the cascade up to $r\approx 1.2\,\fm$, a distance
around which string breaking
will occur~\cite{Bali:2005fu,Koch:2015qxr}. The decrease of the slope
appears to be robust and all large distance data points are consistent with our
parametrizations. However, for the more compact pseudoscalar mesons and
in particular the kaon the data suggests that above $r\approx 0.8\,\fm$
some saturation may set in.

\section{Modification of charmonium binding energies}\label{sec_binding}
We have investigated how the static quark-antiquark potential
changes in the
presence of a light hadron. This is a well-defined observable and
the results by themselves are already interesting. However,
we wish to go one step further and
address possible phenomenological consequences. We start with a few
words of caution. When it comes to charmonia (and
even for bottomonia), relativistic corrections are not small.
Moreover, baryons are not particularly light
in comparison to the charm quark. Therefore, for charmonia it
may be doubtful if their effect can be completely integrated
out in a Born-Oppenheimer or adiabatic spirit and put into
the quark-antiquark interaction potential. This is less
of a problem for the pion and the kaon since
$\mK/m_c$ and $\mpi/m_c$ are of similar sizes as the
squared velocity $v^2\sim 0.3$. In what follows, we will
neglect these effects.

We start from the Schr\"odinger equation
\be
\left[-\frac{\boldsymbol{\nabla}^2}{m_c}+E_H(r)\right]\psi^{(H)}_{nL}(r,\theta,\phi)
=M_{nL}^{(H)}\psi^{(H)}_{nL}(r,\theta,\phi)\,,
\ee
where the reduced mass is $m_c/2$ and
\begin{align}
E_H(r)&=2(m_c-\delta m)+V_H(r)\\
&=2m_c+v_0+\Delta \mu_H-\frac{c_H}{r}+\sigma_Hr\,.
\end{align}
In the second step, we have assumed the Cornell
parametrization given by Eqs.~\eqref{eq:cornell} and \eqref{eq:cornell2},
where we set
$c_H=c+\Delta c_H$ and $\sigma_H=\sigma+\Delta \sigma_H$.
The parameters $\Delta\mu_H$, $\Delta c_H$ and $\Delta\sigma_H$
specify the modifications of
the constant, the Coulomb and the linear terms, respectively,
obtained from the Cornell fits to $\Delta V_H(r)=V_H(r)-V_0(r)$
carried out in the previous section.

The Cornell parametrization is not valid at large distances due to
string breaking effects~\cite{Bali:2005fu,Koch:2015qxr} or
at small distances where one would expect the coefficient
$c_{H}$ to run with the scale $r$. However,
we are only interested in mass differences $\Delta M_{nL}^{(H)}=
M_{nL}^{(H)}-M^{(0)}_{nL}$ between a charmonium state with radial
and angular momentum quantum numbers $n$ and $L$ respectively,
in the presence of a hadron $H$, relative to the same
state in the vacuum. We expect such corrections to
affect both masses in similar ways, and therefore to cancel from
these differences.
The coefficients $\Delta\mu_H$, $c_H$ and $\sigma_H$
are taken from the fits performed in the previous section, while the
mass parameter $m_c$ and the
offset $v_0=\mu-2\delta m$ have to be fixed by matching the
energy levels $M_{nL}=M^{(0)}_{nL}$, obtained from solving the
above Schr\"odinger equation, to experiment.

Due to the approximations made, our discussion
can only be qualitative and hence we neglect our statistical
and systematic uncertainties.
The central values for the parameters from the Cornell
fit to the static potential in the vacuum read (see also
Eq.~\eqref{eq:fitparamspot}),
\be
\label{eq:statp}
\sigma=0.0335a^{-2}\approx \left(423\mev\right)^2\,,\quad
c=0.468\,.
\ee
Numerically solving the Schr\"odinger equation and adjusting $m_c$
and $v_0$ so that we reproduce the spin averaged
$1S$ and $2S$ charmonium levels, we find
\be
m_c=1269\mev\,,\quad v_0=113\mev\,.
\label{eq:mc}
\ee
From Table~\ref{tab_schroed}, we see that the above
parameters indeed reproduce the experimental $1S$ and $2S$ levels,
however, we underestimate the $1P$ mass by $42\mev$.
This is due to a combination of overestimating the value of the
wave function at the origin, as we neglected running coupling
effects, and relativistic corrections~\cite{Bali:1998pi}; within
our approximations, it is not possible to simultaneously reproduce all
spin-independent mass splittings within an accuracy better than 
about $10\%$.

A negative value of $\Delta M^{(H)}_{nL}$
means that embedding a charmonium state within the hadron $H$ is
energetically favourable, which we interpret as attraction.
Unlike in the hydrogen case, the potential is only bound from above by the
$D\overline{D}$ threshold and so it may not be entirely obvious
whether a negative $\Delta V_H(r)$ results
in a positive or a negative shift of the charmonium mass.
On one hand, a lower $V_H$ results in a lower $E_H$ and therefore in a
smaller $M^{(H)}_{nL}$ mass. On the other hand, the slope is reduced,
resulting in a more extended and less strongly bound wave function.

Before numerically solving the Schr\"odinger equation
we investigate a toy model with a purely linear potential
$V(r)=\sigma r$. The virial theorem then gives a kinetic energy
\be
2\langle T\rangle =\langle r\, {\rm d}V/{\rm d}r\rangle
=\sigma\langle r\rangle=2M -2\sigma\langle r\rangle\,,
\ee
where we used
$M=\langle T\rangle +\langle V\rangle=\langle T\rangle+\sigma \langle r\rangle$.
This means that $\langle r\rangle = 2M/(3\sigma)$.
The Feynman--Hellmann theorem then gives
\be
\frac{\partial M}{\partial \sigma}=
\left\langle\frac{\partial H}{\partial\sigma}\right\rangle
=\langle r\rangle =\frac{2M}{3\sigma}\,,
\ee
i.e.\
\be
\Delta M^{(H)}=(\sigma_H-\sigma)\left.
\frac{\partial M}{\partial\sigma}\right|_{\sigma=\sigma_0}
=\frac{2\sigma_H}{3\sigma_0}M^{(0)}\,,
\ee
where $M^{(H)}=M(\sigma_H)$. Therefore, we expect the part of
the mass which is due to the interaction, $M-2(m_c-\delta m)$,
to be lowered by a factor $2\sigma_{H}/(3\sigma)$, which
for our data typically amounts to about $0.4\%$.
As we have neglected Coulomb interactions, we should also
neglect the self-energy $\delta m$. Then, using the
$m_c$ value of Eq.~\eqref{eq:mc} and $M_{1S}=3069\mev$, this difference
gives $530\mev$. So, for the $1S$ state,
we expect an attraction $\Delta M^{(H)}_{1S}\approx -2\mev$.
Using the experimental $1P$--$1S$ and $2S$--$1S$ differences
lowers this to $\Delta M^{(H)}_{1P}\approx -3.9\mev$ and
$\Delta M^{(H)}_{2S}\approx -4.5\mev$, respectively.

\begin{table}
\caption{Masses and mass differences of spin-averaged states
in MeV taken from experiment~\cite{Agashe:2014kda} and from solving the
Schr\"odinger equation using the Cornell parametrization of our
lattice results.\label{tab_schroed}}
\begin{ruledtabular}
\begin{tabular}{cccc}
Mass/Mass difference&$1S\,[\mev]$&$1P\,[\mev]$&$2S\,[\mev]$\\\hline
$M_{nL}$ (experiment)&3068.6&3525.3&3674.4\\
$M_{nL}$ (Schr\"odinger)&3068.6&3483.3&3674.4\\
$\Delta M^{(\pi)}$&-1.7&-3.1&-4.0\\
$\Delta M^{(K)}$&-1.5&-2.9&-3.8\\
$\Delta M^{(\rho)}$&-2.5&-4.9&-6.5\\
$\Delta M^{(K^*)}$&-1.6&-3.2&-4.2\\
$\Delta M^{(\phi)}$&-1.6&-3.2&-4.3\\
$\Delta M^{(N)}$&-2.4&-4.3&-5.5\\
$\Delta M^{(\Xi)}$&-2.0&-3.9&-5.1\\
$\Delta M^{(\Delta)}$&-0.9&-1.0&-1.0\\
$\Delta M^{(\Xi^*)}$&-2.6&-4.8&-6.3
\end{tabular}
\end{ruledtabular}
\end{table}

We now solve the Schr\"odinger equation numerically
for the mesons and for some of the positive parity baryons,
using the parameter values of
Eqs.~\eqref{eq:statp} and \eqref{eq:mc}, together with
$\Delta \mu_H$, $\Delta c_H$ and $\Delta \sigma_H$ obtained
from the fits to $\Delta V_H(r)$, see Tables~\ref{t:shift_fit_mesons} and
\ref{t:shift_fit_octet_p}. The results are collected
in Table~\ref{tab_schroed}. Indeed, the masses in all the channels shown
are lowered by amounts that are in qualitative agreement with the
considerations from the virial and Feynman--Hellmann theorems above, and the
effect becomes larger for spatially more extended charmonia.
Note that the potentials for the $\rho$ meson and the $\Delta$ baryon
have relatively large errors. Therefore, the resulting
mass shifts statistically agree with those shown for the
$K^*$ and the $\Xi^*$, respectively.

In Ref.~\cite{Beane:2014sda}, a charmonium-nucleon
bound state energy of $-20\mev$ was reported --- a factor
of eight larger than our result.
The light quark mass employed in that study was
approximately 13 times larger than the one we use here. However, as one can
see from Table~\ref{tab_schroed}, if we replace
the nucleon by the cascade that contains two strange quarks, which
are eight times heavier than our light quark, the binding
appears to become even weaker, albeit by a statistically insignificant
difference.

We found that, within the approximations made, the binding of the
charmonium $1S$ state becomes stronger by values ranging from
$-1\mev$ to $-2.5\mev$.
For the $2S$ state this effect increases to $-1\mev$ to $-6.5\mev$.
Such estimates will be more reliable for bottomonia where relativistic
and $m_H/m_b$ corrections are smaller. However, these
states are also less extended spatially and $V_0(r)$
is most strongly modified towards large distances.
This means that the mass shifts induced by the
presence of a light hadron will be even smaller in the bottomonium case
since charmonium and bottomonium binding energies $\sim m_Qv^2$
are of similar sizes.

\section{Summary and outlook}\label{sec_summary}
Studying charmonium resonances above strong decay
thresholds poses a considerable challenge to lattice QCD.
In most cases not only radial excitations of the charm
quark-antiquark system need to be resolved but also
several decay channels open up, at least near the physical
values of the light quark mass. Some of the relevant
thresholds involve the scattering of three and more hadrons.
In this case
even the required methodology is under active development
--- for recent progress in this direction,
see Refs.~\cite{Kreuzer:2009jp,Polejaeva:2012ut,Briceno:2012rv,Meissner:2014dea,Hansen:2015zga,Hansen:2016fzj}.
In view of this, testing specific models or making assumptions in certain
limiting cases represents a viable alternative and may provide at least some
first principles insight into the nature of exotic bound states
containing hidden charm.

Here we have investigated in the heavy quark limit
the hadro-quarkonium picture~\cite{Dubynskiy:2008mq},
which assumes quarkonium can be bound inside the core of a
light hadron.
We employed a single CLS~\cite{Bruno:2014jqa} ensemble with
$\nf= 2 + 1$ flavours of non-perturbatively order-$a$ improved
Wilson quarks at a lattice spacing $a\approx 0.085\fm$.
The pion and kaon masses are approximately $223\mev$ and
$476\mev$, respectively, i.e.\ the light quark mass is
by a factor of about 2.7 larger than in nature.
Our approach for testing this picture
was first to determine the potential between a pair of static
sources, approximating a heavy quark-antiquark pair, in the
absence of the hadron. Assuming
the non-relativistic limit, the Schr\"odinger equation
can then be solved with this potential in order to obtain (spin-averaged)
quarkonium energy levels. This
approach can be extended systematically, adding $v^2$ corrections, to
include heavy quark spin and momentum dependent
effects~\cite{Eichten:1980mw,Barchielli:1988zp,Bali:1997am,Bali:2000gf,Brambilla:2004jw,Koma:2006fw}.
Making the additional assumption that
the heavy quark mass is much larger than the mass of the
light hadron, the effect of the light hadron onto the
quarkonium can also be integrated out adiabatically and
cast into the quark-antiquark interaction potential.

We calculated such potentials
in the background of a hadron $H$ for a variety of
pseudoscalar and vector mesons, octet and decuplet baryons and
their negative parity partners. Of particular interest are the
differences $\Delta V_H(r)$, relative to the potential in
the vacuum. Solving the Schr\"odinger equation with the
modified potential and comparing the outcome to the
results obtained {\em in vacuo}
provides an indication of the strength of the binding
between the host hadron and the quarkonium, at least in
the heavy quark limit. In principle this approach can also be
extended, including mass dependent corrections and
interactions between the spins of the hadron and the heavy quarks.
As the effects
we detected were quite small, we have however no immediate
plans of pursuing this line of research.

Resolving very small energy differences was possible by
employing a large number of sources on 1552 gauge configurations,
corresponding to over 6000 molecular dynamics units of the
hybrid Monte Carlo algorithm.
For all the light mesons, namely the $\pi$, $K$,
$\rho$, $K^{\star}$ and $\phi$, as well as the baryons we considered,
namely the $N$, $\Sigma$, $\Lambda$, $\Xi$, $\Delta$, $\Sigma^*$,
$\Xi^*$ and $\Omega$ of both parities,
we found $\Delta V_{H}(r) < 0$, suggesting a tendency to bind.
The main effect could be quantified as a reduction of the linear
slope of the potential. At a distance of $0.5\fm$ the
potential was lowered by only $2$--$3\mev$ for all these
hadrons. Increasing the strangeness resulted in smaller statistical
errors but differences between the investigated hadrons were not
significant. Translating the modification
of the potential into energy levels by solving the Schr\"odinger equation
suggested values for the finite volume binding energy
of $1S$ charmonium ranging
from $-1\mev$ to $-2.5\mev$ and $2S$ charmonium
from $-1$ to $-6.5\mev$, see Table~\ref{tab_schroed}.
These effects should be even smaller for bottomonia that are
most sensitive to modifications of the potential
at very short distances.

These binding energies are similar in size to that of deuterium
and may be hard to reconcile with the hadro-quarkonium
picture where the quarkonium is thought to be localized inside
the light hadron which has a size $\lesssim 1\fm$.
Therefore, in the heavy quark limit, that should at least apply to
bottomonia, this may not be a viable picture.
We cannot exclude, however, different mechanisms to
stabilize hadro-charmonia such as relativistic corrections or
corrections due to the mass of the hosting hadron.

The spatial
lattice extent $L\approx 4.6/\mpi\approx 4.1\fm$ was not only large
relative to the inverse pion mass but also in comparison to the
size of a light hadron or a quarkonium state; however, the observed
effects were very small. Hence, a finite volume
study (see, e.g., Ref.~\cite{Yokokawa:2006td})
is required to establish if the reported binding
energies survive the infinite volume limit.
Simulations on different volumes, and also
injecting momentum to enable a scattering study, are ongoing,
see Ref.~\cite{Knechtli:2016eqx} for preliminary results.
Until these more extensive investigations are concluded,
we cannot exclude the possibility that
no bound state or resonance exists. Therefore,
the binding energies presented here should only be considered as
upper limits.

\acknowledgments
This work was supported by the Deutsche Forschungsgemeinschaft (DFG) Grant
No.\ SFB/TRR 55. G.M.\ acknowledges support from the Herchel Smith Fund at
the University of Cambridge and the DFG
under Contract No.\ KN 947/1-2. 
The ensemble C101 was generated by CLS~\cite{Bruno:2014jqa}
employing the {\sc openQCD}~\cite{ddopenqcd,Luscher:2012av} software and using
computer time provided by PRACE (Partnership for Advanced
Computing in Europe, \url{http://www.prace-ri.eu}) on Fermi at CINECA Bologna
and on SuperMUC at Leibniz Supercomputing Centre Munich.
An additional stream was generated on the ``Clover''
Cluster of the Mainz Helmholtz Institute. Analysis was performed on
the SFB/TRR~55 QPACE~2~\cite{Arts:2015jia} Xeon-Phi installation at Regensburg
and on the Stromboli cluster in Wuppertal.
We used the {\sc CHROMA}~\cite{Edwards:2004sx} software package along with the
{\sc LibHadronAnalysis} library and the multigrid solver implementation
of Ref.~\cite{Heybrock:2015kpy} (see also
Refs.~\cite{Richtmann:2016kcq,Heybrock:2014iga,Frommer:2013fsa})
to generate hadronic two-point functions.
Wilson loops were generated using Bj\"orn Leder's
{\sc wloop} package~\cite{wloop,Donnellan:2010mx}.
For the error analysis we used the software of the ALPHA collaboration
\cite{Wolff:2003sm,algo:csd}
available at \url{http://www-zeuthen.desy.de/alpha/}.
We thank Peter Georg, Benjamin Gl\"a\ss{}le and Daniel Richtmann for
support. Last but not least we thank all our CLS colleagues and
in particular Dalibor Djukanovic, Georg Engel and Leonardo Giusti
for generating the C101 ensemble and Stefan Schaefer for
coordinating the CLS simulations.
\bibliography{hadro2}

\begin{thebibliography}{82}%
\makeatletter
\providecommand \@ifxundefined [1]{%
 \@ifx{#1\undefined}
}%
\providecommand \@ifnum [1]{%
 \ifnum #1\expandafter \@firstoftwo
 \else \expandafter \@secondoftwo
 \fi
}%
\providecommand \@ifx [1]{%
 \ifx #1\expandafter \@firstoftwo
 \else \expandafter \@secondoftwo
 \fi
}%
\providecommand \natexlab [1]{#1}%
\providecommand \enquote  [1]{``#1''}%
\providecommand \bibnamefont  [1]{#1}%
\providecommand \bibfnamefont [1]{#1}%
\providecommand \citenamefont [1]{#1}%
\providecommand \href@noop [0]{\@secondoftwo}%
\providecommand \href [0]{\begingroup \@sanitize@url \@href}%
\providecommand \@href[1]{\@@startlink{#1}\@@href}%
\providecommand \@@href[1]{\endgroup#1\@@endlink}%
\providecommand \@sanitize@url [0]{\catcode `\\12\catcode `\$12\catcode
  `\&12\catcode `\#12\catcode `\^12\catcode `\_12\catcode `\%12\relax}%
\providecommand \@@startlink[1]{}%
\providecommand \@@endlink[0]{}%
\providecommand \url  [0]{\begingroup\@sanitize@url \@url }%
\providecommand \@url [1]{\endgroup\@href {#1}{\urlprefix }}%
\providecommand \urlprefix  [0]{URL }%
\providecommand \Eprint [0]{\href }%
\providecommand \doibase [0]{http://dx.doi.org/}%
\providecommand \selectlanguage [0]{\@gobble}%
\providecommand \bibinfo  [0]{\@secondoftwo}%
\providecommand \bibfield  [0]{\@secondoftwo}%
\providecommand \translation [1]{[#1]}%
\providecommand \BibitemOpen [0]{}%
\providecommand \bibitemStop [0]{}%
\providecommand \bibitemNoStop [0]{.\EOS\space}%
\providecommand \EOS [0]{\spacefactor3000\relax}%
\providecommand \BibitemShut  [1]{\csname bibitem#1\endcsname}%
\let\auto@bib@innerbib\@empty
\bibitem [{\citenamefont {Aaij}\ \emph {et~al.}(2015)\citenamefont {Aaij} \emph
  {et~al.}}]{Aaij:2015tga}%
  \BibitemOpen
  \bibfield  {author} {\bibinfo {author} {\bibfnamefont {Roel}\ \bibnamefont
  {Aaij}} \emph {et~al.} (\bibinfo {collaboration} {LHCb Collaboration}),\
  }\bibfield  {title} {\enquote {\bibinfo {title} {{Observation of
  {$J/\psi\rho$} resonances consistent with pentaquark states in
  {$\Lambda_b^0\rightarrow J/\psi K^-p$} decays}},}\ }\href {\doibase
  10.1103/PhysRevLett.115.072001} {\bibfield  {journal} {\bibinfo  {journal}
  {Phys. Rev. Lett.}\ }\textbf {\bibinfo {volume} {115}},\ \bibinfo {pages}
  {072001} (\bibinfo {year} {2015})},\ \Eprint
  {http://arxiv.org/abs/1507.03414} {arXiv:1507.03414 [hep-ex]} \BibitemShut
  {NoStop}%
\bibitem [{\citenamefont {Aaij}\ \emph {et~al.}(2016)\citenamefont {Aaij} \emph
  {et~al.}}]{Aaij:2016phn}%
  \BibitemOpen
  \bibfield  {author} {\bibinfo {author} {\bibfnamefont {Roel}\ \bibnamefont
  {Aaij}} \emph {et~al.} (\bibinfo {collaboration} {LHCb Collaboration}),\
  }\bibfield  {title} {\enquote {\bibinfo {title} {{Model-independent evidence
  for $J/\psi p$ contributions to $\Lambda_b^0\to J/\psi p K^-$ decays}},}\
  }\href {\doibase 10.1103/PhysRevLett.117.082002} {\bibfield  {journal}
  {\bibinfo  {journal} {Phys. Rev. Lett.}\ }\textbf {\bibinfo {volume} {117}},\
  \bibinfo {pages} {082002} (\bibinfo {year} {2016})},\ \Eprint
  {http://arxiv.org/abs/1604.05708} {arXiv:1604.05708 [hep-ex]} \BibitemShut
  {NoStop}%
\bibitem [{\citenamefont {Szczepaniak}(2016)}]{Szczepaniak:2015hya}%
  \BibitemOpen
  \bibfield  {author} {\bibinfo {author} {\bibfnamefont {Adam~P.}\ \bibnamefont
  {Szczepaniak}},\ }\bibfield  {title} {\enquote {\bibinfo {title} {{Dalitz
  plot distributions in presence of triangle singularities}},}\ }\href
  {\doibase 10.1016/j.physletb.2016.03.064} {\bibfield  {journal} {\bibinfo
  {journal} {Phys. Lett. B}\ }\textbf {\bibinfo {volume} {757}},\ \bibinfo
  {pages} {61} (\bibinfo {year} {2016})},\ \Eprint
  {http://arxiv.org/abs/1510.01789} {arXiv:1510.01789 [hep-ph]} \BibitemShut
  {NoStop}%
\bibitem [{\citenamefont {Guo}\ \emph {et~al.}(2016)\citenamefont {Guo},
  \citenamefont {Mei{\ss}ner}, \citenamefont {Nieves},\ and\ \citenamefont
  {Yang}}]{Guo:2016bkl}%
  \BibitemOpen
  \bibfield  {author} {\bibinfo {author} {\bibfnamefont {Feng-Kun}\
  \bibnamefont {Guo}}, \bibinfo {author} {\bibfnamefont {{Ulf-G.}}\
  \bibnamefont {Mei{\ss}ner}}, \bibinfo {author} {\bibfnamefont {Juan}\
  \bibnamefont {Nieves}}, \ and\ \bibinfo {author} {\bibfnamefont {Zhi}\
  \bibnamefont {Yang}},\ }\bibfield  {title} {\enquote {\bibinfo {title}
  {{Remarks on the $P_c$ structures and triangle singularities}},}\ }\href
  {\doibase 10.1140/epja/i2016-16318-4} {\bibfield  {journal} {\bibinfo
  {journal} {Eur. Phys. J. A}\ }\textbf {\bibinfo {volume} {52}},\ \bibinfo
  {pages} {318} (\bibinfo {year} {2016})},\ \Eprint
  {http://arxiv.org/abs/1605.05113} {arXiv:1605.05113 [hep-ph]} \BibitemShut
  {NoStop}%
\bibitem [{\citenamefont {Brambilla}\ \emph {et~al.}(2011)\citenamefont
  {Brambilla} \emph {et~al.}}]{Brambilla:2010cs}%
  \BibitemOpen
  \bibfield  {author} {\bibinfo {author} {\bibfnamefont {Nora}\ \bibnamefont
  {Brambilla}} \emph {et~al.} (\bibinfo {collaboration} {Quarkonium Working
  Group}),\ }\bibfield  {title} {\enquote {\bibinfo {title} {{Heavy quarkonium:
  progress, puzzles, and opportunities}},}\ }\href {\doibase
  10.1140/epjc/s10052-010-1534-9} {\bibfield  {journal} {\bibinfo  {journal}
  {Eur. Phys. J. C}\ }\textbf {\bibinfo {volume} {71}},\ \bibinfo {pages}
  {1534} (\bibinfo {year} {2011})},\ \Eprint {http://arxiv.org/abs/1010.5827}
  {arXiv:1010.5827 [hep-ph]} \BibitemShut {NoStop}%
\bibitem [{\citenamefont {Chen}\ \emph {et~al.}(2016)\citenamefont {Chen},
  \citenamefont {Chen}, \citenamefont {Liu},\ and\ \citenamefont
  {Zhu}}]{Chen:2016qju}%
  \BibitemOpen
  \bibfield  {author} {\bibinfo {author} {\bibfnamefont {Hua-Xing}\
  \bibnamefont {Chen}}, \bibinfo {author} {\bibfnamefont {Wei}\ \bibnamefont
  {Chen}}, \bibinfo {author} {\bibfnamefont {Xiang}\ \bibnamefont {Liu}}, \
  and\ \bibinfo {author} {\bibfnamefont {Shi-Lin}\ \bibnamefont {Zhu}},\
  }\bibfield  {title} {\enquote {\bibinfo {title} {{The hidden-charm pentaquark
  and tetraquark states}},}\ }\href {\doibase 10.1016/j.physrep.2016.05.004}
  {\bibfield  {journal} {\bibinfo  {journal} {Phys. Rept.}\ }\textbf {\bibinfo
  {volume} {639}},\ \bibinfo {pages} {1} (\bibinfo {year} {2016})},\ \Eprint
  {http://arxiv.org/abs/1601.02092} {arXiv:1601.02092 [hep-ph]} \BibitemShut
  {NoStop}%
\bibitem [{\citenamefont {Jaffe}(1977)}]{Jaffe:1976ig}%
  \BibitemOpen
  \bibfield  {author} {\bibinfo {author} {\bibfnamefont {Robert~L.}\
  \bibnamefont {Jaffe}},\ }\bibfield  {title} {\enquote {\bibinfo {title}
  {{Multi-quark hadrons. 1. The phenomenology of $qq\bar{q}\bar{q}$ mesons}},}\
  }\href {\doibase 10.1103/PhysRevD.15.267} {\bibfield  {journal} {\bibinfo
  {journal} {Phys. Rev. D}\ }\textbf {\bibinfo {volume} {15}},\ \bibinfo
  {pages} {267} (\bibinfo {year} {1977})}\BibitemShut {NoStop}%
\bibitem [{\citenamefont {Weinstein}\ and\ \citenamefont
  {Isgur}(1982)}]{Weinstein:1982gc}%
  \BibitemOpen
  \bibfield  {author} {\bibinfo {author} {\bibfnamefont {John~D.}\ \bibnamefont
  {Weinstein}}\ and\ \bibinfo {author} {\bibfnamefont {Nathan}\ \bibnamefont
  {Isgur}},\ }\bibfield  {title} {\enquote {\bibinfo {title} {{Do multi-quark
  hadrons exist?}}}\ }\href {\doibase 10.1103/PhysRevLett.48.659} {\bibfield
  {journal} {\bibinfo  {journal} {Phys. Rev. Lett.}\ }\textbf {\bibinfo
  {volume} {48}},\ \bibinfo {pages} {659} (\bibinfo {year} {1982})}\BibitemShut
  {NoStop}%
\bibitem [{\citenamefont {Maiani}\ \emph {et~al.}(2007)\citenamefont {Maiani},
  \citenamefont {Polosa},\ and\ \citenamefont {Riquer}}]{Maiani:2007vr}%
  \BibitemOpen
  \bibfield  {author} {\bibinfo {author} {\bibfnamefont {Luciano}\ \bibnamefont
  {Maiani}}, \bibinfo {author} {\bibfnamefont {Antonio~D.}\ \bibnamefont
  {Polosa}}, \ and\ \bibinfo {author} {\bibfnamefont {Veronica}\ \bibnamefont
  {Riquer}},\ }\bibfield  {title} {\enquote {\bibinfo {title} {{Indications of
  a four-quark structure for the $X(3872)$ and $X(3876)$ particles from recent
  Belle and BABAR data}},}\ }\href {\doibase 10.1103/PhysRevLett.99.182003}
  {\bibfield  {journal} {\bibinfo  {journal} {Phys. Rev. Lett.}\ }\textbf
  {\bibinfo {volume} {99}},\ \bibinfo {pages} {182003} (\bibinfo {year}
  {2007})},\ \Eprint {http://arxiv.org/abs/0707.3354} {arXiv:0707.3354
  [hep-ph]} \BibitemShut {NoStop}%
\bibitem [{\citenamefont {Brodsky}\ \emph {et~al.}(2014)\citenamefont
  {Brodsky}, \citenamefont {Hwang},\ and\ \citenamefont
  {Lebed}}]{Brodsky:2014xia}%
  \BibitemOpen
  \bibfield  {author} {\bibinfo {author} {\bibfnamefont {Stanley~J.}\
  \bibnamefont {Brodsky}}, \bibinfo {author} {\bibfnamefont {Dae~Sung}\
  \bibnamefont {Hwang}}, \ and\ \bibinfo {author} {\bibfnamefont {Richard~F.}\
  \bibnamefont {Lebed}},\ }\bibfield  {title} {\enquote {\bibinfo {title}
  {{Dynamical picture for the formation and decay of the exotic XYZ mesons}},}\
  }\href {\doibase 10.1103/PhysRevLett.113.112001} {\bibfield  {journal}
  {\bibinfo  {journal} {Phys. Rev. Lett.}\ }\textbf {\bibinfo {volume} {113}},\
  \bibinfo {pages} {112001} (\bibinfo {year} {2014})},\ \Eprint
  {http://arxiv.org/abs/1406.7281} {arXiv:1406.7281 [hep-ph]} \BibitemShut
  {NoStop}%
\bibitem [{\citenamefont {Lebed}(2016)}]{Lebed:2016epe}%
  \BibitemOpen
  \bibfield  {author} {\bibinfo {author} {\bibfnamefont {Richard~F.}\
  \bibnamefont {Lebed}},\ }\bibfield  {title} {\enquote {\bibinfo {title} {{How
  Often Do Diquarks Form? A Very Simple Model}},}\ }\href {\doibase
  10.1103/PhysRevD.94.034039} {\bibfield  {journal} {\bibinfo  {journal} {Phys.
  Rev. D}\ }\textbf {\bibinfo {volume} {94}},\ \bibinfo {pages} {034039}
  (\bibinfo {year} {2016})},\ \Eprint {http://arxiv.org/abs/1606.07108}
  {arXiv:1606.07108 [hep-ph]} \BibitemShut {NoStop}%
\bibitem [{\citenamefont {Voloshin}\ and\ \citenamefont
  {Okun}(1976)}]{Voloshin:1976ap}%
  \BibitemOpen
  \bibfield  {author} {\bibinfo {author} {\bibfnamefont {Mikhail~B.}\
  \bibnamefont {Voloshin}}\ and\ \bibinfo {author} {\bibfnamefont {Lev~B.}\
  \bibnamefont {Okun}},\ }\bibfield  {title} {\enquote {\bibinfo {title}
  {{Hydronic molecules and the charmonium atom}},}\ }\href
  {http://www.jetpletters.ac.ru/ps/1801/article_27526.shtml} {\bibfield
  {journal} {\bibinfo  {journal} {JETP Lett.}\ }\textbf {\bibinfo {volume}
  {23}},\ \bibinfo {pages} {333} (\bibinfo {year} {1976})},\ \bibinfo {note}
  {[Pisma Zh. Eksp. Teor. Fiz. 23, 369 (1976)]}\BibitemShut {NoStop}%
\bibitem [{\citenamefont {De~R{\'u}jula}\ \emph {et~al.}(1977)\citenamefont
  {De~R{\'u}jula}, \citenamefont {Georgi},\ and\ \citenamefont
  {Glashow}}]{DeRujula:1976zlg}%
  \BibitemOpen
  \bibfield  {author} {\bibinfo {author} {\bibfnamefont {Alvaro}\ \bibnamefont
  {De~R{\'u}jula}}, \bibinfo {author} {\bibfnamefont {Howard}\ \bibnamefont
  {Georgi}}, \ and\ \bibinfo {author} {\bibfnamefont {Sheldon~L.}\ \bibnamefont
  {Glashow}},\ }\bibfield  {title} {\enquote {\bibinfo {title} {{Molecular
  charmonium: A new spectroscopy?}}}\ }\href {\doibase
  10.1103/PhysRevLett.38.317} {\bibfield  {journal} {\bibinfo  {journal} {Phys.
  Rev. Lett.}\ }\textbf {\bibinfo {volume} {38}},\ \bibinfo {pages} {317}
  (\bibinfo {year} {1977})}\BibitemShut {NoStop}%
\bibitem [{\citenamefont {Novikov}\ \emph {et~al.}(1978)\citenamefont
  {Novikov}, \citenamefont {Okun}, \citenamefont {Shifman}, \citenamefont
  {Vainshtein}, \citenamefont {Voloshin},\ and\ \citenamefont
  {Zakharov}}]{Novikov:1977dq}%
  \BibitemOpen
  \bibfield  {author} {\bibinfo {author} {\bibfnamefont {Victor~A.}\
  \bibnamefont {Novikov}}, \bibinfo {author} {\bibfnamefont {Lev~B.}\
  \bibnamefont {Okun}}, \bibinfo {author} {\bibfnamefont {Mikhail~A.}\
  \bibnamefont {Shifman}}, \bibinfo {author} {\bibfnamefont {Arkady~I.}\
  \bibnamefont {Vainshtein}}, \bibinfo {author} {\bibfnamefont {Mikhail~B.}\
  \bibnamefont {Voloshin}}, \ and\ \bibinfo {author} {\bibfnamefont
  {Valentin~I.}\ \bibnamefont {Zakharov}},\ }\bibfield  {title} {\enquote
  {\bibinfo {title} {{Charmonium and gluons: Basic experimental facts and
  theoretical introduction}},}\ }\href {\doibase 10.1016/0370-1573(78)90120-5}
  {\bibfield  {journal} {\bibinfo  {journal} {Phys. Rept.}\ }\textbf {\bibinfo
  {volume} {41}},\ \bibinfo {pages} {1} (\bibinfo {year} {1978})}\BibitemShut
  {NoStop}%
\bibitem [{\citenamefont {T{\"o}rnqvist}(1994)}]{Tornqvist:1993ng}%
  \BibitemOpen
  \bibfield  {author} {\bibinfo {author} {\bibfnamefont {Nils~A.}\ \bibnamefont
  {T{\"o}rnqvist}},\ }\bibfield  {title} {\enquote {\bibinfo {title} {{From the
  deuteron to deusons, an analysis of deuteron-like meson meson bound
  states}},}\ }\href {\doibase 10.1007/BF01413192} {\bibfield  {journal}
  {\bibinfo  {journal} {Z. Phys. C.}\ }\textbf {\bibinfo {volume} {61}},\
  \bibinfo {pages} {525} (\bibinfo {year} {1994})},\ \Eprint
  {http://arxiv.org/abs/hep-ph/9310247} {arXiv:hep-ph/9310247 [hep-ph]}
  \BibitemShut {NoStop}%
\bibitem [{\citenamefont {Close}\ \emph {et~al.}(2010)\citenamefont {Close},
  \citenamefont {Downum},\ and\ \citenamefont {Thomas}}]{Close:2010wq}%
  \BibitemOpen
  \bibfield  {author} {\bibinfo {author} {\bibfnamefont {Frank}\ \bibnamefont
  {Close}}, \bibinfo {author} {\bibfnamefont {Clark}\ \bibnamefont {Downum}}, \
  and\ \bibinfo {author} {\bibfnamefont {Christopher~E.}\ \bibnamefont
  {Thomas}},\ }\bibfield  {title} {\enquote {\bibinfo {title} {{Novel
  charmonium and bottomonium spectroscopies due to deeply bound hadronic
  molecules from single pion exchange}},}\ }\href {\doibase
  10.1103/PhysRevD.81.074033} {\bibfield  {journal} {\bibinfo  {journal} {Phys.
  Rev. D}\ }\textbf {\bibinfo {volume} {81}},\ \bibinfo {pages} {074033}
  (\bibinfo {year} {2010})},\ \Eprint {http://arxiv.org/abs/1001.2553}
  {arXiv:1001.2553 [hep-ph]} \BibitemShut {NoStop}%
\bibitem [{\citenamefont {Barnes}(1979)}]{Barnes:1977hg}%
  \BibitemOpen
  \bibfield  {author} {\bibinfo {author} {\bibfnamefont {Ted}\ \bibnamefont
  {Barnes}},\ }\bibfield  {title} {\enquote {\bibinfo {title} {{Colored quark
  and gluon constituents in the MIT bag model}},}\ }\href {\doibase
  10.1016/0550-3213(79)90194-9} {\bibfield  {journal} {\bibinfo  {journal}
  {Nucl. Phys. B}\ }\textbf {\bibinfo {volume} {158}},\ \bibinfo {pages} {171}
  (\bibinfo {year} {1979})}\BibitemShut {NoStop}%
\bibitem [{\citenamefont {Barnes}\ and\ \citenamefont
  {Close}(1982)}]{Barnes:1982zs}%
  \BibitemOpen
  \bibfield  {author} {\bibinfo {author} {\bibfnamefont {Ted}\ \bibnamefont
  {Barnes}}\ and\ \bibinfo {author} {\bibfnamefont {Frank~E.}\ \bibnamefont
  {Close}},\ }\bibfield  {title} {\enquote {\bibinfo {title} {{A light exotic
  $q\bar{q}g$ hermaphrodite meson?}}}\ }\href {\doibase
  10.1016/0370-2693(82)90301-X} {\bibfield  {journal} {\bibinfo  {journal}
  {Phys. Lett.}\ }\textbf {\bibinfo {volume} {116B}},\ \bibinfo {pages} {365}
  (\bibinfo {year} {1982})}\BibitemShut {NoStop}%
\bibitem [{\citenamefont {Chanowitz}\ and\ \citenamefont
  {Sharpe}(1983)}]{Chanowitz:1982qj}%
  \BibitemOpen
  \bibfield  {author} {\bibinfo {author} {\bibfnamefont {Michael~S.}\
  \bibnamefont {Chanowitz}}\ and\ \bibinfo {author} {\bibfnamefont
  {Stephen~R.}\ \bibnamefont {Sharpe}},\ }\bibfield  {title} {\enquote
  {\bibinfo {title} {{Hybrids: Mixed states of quarks and gluons}},}\ }\href
  {\doibase 10.1016/0550-3213(83)90635-1} {\bibfield  {journal} {\bibinfo
  {journal} {Nucl. Phys. B}\ }\textbf {\bibinfo {volume} {222}},\ \bibinfo
  {pages} {211} (\bibinfo {year} {1983})},\ \bibinfo {note} {[Erratum: Nucl.
  Phys. B {\bf 228}, 588 (1983)]}\BibitemShut {NoStop}%
\bibitem [{\citenamefont {Isgur}\ \emph {et~al.}(1985)\citenamefont {Isgur},
  \citenamefont {Kokoski},\ and\ \citenamefont {Paton}}]{Isgur:1985vy}%
  \BibitemOpen
  \bibfield  {author} {\bibinfo {author} {\bibfnamefont {Nathan}\ \bibnamefont
  {Isgur}}, \bibinfo {author} {\bibfnamefont {Richard}\ \bibnamefont
  {Kokoski}}, \ and\ \bibinfo {author} {\bibfnamefont {Jack}\ \bibnamefont
  {Paton}},\ }\bibfield  {title} {\enquote {\bibinfo {title} {{Gluonic
  excitations of mesons: Why they are missing and where to find them}},}\
  }\bibfield  {booktitle} {\emph {\bibinfo {booktitle} {{Proceedings,
  International Conference on Hadron Spectroscopy}}},\ }\href {\doibase
  10.1103/PhysRevLett.54.869, 10.1063/1.35357} {\bibfield  {journal} {\bibinfo
  {journal} {Phys. Rev. Lett.}\ }\textbf {\bibinfo {volume} {54}},\ \bibinfo
  {pages} {869} (\bibinfo {year} {1985})},\ \bibinfo {note} {[AIP Conf. Proc.
  132, 242 (1985)]}\BibitemShut {NoStop}%
\bibitem [{\citenamefont {Dubynskiy}\ and\ \citenamefont
  {Voloshin}(2008)}]{Dubynskiy:2008mq}%
  \BibitemOpen
  \bibfield  {author} {\bibinfo {author} {\bibfnamefont {S.}~\bibnamefont
  {Dubynskiy}}\ and\ \bibinfo {author} {\bibfnamefont {Mikhail~B.}\
  \bibnamefont {Voloshin}},\ }\bibfield  {title} {\enquote {\bibinfo {title}
  {{Hadro-charmonium}},}\ }\href {\doibase 10.1016/j.physletb.2008.07.086}
  {\bibfield  {journal} {\bibinfo  {journal} {Phys. Lett. B}\ }\textbf
  {\bibinfo {volume} {666}},\ \bibinfo {pages} {344} (\bibinfo {year}
  {2008})},\ \Eprint {http://arxiv.org/abs/0803.2224} {arXiv:0803.2224
  [hep-ph]} \BibitemShut {NoStop}%
\bibitem [{\citenamefont {Li}\ and\ \citenamefont
  {Voloshin}(2014)}]{Li:2013ssa}%
  \BibitemOpen
  \bibfield  {author} {\bibinfo {author} {\bibfnamefont {Xin}\ \bibnamefont
  {Li}}\ and\ \bibinfo {author} {\bibfnamefont {Mikhail~B.}\ \bibnamefont
  {Voloshin}},\ }\bibfield  {title} {\enquote {\bibinfo {title} {{$Y$(4260) and
  $Y$(4360) as mixed hadrocharmonium}},}\ }\href {\doibase
  10.1142/S0217732314500606} {\bibfield  {journal} {\bibinfo  {journal} {Mod.
  Phys. Lett. A}\ }\textbf {\bibinfo {volume} {29}},\ \bibinfo {pages}
  {1450060} (\bibinfo {year} {2014})},\ \Eprint
  {http://arxiv.org/abs/1309.1681} {arXiv:1309.1681 [hep-ph]} \BibitemShut
  {NoStop}%
\bibitem [{\citenamefont {L{\"u}scher}(1991)}]{Luscher:1990ux}%
  \BibitemOpen
  \bibfield  {author} {\bibinfo {author} {\bibfnamefont {Martin}\ \bibnamefont
  {L{\"u}scher}},\ }\bibfield  {title} {\enquote {\bibinfo {title} {{Two
  particle states on a torus and their relation to the scattering matrix}},}\
  }\href {\doibase 10.1016/0550-3213(91)90366-6} {\bibfield  {journal}
  {\bibinfo  {journal} {Nucl. Phys. B}\ }\textbf {\bibinfo {volume} {354}},\
  \bibinfo {pages} {531} (\bibinfo {year} {1991})}\BibitemShut {NoStop}%
\bibitem [{\citenamefont {Prelovsek}(2015)}]{Prelovsek:2015fra}%
  \BibitemOpen
  \bibfield  {author} {\bibinfo {author} {\bibfnamefont {Sasa}\ \bibnamefont
  {Prelovsek}},\ }\bibfield  {title} {\enquote {\bibinfo {title} {{Lattice
  studies of charmonia and exotics}},}\ }in\ \href
  {https://inspirehep.net/record/1390979/files/arXiv:1508.07322.pdf} {\emph
  {\bibinfo {booktitle} {{7th International Workshop on Charm Physics (Charm
  2015) Detroit, MI, USA, May 18-22, 2015}}}}\ (\bibinfo {year} {2015})\
  \Eprint {http://arxiv.org/abs/1508.07322} {arXiv:1508.07322 [hep-lat]}
  \BibitemShut {NoStop}%
\bibitem [{\citenamefont {Caswell}\ and\ \citenamefont
  {Lepage}(1986)}]{Caswell:1985ui}%
  \BibitemOpen
  \bibfield  {author} {\bibinfo {author} {\bibfnamefont {William~E.}\
  \bibnamefont {Caswell}}\ and\ \bibinfo {author} {\bibfnamefont {G.~Peter}\
  \bibnamefont {Lepage}},\ }\bibfield  {title} {\enquote {\bibinfo {title}
  {{Effective Lagrangians for bound state problems in QED, QCD, and other field
  theories}},}\ }\href {\doibase 10.1016/0370-2693(86)91297-9} {\bibfield
  {journal} {\bibinfo  {journal} {Phys. Lett. B}\ }\textbf {\bibinfo {volume}
  {167}},\ \bibinfo {pages} {437} (\bibinfo {year} {1986})}\BibitemShut
  {NoStop}%
\bibitem [{\citenamefont {Pineda}\ and\ \citenamefont
  {Soto}(1998)}]{Pineda:1997bj}%
  \BibitemOpen
  \bibfield  {author} {\bibinfo {author} {\bibfnamefont {Antonio}\ \bibnamefont
  {Pineda}}\ and\ \bibinfo {author} {\bibfnamefont {Joan}\ \bibnamefont
  {Soto}},\ }\bibfield  {title} {\enquote {\bibinfo {title} {{Effective field
  theory for ultrasoft momenta in NRQCD and NRQED}},}\ }\bibfield  {booktitle}
  {\emph {\bibinfo {booktitle} {{Quantum chromodynamics. Proceedings,
  Conference, QCD'97, Montpellier, France, July 3-9, 1997}}},\ }\href {\doibase
  10.1016/S0920-5632(97)01102-X} {\bibfield  {journal} {\bibinfo  {journal}
  {Nucl. Phys. Proc. Suppl.}\ }\textbf {\bibinfo {volume} {64}},\ \bibinfo
  {pages} {428} (\bibinfo {year} {1998})},\ \Eprint
  {http://arxiv.org/abs/hep-ph/9707481} {arXiv:hep-ph/9707481 [hep-ph]}
  \BibitemShut {NoStop}%
\bibitem [{\citenamefont {Brambilla}\ \emph {et~al.}(2000)\citenamefont
  {Brambilla}, \citenamefont {Pineda}, \citenamefont {Soto},\ and\
  \citenamefont {Vairo}}]{Brambilla:1999xf}%
  \BibitemOpen
  \bibfield  {author} {\bibinfo {author} {\bibfnamefont {Nora}\ \bibnamefont
  {Brambilla}}, \bibinfo {author} {\bibfnamefont {Antonio}\ \bibnamefont
  {Pineda}}, \bibinfo {author} {\bibfnamefont {Joan}\ \bibnamefont {Soto}}, \
  and\ \bibinfo {author} {\bibfnamefont {Antonio}\ \bibnamefont {Vairo}},\
  }\bibfield  {title} {\enquote {\bibinfo {title} {{Potential NRQCD: An
  effective theory for heavy quarkonium}},}\ }\href {\doibase
  10.1016/S0550-3213(99)00693-8} {\bibfield  {journal} {\bibinfo  {journal}
  {Nucl. Phys. B}\ }\textbf {\bibinfo {volume} {566}},\ \bibinfo {pages} {275}
  (\bibinfo {year} {2000})},\ \Eprint {http://arxiv.org/abs/hep-ph/9907240}
  {arXiv:hep-ph/9907240 [hep-ph]} \BibitemShut {NoStop}%
\bibitem [{\citenamefont {Brambilla}\ \emph {et~al.}(2016)\citenamefont
  {Brambilla}, \citenamefont {Krein}, \citenamefont {Tarr{\'u}s~Castell{\`a}},\
  and\ \citenamefont {Vairo}}]{Brambilla:2015rqa}%
  \BibitemOpen
  \bibfield  {author} {\bibinfo {author} {\bibfnamefont {Nora}\ \bibnamefont
  {Brambilla}}, \bibinfo {author} {\bibfnamefont {Gast{\~a}o}\ \bibnamefont
  {Krein}}, \bibinfo {author} {\bibfnamefont {Jaume}\ \bibnamefont
  {Tarr{\'u}s~Castell{\`a}}}, \ and\ \bibinfo {author} {\bibfnamefont
  {Antonio}\ \bibnamefont {Vairo}},\ }\bibfield  {title} {\enquote {\bibinfo
  {title} {{Long-range properties of $1S$ bottomonium states}},}\ }\href
  {\doibase 10.1103/PhysRevD.93.054002} {\bibfield  {journal} {\bibinfo
  {journal} {Phys. Rev. D}\ }\textbf {\bibinfo {volume} {93}},\ \bibinfo
  {pages} {054002} (\bibinfo {year} {2016})},\ \Eprint
  {http://arxiv.org/abs/1510.05895} {arXiv:1510.05895 [hep-ph]} \BibitemShut
  {NoStop}%
\bibitem [{\citenamefont {Brodsky}\ \emph {et~al.}(1990)\citenamefont
  {Brodsky}, \citenamefont {Schmidt},\ and\ \citenamefont
  {de~T{\'e}ramond}}]{Brodsky:1989jd}%
  \BibitemOpen
  \bibfield  {author} {\bibinfo {author} {\bibfnamefont {Stanley~J.}\
  \bibnamefont {Brodsky}}, \bibinfo {author} {\bibfnamefont {Ivan~A.}\
  \bibnamefont {Schmidt}}, \ and\ \bibinfo {author} {\bibfnamefont {Guy~F.}\
  \bibnamefont {de~T{\'e}ramond}},\ }\bibfield  {title} {\enquote {\bibinfo
  {title} {{Nuclear bound quarkonium}},}\ }\href {\doibase
  10.1103/PhysRevLett.64.1011} {\bibfield  {journal} {\bibinfo  {journal}
  {Phys. Rev. Lett.}\ }\textbf {\bibinfo {volume} {64}},\ \bibinfo {pages}
  {1011} (\bibinfo {year} {1990})}\BibitemShut {NoStop}%
\bibitem [{\citenamefont {Kaidalov}\ and\ \citenamefont
  {Volkovitsky}(1992)}]{Kaidalov:1992hd}%
  \BibitemOpen
  \bibfield  {author} {\bibinfo {author} {\bibfnamefont {Alexei~B.}\
  \bibnamefont {Kaidalov}}\ and\ \bibinfo {author} {\bibfnamefont {Peter~E.}\
  \bibnamefont {Volkovitsky}},\ }\bibfield  {title} {\enquote {\bibinfo {title}
  {{Heavy quarkonia interactions with nucleons and nuclei}},}\ }\href {\doibase
  10.1103/PhysRevLett.69.3155} {\bibfield  {journal} {\bibinfo  {journal}
  {Phys. Rev. Lett.}\ }\textbf {\bibinfo {volume} {69}},\ \bibinfo {pages}
  {3155} (\bibinfo {year} {1992})}\BibitemShut {NoStop}%
\bibitem [{\citenamefont {Wasson}(1991)}]{Wasson:1991fb}%
  \BibitemOpen
  \bibfield  {author} {\bibinfo {author} {\bibfnamefont {David~A.}\
  \bibnamefont {Wasson}},\ }\bibfield  {title} {\enquote {\bibinfo {title}
  {{Comment on `Nuclear bound quarkonium'}},}\ }\href {\doibase
  10.1103/PhysRevLett.67.2237} {\bibfield  {journal} {\bibinfo  {journal}
  {Phys. Rev. Lett.}\ }\textbf {\bibinfo {volume} {67}},\ \bibinfo {pages}
  {2237} (\bibinfo {year} {1991})}\BibitemShut {NoStop}%
\bibitem [{\citenamefont {Luke}\ \emph {et~al.}(1992)\citenamefont {Luke},
  \citenamefont {Manohar},\ and\ \citenamefont {Savage}}]{Luke:1992tm}%
  \BibitemOpen
  \bibfield  {author} {\bibinfo {author} {\bibfnamefont {Michael~E.}\
  \bibnamefont {Luke}}, \bibinfo {author} {\bibfnamefont {Aneesh~V.}\
  \bibnamefont {Manohar}}, \ and\ \bibinfo {author} {\bibfnamefont {Martin~J.}\
  \bibnamefont {Savage}},\ }\bibfield  {title} {\enquote {\bibinfo {title} {{A
  QCD calculation of the interaction of quarkonium with nuclei}},}\ }\href
  {\doibase 10.1016/0370-2693(92)91114-O} {\bibfield  {journal} {\bibinfo
  {journal} {Phys. Lett. B}\ }\textbf {\bibinfo {volume} {288}},\ \bibinfo
  {pages} {355} (\bibinfo {year} {1992})},\ \Eprint
  {http://arxiv.org/abs/hep-ph/9204219} {arXiv:hep-ph/9204219 [hep-ph]}
  \BibitemShut {NoStop}%
\bibitem [{\citenamefont {Yokokawa}\ \emph {et~al.}(2006)\citenamefont
  {Yokokawa}, \citenamefont {Sasaki}, \citenamefont {Hatsuda},\ and\
  \citenamefont {Hayashigaki}}]{Yokokawa:2006td}%
  \BibitemOpen
  \bibfield  {author} {\bibinfo {author} {\bibfnamefont {Kazuo}\ \bibnamefont
  {Yokokawa}}, \bibinfo {author} {\bibfnamefont {Shoichi}\ \bibnamefont
  {Sasaki}}, \bibinfo {author} {\bibfnamefont {Tetsuo}\ \bibnamefont
  {Hatsuda}}, \ and\ \bibinfo {author} {\bibfnamefont {Arata}\ \bibnamefont
  {Hayashigaki}},\ }\bibfield  {title} {\enquote {\bibinfo {title} {{First
  lattice study of low-energy charmonium-hadron interaction}},}\ }\href
  {\doibase 10.1103/PhysRevD.74.034504} {\bibfield  {journal} {\bibinfo
  {journal} {Phys. Rev. D}\ }\textbf {\bibinfo {volume} {74}},\ \bibinfo
  {pages} {034504} (\bibinfo {year} {2006})},\ \Eprint
  {http://arxiv.org/abs/hep-lat/0605009} {arXiv:hep-lat/0605009 [hep-lat]}
  \BibitemShut {NoStop}%
\bibitem [{\citenamefont {Liu}\ \emph {et~al.}(2008)\citenamefont {Liu},
  \citenamefont {Lin},\ and\ \citenamefont {Orginos}}]{Liu:2008rza}%
  \BibitemOpen
  \bibfield  {author} {\bibinfo {author} {\bibfnamefont {Liuming}\ \bibnamefont
  {Liu}}, \bibinfo {author} {\bibfnamefont {Huey-Wen}\ \bibnamefont {Lin}}, \
  and\ \bibinfo {author} {\bibfnamefont {Kostas}\ \bibnamefont {Orginos}},\
  }\bibfield  {title} {\enquote {\bibinfo {title} {{Charmed Hadron
  Interactions}},}\ }\bibfield  {booktitle} {\emph {\bibinfo {booktitle}
  {{Proceedings, 26th International Symposium on Lattice field theory (Lattice
  2008)}}},\ }\href@noop {} {\bibfield  {journal} {\bibinfo  {journal} {PoS}\
  }\textbf {\bibinfo {volume} {LATTICE2008}},\ \bibinfo {pages} {112} (\bibinfo
  {year} {2008})},\ \Eprint {http://arxiv.org/abs/0810.5412} {arXiv:0810.5412
  [hep-lat]} \BibitemShut {NoStop}%
\bibitem [{\citenamefont {Beane}\ \emph {et~al.}(2015)\citenamefont {Beane},
  \citenamefont {Chang}, \citenamefont {Cohen}, \citenamefont {Detmold},
  \citenamefont {Lin}, \citenamefont {Orginos}, \citenamefont {Parre{\~n}o},\
  and\ \citenamefont {Savage}}]{Beane:2014sda}%
  \BibitemOpen
  \bibfield  {author} {\bibinfo {author} {\bibfnamefont {Silas~R.}\
  \bibnamefont {Beane}}, \bibinfo {author} {\bibfnamefont {Emmanuel}\
  \bibnamefont {Chang}}, \bibinfo {author} {\bibfnamefont {Saul~D.}\
  \bibnamefont {Cohen}}, \bibinfo {author} {\bibfnamefont {William}\
  \bibnamefont {Detmold}}, \bibinfo {author} {\bibfnamefont {Huey-Wen}\
  \bibnamefont {Lin}}, \bibinfo {author} {\bibfnamefont {Kostas}\ \bibnamefont
  {Orginos}}, \bibinfo {author} {\bibfnamefont {Assumpta}\ \bibnamefont
  {Parre{\~n}o}}, \ and\ \bibinfo {author} {\bibfnamefont {Martin~J.}\
  \bibnamefont {Savage}} (\bibinfo {collaboration} {NPLQCD Collaboration}),\
  }\bibfield  {title} {\enquote {\bibinfo {title} {{Quarkonium-nucleus bound
  states from Lattice QCD}},}\ }\href {\doibase 10.1103/PhysRevD.91.114503}
  {\bibfield  {journal} {\bibinfo  {journal} {Phys. Rev. D}\ }\textbf {\bibinfo
  {volume} {91}},\ \bibinfo {pages} {114503} (\bibinfo {year} {2015})},\
  \Eprint {http://arxiv.org/abs/1410.7069} {arXiv:1410.7069 [hep-lat]}
  \BibitemShut {NoStop}%
\bibitem [{\citenamefont {Kawanai}\ and\ \citenamefont
  {Sasaki}(2010)}]{Kawanai:2010ev}%
  \BibitemOpen
  \bibfield  {author} {\bibinfo {author} {\bibfnamefont {Taichi}\ \bibnamefont
  {Kawanai}}\ and\ \bibinfo {author} {\bibfnamefont {Shoichi}\ \bibnamefont
  {Sasaki}},\ }\bibfield  {title} {\enquote {\bibinfo {title}
  {{Charmonium-nucleon potential from Lattice QCD}},}\ }\href {\doibase
  10.1103/PhysRevD.82.091501} {\bibfield  {journal} {\bibinfo  {journal} {Phys.
  Rev. D}\ }\textbf {\bibinfo {volume} {82}},\ \bibinfo {pages} {091501}
  (\bibinfo {year} {2010})},\ \Eprint {http://arxiv.org/abs/1009.3332}
  {arXiv:1009.3332 [hep-lat]} \BibitemShut {NoStop}%
\bibitem [{\citenamefont {Richards}\ \emph {et~al.}(1990)\citenamefont
  {Richards}, \citenamefont {Sinclair},\ and\ \citenamefont
  {Sivers}}]{Richards:1990xf}%
  \BibitemOpen
  \bibfield  {author} {\bibinfo {author} {\bibfnamefont {David~G.}\
  \bibnamefont {Richards}}, \bibinfo {author} {\bibfnamefont {Don~K.}\
  \bibnamefont {Sinclair}}, \ and\ \bibinfo {author} {\bibfnamefont
  {Dennis~W.}\ \bibnamefont {Sivers}},\ }\bibfield  {title} {\enquote {\bibinfo
  {title} {{Lattice QCD simulation of meson exchange forces}},}\ }\href
  {\doibase 10.1103/PhysRevD.42.3191} {\bibfield  {journal} {\bibinfo
  {journal} {Phys. Rev. D}\ }\textbf {\bibinfo {volume} {42}},\ \bibinfo
  {pages} {3191} (\bibinfo {year} {1990})}\BibitemShut {NoStop}%
\bibitem [{\citenamefont {Pennanen}\ \emph {et~al.}(2000)\citenamefont
  {Pennanen}, \citenamefont {Michael},\ and\ \citenamefont
  {Green}}]{Pennanen:1999xi}%
  \BibitemOpen
  \bibfield  {author} {\bibinfo {author} {\bibfnamefont {Petrus}\ \bibnamefont
  {Pennanen}}, \bibinfo {author} {\bibfnamefont {Christopher}\ \bibnamefont
  {Michael}}, \ and\ \bibinfo {author} {\bibfnamefont {Anthony~M.}\
  \bibnamefont {Green}} (\bibinfo {collaboration} {UKQCD Collaboration}),\
  }\bibfield  {title} {\enquote {\bibinfo {title} {{Interactions of heavy light
  mesons}},}\ }\bibfield  {booktitle} {\emph {\bibinfo {booktitle} {{Lattice
  field theory. Proceedings, 17th International Symposium, Lattice'99, Pisa,
  Italy, June 29-July 3, 1999}}},\ }\href {\doibase
  10.1016/S0920-5632(00)91622-0} {\bibfield  {journal} {\bibinfo  {journal}
  {Nucl. Phys. Proc. Suppl.}\ }\textbf {\bibinfo {volume} {83}},\ \bibinfo
  {pages} {200} (\bibinfo {year} {2000})},\ \Eprint
  {http://arxiv.org/abs/hep-lat/9908032} {arXiv:hep-lat/9908032 [hep-lat]}
  \BibitemShut {NoStop}%
\bibitem [{\citenamefont {Bali}\ \emph {et~al.}(2005)\citenamefont {Bali},
  \citenamefont {Neff}, \citenamefont {D{\"u}ssel}, \citenamefont {Lippert},\
  and\ \citenamefont {Schilling}}]{Bali:2005fu}%
  \BibitemOpen
  \bibfield  {author} {\bibinfo {author} {\bibfnamefont {Gunnar~S.}\
  \bibnamefont {Bali}}, \bibinfo {author} {\bibfnamefont {Hartmut}\
  \bibnamefont {Neff}}, \bibinfo {author} {\bibfnamefont {Thomas}\ \bibnamefont
  {D{\"u}ssel}}, \bibinfo {author} {\bibfnamefont {Thomas}\ \bibnamefont
  {Lippert}}, \ and\ \bibinfo {author} {\bibfnamefont {Klaus}\ \bibnamefont
  {Schilling}} (\bibinfo {collaboration} {SESAM Collaboration}),\ }\bibfield
  {title} {\enquote {\bibinfo {title} {{Observation of string breaking in
  QCD}},}\ }\href {\doibase 10.1103/PhysRevD.71.114513} {\bibfield  {journal}
  {\bibinfo  {journal} {Phys. Rev. D}\ }\textbf {\bibinfo {volume} {71}},\
  \bibinfo {pages} {114513} (\bibinfo {year} {2005})},\ \Eprint
  {http://arxiv.org/abs/hep-lat/0505012} {arXiv:hep-lat/0505012 [hep-lat]}
  \BibitemShut {NoStop}%
\bibitem [{\citenamefont {Bali}\ and\ \citenamefont
  {Hetzenegger}(2010)}]{Bali:2010xa}%
  \BibitemOpen
  \bibfield  {author} {\bibinfo {author} {\bibfnamefont {Gunnar}\ \bibnamefont
  {Bali}}\ and\ \bibinfo {author} {\bibfnamefont {Martin}\ \bibnamefont
  {Hetzenegger}} (\bibinfo {collaboration} {QCDSF Collaboration}),\ }\bibfield
  {title} {\enquote {\bibinfo {title} {{Static-light meson-meson
  potentials}},}\ }\bibfield  {booktitle} {\emph {\bibinfo {booktitle}
  {{Proceedings, 28th International Symposium on Lattice field theory (Lattice
  2010)}}},\ }\href@noop {} {\bibfield  {journal} {\bibinfo  {journal} {Proc.
  Science}\ }\textbf {\bibinfo {volume} {LATTICE2010}},\ \bibinfo {pages} {142}
  (\bibinfo {year} {2010})},\ \Eprint {http://arxiv.org/abs/1011.0571}
  {arXiv:1011.0571 [hep-lat]} \BibitemShut {NoStop}%
\bibitem [{\citenamefont {Peters}\ \emph {et~al.}(2016)\citenamefont {Peters},
  \citenamefont {Bicudo}, \citenamefont {Cichy},\ and\ \citenamefont
  {Wagner}}]{Peters:2016wjm}%
  \BibitemOpen
  \bibfield  {author} {\bibinfo {author} {\bibfnamefont {Antje}\ \bibnamefont
  {Peters}}, \bibinfo {author} {\bibfnamefont {Pedro}\ \bibnamefont {Bicudo}},
  \bibinfo {author} {\bibfnamefont {Krzysztof}\ \bibnamefont {Cichy}}, \ and\
  \bibinfo {author} {\bibfnamefont {Marc}\ \bibnamefont {Wagner}},\ }\bibfield
  {title} {\enquote {\bibinfo {title} {{Investigation of $B\overline{B}$
  four-quark systems using Lattice QCD}},}\ }in\ \href
  {https://inspirehep.net/record/1423304/files/arXiv:1602.07621.pdf} {\emph
  {\bibinfo {booktitle} {{4th FAIR NExt generation ScientistS (FAIRNESS 2016)
  Garmisch-Partenkirchen, Germany, February 14-19, 2016}}}}\ (\bibinfo {year}
  {2016})\ \Eprint {http://arxiv.org/abs/1602.07621} {arXiv:1602.07621
  [hep-lat]} \BibitemShut {NoStop}%
\bibitem [{\citenamefont {Bali}\ and\ \citenamefont
  {Hetzenegger}(2011)}]{Bali:2011gq}%
  \BibitemOpen
  \bibfield  {author} {\bibinfo {author} {\bibfnamefont {Gunnar}\ \bibnamefont
  {Bali}}\ and\ \bibinfo {author} {\bibfnamefont {Martin}\ \bibnamefont
  {Hetzenegger}} (\bibinfo {collaboration} {QCDSF Collaboration}),\ }\bibfield
  {title} {\enquote {\bibinfo {title} {{Potentials between pairs of
  static-light mesons}},}\ }\bibfield  {booktitle} {\emph {\bibinfo {booktitle}
  {{Proceedings, 29th International Symposium on Lattice field theory (Lattice
  2011)}}},\ }\href@noop {} {\bibfield  {journal} {\bibinfo  {journal} {Proc.
  Science}\ }\textbf {\bibinfo {volume} {LATTICE2011}},\ \bibinfo {pages} {123}
  (\bibinfo {year} {2011})},\ \Eprint {http://arxiv.org/abs/1111.2222}
  {arXiv:1111.2222 [hep-lat]} \BibitemShut {NoStop}%
\bibitem [{\citenamefont {Olive}\ \emph {et~al.}(2014)\citenamefont {Olive}
  \emph {et~al.}}]{Agashe:2014kda}%
  \BibitemOpen
  \bibfield  {author} {\bibinfo {author} {\bibfnamefont {Keith~A.}\
  \bibnamefont {Olive}} \emph {et~al.} (\bibinfo {collaboration} {Particle Data
  Group}),\ }\bibfield  {title} {\enquote {\bibinfo {title} {{Review of
  Particle Physics}},}\ }\href {\doibase 10.1088/1674-1137/38/9/090001}
  {\bibfield  {journal} {\bibinfo  {journal} {Chin. Phys. C}\ }\textbf
  {\bibinfo {volume} {38}},\ \bibinfo {pages} {090001} (\bibinfo {year}
  {2014})}\BibitemShut {NoStop}%
\bibitem [{\citenamefont {Bruno}\ \emph {et~al.}(2015)\citenamefont {Bruno}
  \emph {et~al.}}]{Bruno:2014jqa}%
  \BibitemOpen
  \bibfield  {author} {\bibinfo {author} {\bibfnamefont {Mattia}\ \bibnamefont
  {Bruno}} \emph {et~al.} (\bibinfo {collaboration} {CLS}),\ }\bibfield
  {title} {\enquote {\bibinfo {title} {{Simulation of QCD with $N_{f} = 2 + 1$
  flavors of non-perturbatively improved Wilson fermions}},}\ }\href {\doibase
  10.1007/JHEP02(2015)043} {\bibfield  {journal} {\bibinfo  {journal} {J. High
  Energy Phys.}\ }\textbf {\bibinfo {volume} {02}},\ \bibinfo {pages} {043}
  (\bibinfo {year} {2015})},\ \Eprint {http://arxiv.org/abs/1411.3982}
  {arXiv:1411.3982 [hep-lat]} \BibitemShut {NoStop}%
\bibitem [{\citenamefont {{Martin L\"uscher {\em et al.},
  \url{http://luscher.web.cern.ch/luscher/openQCD/}, accessed
  2016}}()}]{ddopenqcd}%
  \BibitemOpen
  \bibfield  {author} {\bibinfo {author} {\bibnamefont {{Martin L\"uscher {\em
  et al.}, \url{http://luscher.web.cern.ch/luscher/openQCD/}, accessed
  2016}}}\ }\href@noop {} {}\BibitemShut {NoStop}%
\bibitem [{\citenamefont {L{\"u}scher}\ and\ \citenamefont
  {Schaefer}(2013)}]{Luscher:2012av}%
  \BibitemOpen
  \bibfield  {author} {\bibinfo {author} {\bibfnamefont {Martin}\ \bibnamefont
  {L{\"u}scher}}\ and\ \bibinfo {author} {\bibfnamefont {Stefan}\ \bibnamefont
  {Schaefer}},\ }\bibfield  {title} {\enquote {\bibinfo {title} {{Lattice QCD
  with open boundary conditions and twisted-mass reweighting}},}\ }\href
  {\doibase 10.1016/j.cpc.2012.10.003} {\bibfield  {journal} {\bibinfo
  {journal} {Comput. Phys. Commun.}\ }\textbf {\bibinfo {volume} {184}},\
  \bibinfo {pages} {519} (\bibinfo {year} {2013})},\ \Eprint
  {http://arxiv.org/abs/1206.2809} {arXiv:1206.2809 [hep-lat]} \BibitemShut
  {NoStop}%
\bibitem [{\citenamefont {L{\"u}scher}(2010)}]{Luscher:2010iy}%
  \BibitemOpen
  \bibfield  {author} {\bibinfo {author} {\bibfnamefont {Martin}\ \bibnamefont
  {L{\"u}scher}},\ }\bibfield  {title} {\enquote {\bibinfo {title} {{Properties
  and uses of the Wilson flow in Lattice QCD}},}\ }\href {\doibase
  10.1007/JHEP08(2010)071, 10.1007/JHEP03(2014)092} {\bibfield  {journal}
  {\bibinfo  {journal} {J. High Energy Phys.}\ }\textbf {\bibinfo {volume}
  {08}},\ \bibinfo {pages} {071} (\bibinfo {year} {2010})},\ \bibinfo {note}
  {[Erratum: J. High Energy Phys. {\bf 03}, 092 (2014)]},\ \Eprint
  {http://arxiv.org/abs/1006.4518} {arXiv:1006.4518 [hep-lat]} \BibitemShut
  {NoStop}%
\bibitem [{\citenamefont {Bali}\ \emph {et~al.}(2016)\citenamefont {Bali},
  \citenamefont {Scholz}, \citenamefont {Simeth},\ and\ \citenamefont
  {S{\"o}ldner}}]{Bali:2016umi}%
  \BibitemOpen
  \bibfield  {author} {\bibinfo {author} {\bibfnamefont {Gunnar~S.}\
  \bibnamefont {Bali}}, \bibinfo {author} {\bibfnamefont {Enno~E.}\
  \bibnamefont {Scholz}}, \bibinfo {author} {\bibfnamefont {Jakob}\
  \bibnamefont {Simeth}}, \ and\ \bibinfo {author} {\bibfnamefont {Wolfgang}\
  \bibnamefont {S{\"o}ldner}} (\bibinfo {collaboration} {RQCD Collaboration}),\
  }\bibfield  {title} {\enquote {\bibinfo {title} {{Lattice simulations with
  $N_f=2+1$ improved Wilson fermions at a fixed strange quark mass}},}\ }\href
  {\doibase 10.1103/PhysRevD.94.074501} {\bibfield  {journal} {\bibinfo
  {journal} {Phys. Rev. D}\ }\textbf {\bibinfo {volume} {94}},\ \bibinfo
  {pages} {074501} (\bibinfo {year} {2016})},\ \Eprint
  {http://arxiv.org/abs/1606.09039} {arXiv:1606.09039 [hep-lat]} \BibitemShut
  {NoStop}%
\bibitem [{\citenamefont {Bors{\'a}nyi}\ \emph {et~al.}(2012)\citenamefont
  {Bors{\'a}nyi} \emph {et~al.}}]{Borsanyi:2012zs}%
  \BibitemOpen
  \bibfield  {author} {\bibinfo {author} {\bibfnamefont {Szabolcs}\
  \bibnamefont {Bors{\'a}nyi}} \emph {et~al.} (\bibinfo {collaboration}
  {BMW-c}),\ }\bibfield  {title} {\enquote {\bibinfo {title} {{High-precision
  scale setting in Lattice QCD}},}\ }\href {\doibase 10.1007/JHEP09(2012)010}
  {\bibfield  {journal} {\bibinfo  {journal} {J. High Energy Phys.}\ }\textbf
  {\bibinfo {volume} {09}},\ \bibinfo {pages} {010} (\bibinfo {year} {2012})},\
  \Eprint {http://arxiv.org/abs/1203.4469} {arXiv:1203.4469 [hep-lat]}
  \BibitemShut {NoStop}%
\bibitem [{\citenamefont {{Bj\"orn Leder,
  \url{https://github.com/bjoern-leder/wloop/}, accessed 2016}}()}]{wloop}%
  \BibitemOpen
  \bibfield  {author} {\bibinfo {author} {\bibnamefont {{Bj\"orn Leder,
  \url{https://github.com/bjoern-leder/wloop/}, accessed 2016}}}\ }\href@noop
  {} {}\BibitemShut {NoStop}%
\bibitem [{\citenamefont {Donnellan}\ \emph {et~al.}(2011)\citenamefont
  {Donnellan}, \citenamefont {Knechtli}, \citenamefont {Leder},\ and\
  \citenamefont {Sommer}}]{Donnellan:2010mx}%
  \BibitemOpen
  \bibfield  {author} {\bibinfo {author} {\bibfnamefont {Michael}\ \bibnamefont
  {Donnellan}}, \bibinfo {author} {\bibfnamefont {Francesco}\ \bibnamefont
  {Knechtli}}, \bibinfo {author} {\bibfnamefont {Bj{\"o}rn}\ \bibnamefont
  {Leder}}, \ and\ \bibinfo {author} {\bibfnamefont {Rainer}\ \bibnamefont
  {Sommer}} (\bibinfo {collaboration} {ALPHA Collaboration}),\ }\bibfield
  {title} {\enquote {\bibinfo {title} {{Determination of the static potential
  with dynamical fermions}},}\ }\href {\doibase
  10.1016/j.nuclphysb.2011.03.013} {\bibfield  {journal} {\bibinfo  {journal}
  {Nucl. Phys. B}\ }\textbf {\bibinfo {volume} {849}},\ \bibinfo {pages} {45}
  (\bibinfo {year} {2011})},\ \Eprint {http://arxiv.org/abs/1012.3037}
  {arXiv:1012.3037 [hep-lat]} \BibitemShut {NoStop}%
\bibitem [{\citenamefont {Hasenfratz}\ and\ \citenamefont
  {Knechtli}(2001)}]{Hasenfratz:2001hp}%
  \BibitemOpen
  \bibfield  {author} {\bibinfo {author} {\bibfnamefont {Anna}\ \bibnamefont
  {Hasenfratz}}\ and\ \bibinfo {author} {\bibfnamefont {Francesco}\
  \bibnamefont {Knechtli}},\ }\bibfield  {title} {\enquote {\bibinfo {title}
  {{Flavor symmetry and the static potential with hypercubic blocking}},}\
  }\href {\doibase 10.1103/PhysRevD.64.034504} {\bibfield  {journal} {\bibinfo
  {journal} {Phys. Rev. D}\ }\textbf {\bibinfo {volume} {64}},\ \bibinfo
  {pages} {034504} (\bibinfo {year} {2001})},\ \Eprint
  {http://arxiv.org/abs/hep-lat/0103029} {arXiv:hep-lat/0103029 [hep-lat]}
  \BibitemShut {NoStop}%
\bibitem [{\citenamefont {Della~Morte}\ \emph {et~al.}(2004)\citenamefont
  {Della~Morte}, \citenamefont {D{\"u}rr}, \citenamefont {Heitger},
  \citenamefont {Molke}, \citenamefont {Rolf}, \citenamefont {Shindler},\ and\
  \citenamefont {Sommer}}]{DellaMorte:2003mn}%
  \BibitemOpen
  \bibfield  {author} {\bibinfo {author} {\bibfnamefont {Michele}\ \bibnamefont
  {Della~Morte}}, \bibinfo {author} {\bibfnamefont {Stephan}\ \bibnamefont
  {D{\"u}rr}}, \bibinfo {author} {\bibfnamefont {Jochen}\ \bibnamefont
  {Heitger}}, \bibinfo {author} {\bibfnamefont {Heiko}\ \bibnamefont {Molke}},
  \bibinfo {author} {\bibfnamefont {Juri}\ \bibnamefont {Rolf}}, \bibinfo
  {author} {\bibfnamefont {Andrea}\ \bibnamefont {Shindler}}, \ and\ \bibinfo
  {author} {\bibfnamefont {Rainer}\ \bibnamefont {Sommer}} (\bibinfo
  {collaboration} {ALPHA Collaboration}),\ }\bibfield  {title} {\enquote
  {\bibinfo {title} {{Lattice HQET with exponentially improved statistical
  precision}},}\ }\href {\doibase 10.1016/j.physletb.2005.03.017,
  10.1016/j.physletb.2003.11.064} {\bibfield  {journal} {\bibinfo  {journal}
  {Phys. Lett. B}\ }\textbf {\bibinfo {volume} {581}},\ \bibinfo {pages} {93}
  (\bibinfo {year} {2004})},\ \bibinfo {note} {[Erratum: Phys. Lett. B {\bf
  612}, 313 (2005)]},\ \Eprint {http://arxiv.org/abs/hep-lat/0307021}
  {arXiv:hep-lat/0307021 [hep-lat]} \BibitemShut {NoStop}%
\bibitem [{\citenamefont {Hasenfratz}\ \emph {et~al.}(2002)\citenamefont
  {Hasenfratz}, \citenamefont {Hoffmann},\ and\ \citenamefont
  {Knechtli}}]{Hasenfratz:2001tw}%
  \BibitemOpen
  \bibfield  {author} {\bibinfo {author} {\bibfnamefont {Anna}\ \bibnamefont
  {Hasenfratz}}, \bibinfo {author} {\bibfnamefont {Roland}\ \bibnamefont
  {Hoffmann}}, \ and\ \bibinfo {author} {\bibfnamefont {Francesco}\
  \bibnamefont {Knechtli}},\ }\bibfield  {title} {\enquote {\bibinfo {title}
  {{The static potential with hypercubic blocking}},}\ }\href {\doibase
  10.1016/S0920-5632(01)01733-9} {\bibfield  {journal} {\bibinfo  {journal}
  {Nucl. Phys. Proc. Suppl.}\ }\textbf {\bibinfo {volume} {106}},\ \bibinfo
  {pages} {418} (\bibinfo {year} {2002})},\ \Eprint
  {http://arxiv.org/abs/hep-lat/0110168} {arXiv:hep-lat/0110168 [hep-lat]}
  \BibitemShut {NoStop}%
\bibitem [{\citenamefont {Della~Morte}\ \emph {et~al.}(2005)\citenamefont
  {Della~Morte}, \citenamefont {Shindler},\ and\ \citenamefont
  {Sommer}}]{DellaMorte:2005yc}%
  \BibitemOpen
  \bibfield  {author} {\bibinfo {author} {\bibfnamefont {Michele}\ \bibnamefont
  {Della~Morte}}, \bibinfo {author} {\bibfnamefont {Andrea}\ \bibnamefont
  {Shindler}}, \ and\ \bibinfo {author} {\bibfnamefont {Rainer}\ \bibnamefont
  {Sommer}} (\bibinfo {collaboration} {ALPHA Collaboration}),\ }\bibfield
  {title} {\enquote {\bibinfo {title} {{On lattice actions for static
  quarks}},}\ }\href@noop {} {\bibfield  {journal} {\bibinfo  {journal} {J.
  High Energy Phys.}\ }\textbf {\bibinfo {volume} {08}},\ \bibinfo {pages}
  {051} (\bibinfo {year} {2005})},\ \Eprint
  {http://arxiv.org/abs/hep-lat/0506008} {arXiv:hep-lat/0506008 [hep-lat]}
  \BibitemShut {NoStop}%
\bibitem [{\citenamefont {Grimbach}\ \emph {et~al.}(2008)\citenamefont
  {Grimbach}, \citenamefont {Guazzini}, \citenamefont {Knechtli},\ and\
  \citenamefont {Palombi}}]{Grimbach:2008uy}%
  \BibitemOpen
  \bibfield  {author} {\bibinfo {author} {\bibfnamefont {Alois}\ \bibnamefont
  {Grimbach}}, \bibinfo {author} {\bibfnamefont {Damiano}\ \bibnamefont
  {Guazzini}}, \bibinfo {author} {\bibfnamefont {Francesco}\ \bibnamefont
  {Knechtli}}, \ and\ \bibinfo {author} {\bibfnamefont {Filippo}\ \bibnamefont
  {Palombi}} (\bibinfo {collaboration} {ALPHA Collaboration}),\ }\bibfield
  {title} {\enquote {\bibinfo {title} {{$O(a)$ improvement of the HYP static
  axial and vector currents at one-loop order of perturbation theory}},}\
  }\href {\doibase 10.1088/1126-6708/2008/03/039} {\bibfield  {journal}
  {\bibinfo  {journal} {J. High Energy Phys.}\ }\textbf {\bibinfo {volume}
  {03}},\ \bibinfo {pages} {039} (\bibinfo {year} {2008})},\ \Eprint
  {http://arxiv.org/abs/0802.0862} {arXiv:0802.0862 [hep-lat]} \BibitemShut
  {NoStop}%
\bibitem [{\citenamefont {Wolff}(2004)}]{Wolff:2003sm}%
  \BibitemOpen
  \bibfield  {author} {\bibinfo {author} {\bibfnamefont {Ulli}\ \bibnamefont
  {Wolff}} (\bibinfo {collaboration} {ALPHA Collaboration}),\ }\bibfield
  {title} {\enquote {\bibinfo {title} {{Monte Carlo errors with less
  errors}},}\ }\href {\doibase 10.1016/S0010-4655(03)00467-3,
  10.1016/j.cpc.2006.12.001} {\bibfield  {journal} {\bibinfo  {journal}
  {Comput. Phys. Commun.}\ }\textbf {\bibinfo {volume} {156}},\ \bibinfo
  {pages} {143} (\bibinfo {year} {2004})},\ \Eprint
  {http://arxiv.org/abs/hep-lat/0306017} {arXiv:hep-lat/0306017 [hep-lat]}
  \BibitemShut {NoStop}%
\bibitem [{\citenamefont {Schaefer}\ \emph {et~al.}(2011)\citenamefont
  {Schaefer}, \citenamefont {Sommer},\ and\ \citenamefont
  {Virotta}}]{algo:csd}%
  \BibitemOpen
  \bibfield  {author} {\bibinfo {author} {\bibfnamefont {Stefan}\ \bibnamefont
  {Schaefer}}, \bibinfo {author} {\bibfnamefont {Rainer}\ \bibnamefont
  {Sommer}}, \ and\ \bibinfo {author} {\bibfnamefont {Francesco}\ \bibnamefont
  {Virotta}} (\bibinfo {collaboration} {ALPHA Collaboration}),\ }\bibfield
  {title} {\enquote {\bibinfo {title} {{Critical slowing down and error
  analysis in Lattice QCD simulations}},}\ }\href {\doibase
  10.1016/j.nuclphysb.2010.11.020} {\bibfield  {journal} {\bibinfo  {journal}
  {Nucl. Phys. B}\ }\textbf {\bibinfo {volume} {845}},\ \bibinfo {pages} {93}
  (\bibinfo {year} {2011})},\ \Eprint {http://arxiv.org/abs/1009.5228}
  {arXiv:1009.5228 [hep-lat]} \BibitemShut {NoStop}%
\bibitem [{\citenamefont {Necco}\ and\ \citenamefont
  {Sommer}(2002)}]{Necco:2001xg}%
  \BibitemOpen
  \bibfield  {author} {\bibinfo {author} {\bibfnamefont {Silvia}\ \bibnamefont
  {Necco}}\ and\ \bibinfo {author} {\bibfnamefont {Rainer}\ \bibnamefont
  {Sommer}},\ }\bibfield  {title} {\enquote {\bibinfo {title} {{The $N_f = 0$
  heavy quark potential from short to intermediate distances}},}\ }\href
  {\doibase 10.1016/S0550-3213(01)00582-X} {\bibfield  {journal} {\bibinfo
  {journal} {Nucl. Phys. B}\ }\textbf {\bibinfo {volume} {622}},\ \bibinfo
  {pages} {328} (\bibinfo {year} {2002})},\ \Eprint
  {http://arxiv.org/abs/hep-lat/0108008} {arXiv:hep-lat/0108008} \BibitemShut
  {NoStop}%
\bibitem [{\citenamefont {Eichten}\ \emph {et~al.}(1975)\citenamefont
  {Eichten}, \citenamefont {Gottfried}, \citenamefont {Kinoshita},
  \citenamefont {Kogut}, \citenamefont {Lane},\ and\ \citenamefont
  {Yan}}]{Eichten:1974af}%
  \BibitemOpen
  \bibfield  {author} {\bibinfo {author} {\bibfnamefont {Estia}\ \bibnamefont
  {Eichten}}, \bibinfo {author} {\bibfnamefont {Kurt}\ \bibnamefont
  {Gottfried}}, \bibinfo {author} {\bibfnamefont {Toichiro}\ \bibnamefont
  {Kinoshita}}, \bibinfo {author} {\bibfnamefont {John~B.}\ \bibnamefont
  {Kogut}}, \bibinfo {author} {\bibfnamefont {Kenneth~D.}\ \bibnamefont
  {Lane}}, \ and\ \bibinfo {author} {\bibfnamefont {Tung-Mow}\ \bibnamefont
  {Yan}},\ }\bibfield  {title} {\enquote {\bibinfo {title} {{The spectrum of
  charmonium}},}\ }\href {\doibase 10.1103/PhysRevLett.34.369} {\bibfield
  {journal} {\bibinfo  {journal} {Phys. Rev. Lett.}\ }\textbf {\bibinfo
  {volume} {34}},\ \bibinfo {pages} {369} (\bibinfo {year} {1975})},\ \bibinfo
  {note} {[Erratum: Phys. Rev. Lett. {\bf 36}, 1276 (1976)]}\BibitemShut
  {NoStop}%
\bibitem [{\citenamefont {Koch}\ \emph {et~al.}(2016)\citenamefont {Koch},
  \citenamefont {Bulava}, \citenamefont {H{\"o}rz}, \citenamefont {Knechtli},
  \citenamefont {Moir}, \citenamefont {Morningstar},\ and\ \citenamefont
  {Peardon}}]{Koch:2015qxr}%
  \BibitemOpen
  \bibfield  {author} {\bibinfo {author} {\bibfnamefont {Vanessa}\ \bibnamefont
  {Koch}}, \bibinfo {author} {\bibfnamefont {John}\ \bibnamefont {Bulava}},
  \bibinfo {author} {\bibfnamefont {Ben}\ \bibnamefont {H{\"o}rz}}, \bibinfo
  {author} {\bibfnamefont {Francesco}\ \bibnamefont {Knechtli}}, \bibinfo
  {author} {\bibfnamefont {Graham}\ \bibnamefont {Moir}}, \bibinfo {author}
  {\bibfnamefont {Colin}\ \bibnamefont {Morningstar}}, \ and\ \bibinfo {author}
  {\bibfnamefont {Mike}\ \bibnamefont {Peardon}},\ }\bibfield  {title}
  {\enquote {\bibinfo {title} {{Towards string breaking with $2+1$ dynamical
  fermions using the stochastic LapH method}},}\ }\bibfield  {booktitle} {\emph
  {\bibinfo {booktitle} {{Proceedings, 33rd International Symposium on Lattice
  Field Theory (Lattice 2015): Kobe, Japan, July 14-18, 2015}}},\ }\href@noop
  {} {\bibfield  {journal} {\bibinfo  {journal} {Proc. Science}\ }\textbf
  {\bibinfo {volume} {LATTICE2015}},\ \bibinfo {pages} {100} (\bibinfo {year}
  {2016})},\ \Eprint {http://arxiv.org/abs/1511.04029} {arXiv:1511.04029
  [hep-lat]} \BibitemShut {NoStop}%
\bibitem [{\citenamefont {Sommer}(1994)}]{Sommer:1993ce}%
  \BibitemOpen
  \bibfield  {author} {\bibinfo {author} {\bibfnamefont {Rainer}\ \bibnamefont
  {Sommer}},\ }\bibfield  {title} {\enquote {\bibinfo {title} {{A new way to
  set the energy scale in lattice gauge theories and its applications to the
  static force and $\alpha_s$ in SU(2) Yang-Mills theory}},}\ }\href {\doibase
  10.1016/0550-3213(94)90473-1} {\bibfield  {journal} {\bibinfo  {journal}
  {Nucl. Phys. B}\ }\textbf {\bibinfo {volume} {411}},\ \bibinfo {pages} {839}
  (\bibinfo {year} {1994})},\ \Eprint {http://arxiv.org/abs/hep-lat/9310022}
  {arXiv:hep-lat/9310022 [hep-lat]} \BibitemShut {NoStop}%
\bibitem [{\citenamefont {Bali}\ and\ \citenamefont
  {Boyle}(1999)}]{Bali:1998pi}%
  \BibitemOpen
  \bibfield  {author} {\bibinfo {author} {\bibfnamefont {Gunnar~S.}\
  \bibnamefont {Bali}}\ and\ \bibinfo {author} {\bibfnamefont {Peter}\
  \bibnamefont {Boyle}},\ }\bibfield  {title} {\enquote {\bibinfo {title} {{A
  Lattice potential investigation of quark mass and volume dependence of the
  $\Upsilon$ spectrum}},}\ }\href {\doibase 10.1103/PhysRevD.59.114504}
  {\bibfield  {journal} {\bibinfo  {journal} {Phys. Rev. D}\ }\textbf {\bibinfo
  {volume} {59}},\ \bibinfo {pages} {114504} (\bibinfo {year} {1999})},\
  \Eprint {http://arxiv.org/abs/hep-lat/9809180} {arXiv:hep-lat/9809180
  [hep-lat]} \BibitemShut {NoStop}%
\bibitem [{\citenamefont {Kreuzer}\ and\ \citenamefont
  {Hammer}(2010)}]{Kreuzer:2009jp}%
  \BibitemOpen
  \bibfield  {author} {\bibinfo {author} {\bibfnamefont {Simon}\ \bibnamefont
  {Kreuzer}}\ and\ \bibinfo {author} {\bibfnamefont {Hans-Werner}\ \bibnamefont
  {Hammer}},\ }\bibfield  {title} {\enquote {\bibinfo {title} {{On the
  modification of the Efimov spectrum in a finite cubic box}},}\ }\href
  {\doibase 10.1140/epja/i2010-10910-6} {\bibfield  {journal} {\bibinfo
  {journal} {Eur. Phys. J. A}\ }\textbf {\bibinfo {volume} {43}},\ \bibinfo
  {pages} {229} (\bibinfo {year} {2010})},\ \Eprint
  {http://arxiv.org/abs/0910.2191} {arXiv:0910.2191 [nucl-th]} \BibitemShut
  {NoStop}%
\bibitem [{\citenamefont {Polejaeva}\ and\ \citenamefont
  {Rusetsky}(2012)}]{Polejaeva:2012ut}%
  \BibitemOpen
  \bibfield  {author} {\bibinfo {author} {\bibfnamefont {Kathryn}\ \bibnamefont
  {Polejaeva}}\ and\ \bibinfo {author} {\bibfnamefont {Akaki}\ \bibnamefont
  {Rusetsky}},\ }\bibfield  {title} {\enquote {\bibinfo {title} {{Three
  particles in a finite volume}},}\ }\href {\doibase
  10.1140/epja/i2012-12067-8} {\bibfield  {journal} {\bibinfo  {journal} {Eur.
  Phys. J.}\ }\textbf {\bibinfo {volume} {A48}},\ \bibinfo {pages} {67}
  (\bibinfo {year} {2012})},\ \Eprint {http://arxiv.org/abs/1203.1241}
  {arXiv:1203.1241 [hep-lat]} \BibitemShut {NoStop}%
\bibitem [{\citenamefont {Brice{\~n}o}\ and\ \citenamefont
  {Davoudi}(2013)}]{Briceno:2012rv}%
  \BibitemOpen
  \bibfield  {author} {\bibinfo {author} {\bibfnamefont {Ra{\'u}l~A.}\
  \bibnamefont {Brice{\~n}o}}\ and\ \bibinfo {author} {\bibfnamefont {Zohreh}\
  \bibnamefont {Davoudi}},\ }\bibfield  {title} {\enquote {\bibinfo {title}
  {{Three-particle scattering amplitudes from a finite volume formalism}},}\
  }\href {\doibase 10.1103/PhysRevD.87.094507} {\bibfield  {journal} {\bibinfo
  {journal} {Phys. Rev. D}\ }\textbf {\bibinfo {volume} {87}},\ \bibinfo
  {pages} {094507} (\bibinfo {year} {2013})},\ \Eprint
  {http://arxiv.org/abs/1212.3398} {arXiv:1212.3398 [hep-lat]} \BibitemShut
  {NoStop}%
\bibitem [{\citenamefont {Mei{\ss}ner}\ \emph {et~al.}(2015)\citenamefont
  {Mei{\ss}ner}, \citenamefont {R{\'{\i}}os},\ and\ \citenamefont
  {Rusetsky}}]{Meissner:2014dea}%
  \BibitemOpen
  \bibfield  {author} {\bibinfo {author} {\bibfnamefont {Ulf-G.}\ \bibnamefont
  {Mei{\ss}ner}}, \bibinfo {author} {\bibfnamefont {Guillermo}\ \bibnamefont
  {R{\'{\i}}os}}, \ and\ \bibinfo {author} {\bibfnamefont {Akaki}\ \bibnamefont
  {Rusetsky}},\ }\bibfield  {title} {\enquote {\bibinfo {title} {{Spectrum of
  three-body bound states in a finite volume}},}\ }\href {\doibase
  10.1103/PhysRevLett.117.069902, 10.1103/PhysRevLett.114.091602} {\bibfield
  {journal} {\bibinfo  {journal} {Phys. Rev. Lett.}\ }\textbf {\bibinfo
  {volume} {114}},\ \bibinfo {pages} {091602} (\bibinfo {year} {2015})},\
  \bibinfo {note} {[Erratum: Phys. Rev. Lett. {\bf 117}, 069902 (2016)]},\
  \Eprint {http://arxiv.org/abs/1412.4969} {arXiv:1412.4969 [hep-lat]}
  \BibitemShut {NoStop}%
\bibitem [{\citenamefont {Hansen}\ and\ \citenamefont
  {Sharpe}(2015)}]{Hansen:2015zga}%
  \BibitemOpen
  \bibfield  {author} {\bibinfo {author} {\bibfnamefont {Maxwell~T.}\
  \bibnamefont {Hansen}}\ and\ \bibinfo {author} {\bibfnamefont {Stephen~R.}\
  \bibnamefont {Sharpe}},\ }\bibfield  {title} {\enquote {\bibinfo {title}
  {{Expressing the three-particle finite-volume spectrum in terms of the
  three-to-three scattering amplitude}},}\ }\href {\doibase
  10.1103/PhysRevD.92.114509} {\bibfield  {journal} {\bibinfo  {journal} {Phys.
  Rev. D}\ }\textbf {\bibinfo {volume} {92}},\ \bibinfo {pages} {114509}
  (\bibinfo {year} {2015})},\ \Eprint {http://arxiv.org/abs/1504.04248}
  {arXiv:1504.04248 [hep-lat]} \BibitemShut {NoStop}%
\bibitem [{\citenamefont {Hansen}\ and\ \citenamefont
  {Sharpe}(2016)}]{Hansen:2016fzj}%
  \BibitemOpen
  \bibfield  {author} {\bibinfo {author} {\bibfnamefont {Maxwell~T.}\
  \bibnamefont {Hansen}}\ and\ \bibinfo {author} {\bibfnamefont {Stephen~R.}\
  \bibnamefont {Sharpe}},\ }\bibfield  {title} {\enquote {\bibinfo {title}
  {{Threshold expansion of the three-particle quantization condition}},}\
  }\href {\doibase 10.1103/PhysRevD.93.096006} {\bibfield  {journal} {\bibinfo
  {journal} {Phys. Rev. D}\ }\textbf {\bibinfo {volume} {93}},\ \bibinfo
  {pages} {096006} (\bibinfo {year} {2016})},\ \Eprint
  {http://arxiv.org/abs/1602.00324} {arXiv:1602.00324 [hep-lat]} \BibitemShut
  {NoStop}%
\bibitem [{\citenamefont {Eichten}\ and\ \citenamefont
  {Feinberg}(1981)}]{Eichten:1980mw}%
  \BibitemOpen
  \bibfield  {author} {\bibinfo {author} {\bibfnamefont {Estia}\ \bibnamefont
  {Eichten}}\ and\ \bibinfo {author} {\bibfnamefont {Frank~L.}\ \bibnamefont
  {Feinberg}},\ }\bibfield  {title} {\enquote {\bibinfo {title} {{Spin
  dependent forces in QCD}},}\ }\href {\doibase 10.1103/PhysRevD.23.2724}
  {\bibfield  {journal} {\bibinfo  {journal} {Phys. Rev. D}\ }\textbf {\bibinfo
  {volume} {23}},\ \bibinfo {pages} {2724} (\bibinfo {year}
  {1981})}\BibitemShut {NoStop}%
\bibitem [{\citenamefont {Barchielli}\ \emph {et~al.}(1990)\citenamefont
  {Barchielli}, \citenamefont {Brambilla},\ and\ \citenamefont
  {Prosperi}}]{Barchielli:1988zp}%
  \BibitemOpen
  \bibfield  {author} {\bibinfo {author} {\bibfnamefont {Alberto}\ \bibnamefont
  {Barchielli}}, \bibinfo {author} {\bibfnamefont {Nora}\ \bibnamefont
  {Brambilla}}, \ and\ \bibinfo {author} {\bibfnamefont {Giovanni~M.}\
  \bibnamefont {Prosperi}},\ }\bibfield  {title} {\enquote {\bibinfo {title}
  {{Relativistic corrections to the quark-antiquark potential and the
  quarkonium spectrum}},}\ }\href {\doibase 10.1007/BF02902620} {\bibfield
  {journal} {\bibinfo  {journal} {Nuovo Cim. A}\ }\textbf {\bibinfo {volume}
  {103}},\ \bibinfo {pages} {59} (\bibinfo {year} {1990})}\BibitemShut
  {NoStop}%
\bibitem [{\citenamefont {Bali}\ \emph {et~al.}(1997)\citenamefont {Bali},
  \citenamefont {Schilling},\ and\ \citenamefont {Wachter}}]{Bali:1997am}%
  \BibitemOpen
  \bibfield  {author} {\bibinfo {author} {\bibfnamefont {Gunnar~S.}\
  \bibnamefont {Bali}}, \bibinfo {author} {\bibfnamefont {Klaus}\ \bibnamefont
  {Schilling}}, \ and\ \bibinfo {author} {\bibfnamefont {Armin}\ \bibnamefont
  {Wachter}},\ }\bibfield  {title} {\enquote {\bibinfo {title} {{Complete
  $O(v^2)$ corrections to the static interquark potential from SU(3) gauge
  theory}},}\ }\href {\doibase 10.1103/PhysRevD.56.2566} {\bibfield  {journal}
  {\bibinfo  {journal} {Phys. Rev. D}\ }\textbf {\bibinfo {volume} {56}},\
  \bibinfo {pages} {2566} (\bibinfo {year} {1997})},\ \Eprint
  {http://arxiv.org/abs/hep-lat/9703019} {arXiv:hep-lat/9703019 [hep-lat]}
  \BibitemShut {NoStop}%
\bibitem [{\citenamefont {Bali}(2001)}]{Bali:2000gf}%
  \BibitemOpen
  \bibfield  {author} {\bibinfo {author} {\bibfnamefont {Gunnar~S.}\
  \bibnamefont {Bali}},\ }\bibfield  {title} {\enquote {\bibinfo {title} {{QCD
  forces and heavy quark bound states}},}\ }\href {\doibase
  10.1016/S0370-1573(00)00079-X} {\bibfield  {journal} {\bibinfo  {journal}
  {Phys. Rept.}\ }\textbf {\bibinfo {volume} {343}},\ \bibinfo {pages} {1}
  (\bibinfo {year} {2001})},\ \Eprint {http://arxiv.org/abs/hep-ph/0001312}
  {arXiv:hep-ph/0001312 [hep-ph]} \BibitemShut {NoStop}%
\bibitem [{\citenamefont {Brambilla}\ \emph {et~al.}(2005)\citenamefont
  {Brambilla}, \citenamefont {Pineda}, \citenamefont {Soto},\ and\
  \citenamefont {Vairo}}]{Brambilla:2004jw}%
  \BibitemOpen
  \bibfield  {author} {\bibinfo {author} {\bibfnamefont {Nora}\ \bibnamefont
  {Brambilla}}, \bibinfo {author} {\bibfnamefont {Antonio}\ \bibnamefont
  {Pineda}}, \bibinfo {author} {\bibfnamefont {Joan}\ \bibnamefont {Soto}}, \
  and\ \bibinfo {author} {\bibfnamefont {Antonio}\ \bibnamefont {Vairo}},\
  }\bibfield  {title} {\enquote {\bibinfo {title} {{Effective field theories
  for heavy quarkonium}},}\ }\href {\doibase 10.1103/RevModPhys.77.1423}
  {\bibfield  {journal} {\bibinfo  {journal} {Rev. Mod. Phys.}\ }\textbf
  {\bibinfo {volume} {77}},\ \bibinfo {pages} {1423} (\bibinfo {year}
  {2005})},\ \Eprint {http://arxiv.org/abs/hep-ph/0410047}
  {arXiv:hep-ph/0410047 [hep-ph]} \BibitemShut {NoStop}%
\bibitem [{\citenamefont {Koma}\ and\ \citenamefont
  {Koma}(2007)}]{Koma:2006fw}%
  \BibitemOpen
  \bibfield  {author} {\bibinfo {author} {\bibfnamefont {Yoshiaki}\
  \bibnamefont {Koma}}\ and\ \bibinfo {author} {\bibfnamefont {Miho}\
  \bibnamefont {Koma}},\ }\bibfield  {title} {\enquote {\bibinfo {title}
  {{Spin-dependent potentials from lattice QCD}},}\ }\href {\doibase
  10.1016/j.nuclphysb.2007.01.033} {\bibfield  {journal} {\bibinfo  {journal}
  {Nucl. Phys. B}\ }\textbf {\bibinfo {volume} {769}},\ \bibinfo {pages} {79}
  (\bibinfo {year} {2007})},\ \Eprint {http://arxiv.org/abs/hep-lat/0609078}
  {arXiv:hep-lat/0609078 [hep-lat]} \BibitemShut {NoStop}%
\bibitem [{\citenamefont {Knechtli}\ \emph {et~al.}(2017)\citenamefont
  {Knechtli}, \citenamefont {Alberti}, \citenamefont {Bali}, \citenamefont
  {Collins}, \citenamefont {Moir},\ and\ \citenamefont
  {S{\"o}ldner}}]{Knechtli:2016eqx}%
  \BibitemOpen
  \bibfield  {author} {\bibinfo {author} {\bibfnamefont {Francesco}\
  \bibnamefont {Knechtli}}, \bibinfo {author} {\bibfnamefont {Maurizio}\
  \bibnamefont {Alberti}}, \bibinfo {author} {\bibfnamefont {Gunnar~S.}\
  \bibnamefont {Bali}}, \bibinfo {author} {\bibfnamefont {Sara}\ \bibnamefont
  {Collins}}, \bibinfo {author} {\bibfnamefont {Graham}\ \bibnamefont {Moir}},
  \ and\ \bibinfo {author} {\bibfnamefont {Wolfgang}\ \bibnamefont
  {S{\"o}ldner}},\ }\bibfield  {title} {\enquote {\bibinfo {title} {{Testing
  the hadro-quarkonium model on the lattice}},}\ }\bibfield  {booktitle} {\emph
  {\bibinfo {booktitle} {{Proceedings, 34th International Symposium on Lattice
  Field Theory (Lattice 2016)}}},\ }\href@noop {} {\bibfield  {journal}
  {\bibinfo  {journal} {Proc. Science}\ }\textbf {\bibinfo {volume}
  {LATTICE2016}},\ \bibinfo {pages} {113} (\bibinfo {year} {2017})},\ \Eprint
  {http://arxiv.org/abs/1611.00912} {arXiv:1611.00912 [hep-lat]} \BibitemShut
  {NoStop}%
\bibitem [{\citenamefont {Arts}\ \emph {et~al.}(2015)\citenamefont {Arts} \emph
  {et~al.}}]{Arts:2015jia}%
  \BibitemOpen
  \bibfield  {author} {\bibinfo {author} {\bibfnamefont {Paul}\ \bibnamefont
  {Arts}} \emph {et~al.},\ }\bibfield  {title} {\enquote {\bibinfo {title}
  {{QPACE 2 and domain decomposition on the Intel Xeon Phi}},}\ }\bibfield
  {booktitle} {\emph {\bibinfo {booktitle} {{Proceedings, 32nd International
  Symposium on Lattice Field Theory (Lattice 2014)}}},\ }\href@noop {}
  {\bibfield  {journal} {\bibinfo  {journal} {Proc. Science}\ }\textbf
  {\bibinfo {volume} {LATTICE2014}},\ \bibinfo {pages} {021} (\bibinfo {year}
  {2015})},\ \Eprint {http://arxiv.org/abs/1502.04025} {arXiv:1502.04025
  [cs.DC]} \BibitemShut {NoStop}%
\bibitem [{\citenamefont {Edwards}\ and\ \citenamefont
  {Jo{\'o}}(2005)}]{Edwards:2004sx}%
  \BibitemOpen
  \bibfield  {author} {\bibinfo {author} {\bibfnamefont {Robert~G.}\
  \bibnamefont {Edwards}}\ and\ \bibinfo {author} {\bibfnamefont {Balint}\
  \bibnamefont {Jo{\'o}}} (\bibinfo {collaboration} {SciDAC, LHP Collaboration
  and UKQCD Collaboration}),\ }\bibfield  {title} {\enquote {\bibinfo {title}
  {{The Chroma software system for lattice QCD}},}\ }\href {\doibase
  10.1016/j.nuclphysbps.2004.11.254} {\bibfield  {journal} {\bibinfo  {journal}
  {Nucl. Phys. B Proc. Suppl.}\ }\textbf {\bibinfo {volume} {140}},\ \bibinfo
  {pages} {832} (\bibinfo {year} {2005})},\ \Eprint
  {http://arxiv.org/abs/hep-lat/0409003} {arXiv:hep-lat/0409003} \BibitemShut
  {NoStop}%
\bibitem [{\citenamefont {Heybrock}\ \emph {et~al.}(2016)\citenamefont
  {Heybrock}, \citenamefont {Rottmann}, \citenamefont {Georg},\ and\
  \citenamefont {Wettig}}]{Heybrock:2015kpy}%
  \BibitemOpen
  \bibfield  {author} {\bibinfo {author} {\bibfnamefont {Simon}\ \bibnamefont
  {Heybrock}}, \bibinfo {author} {\bibfnamefont {Matthias}\ \bibnamefont
  {Rottmann}}, \bibinfo {author} {\bibfnamefont {Peter}\ \bibnamefont {Georg}},
  \ and\ \bibinfo {author} {\bibfnamefont {Tilo}\ \bibnamefont {Wettig}},\
  }\bibfield  {title} {\enquote {\bibinfo {title} {{Adaptive algebraic
  multigrid on SIMD architectures}},}\ }\bibfield  {booktitle} {\emph {\bibinfo
  {booktitle} {{Proceedings, 33rd International Symposium on Lattice Field
  Theory (Lattice 2015)}}},\ }\href
  {http://inspirehep.net/record/1409505/files/arXiv:1512.04506.pdf} {\bibfield
  {journal} {\bibinfo  {journal} {Proc. Science}\ }\textbf {\bibinfo {volume}
  {LATTICE2015}},\ \bibinfo {pages} {036} (\bibinfo {year} {2016})},\ \Eprint
  {http://arxiv.org/abs/1512.04506} {arXiv:1512.04506 [physics.comp-ph]}
  \BibitemShut {NoStop}%
\bibitem [{\citenamefont {Richtmann}\ \emph {et~al.}(2016)\citenamefont
  {Richtmann}, \citenamefont {Heybrock},\ and\ \citenamefont
  {Wettig}}]{Richtmann:2016kcq}%
  \BibitemOpen
  \bibfield  {author} {\bibinfo {author} {\bibfnamefont {Daniel}\ \bibnamefont
  {Richtmann}}, \bibinfo {author} {\bibfnamefont {Simon}\ \bibnamefont
  {Heybrock}}, \ and\ \bibinfo {author} {\bibfnamefont {Tilo}\ \bibnamefont
  {Wettig}},\ }\bibfield  {title} {\enquote {\bibinfo {title} {{Multiple
  right-hand-side setup for the DD-$\alpha$AMG}},}\ }\bibfield  {booktitle}
  {\emph {\bibinfo {booktitle} {{Proceedings, 33rd International Symposium on
  Lattice Field Theory (Lattice 2015)}}},\ }\href
  {http://inspirehep.net/record/1415122/files/arXiv:1601.03184.pdf} {\bibfield
  {journal} {\bibinfo  {journal} {Proc. Science}\ }\textbf {\bibinfo {volume}
  {LATTICE2015}},\ \bibinfo {pages} {035} (\bibinfo {year} {2016})},\ \Eprint
  {http://arxiv.org/abs/1601.03184} {arXiv:1601.03184 [hep-lat]} \BibitemShut
  {NoStop}%
\bibitem [{\citenamefont {Heybrock}\ \emph {et~al.}(2014)\citenamefont
  {Heybrock}, \citenamefont {Jo{\'o}}, \citenamefont {Kalamkar}, \citenamefont
  {Smelyanskiy}, \citenamefont {Vaidyanathan}, \citenamefont {Wettig},\ and\
  \citenamefont {Dubey}}]{Heybrock:2014iga}%
  \BibitemOpen
  \bibfield  {author} {\bibinfo {author} {\bibfnamefont {Simon}\ \bibnamefont
  {Heybrock}}, \bibinfo {author} {\bibfnamefont {B{\'a}lint}\ \bibnamefont
  {Jo{\'o}}}, \bibinfo {author} {\bibfnamefont {Dhiraj~D.}\ \bibnamefont
  {Kalamkar}}, \bibinfo {author} {\bibfnamefont {Mikhail}\ \bibnamefont
  {Smelyanskiy}}, \bibinfo {author} {\bibfnamefont {Karthikeyan}\ \bibnamefont
  {Vaidyanathan}}, \bibinfo {author} {\bibfnamefont {Tilo}\ \bibnamefont
  {Wettig}}, \ and\ \bibinfo {author} {\bibfnamefont {Pradeep}\ \bibnamefont
  {Dubey}},\ }\bibfield  {title} {\enquote {\bibinfo {title} {{Lattice QCD with
  domain decomposition on Intel Xeon Phi co-processors}},}\ }\bibfield
  {booktitle} {\emph {\bibinfo {booktitle} {{The International Conference for
  High Performance Computing, Networking, Storage, and Analysis, New Orleans,
  LA, USA, November 16-21, 2014}}},\ }\href {\doibase 10.1109/SC.2014.11}
  {\bibfield  {journal} {\bibinfo  {journal} {Proceedings of SC14}\ ,\ \bibinfo
  {pages} {69}} (\bibinfo {year} {2014})},\ \Eprint
  {http://arxiv.org/abs/1412.2629} {arXiv:1412.2629 [hep-lat]} \BibitemShut
  {NoStop}%
\bibitem [{\citenamefont {Frommer}\ \emph {et~al.}(2014)\citenamefont
  {Frommer}, \citenamefont {Kahl}, \citenamefont {Krieg}, \citenamefont
  {Leder},\ and\ \citenamefont {Rottmann}}]{Frommer:2013fsa}%
  \BibitemOpen
  \bibfield  {author} {\bibinfo {author} {\bibfnamefont {Andreas}\ \bibnamefont
  {Frommer}}, \bibinfo {author} {\bibfnamefont {Karsten}\ \bibnamefont {Kahl}},
  \bibinfo {author} {\bibfnamefont {Stefan}\ \bibnamefont {Krieg}}, \bibinfo
  {author} {\bibfnamefont {Bj{\"o}rn}\ \bibnamefont {Leder}}, \ and\ \bibinfo
  {author} {\bibfnamefont {Matthias}\ \bibnamefont {Rottmann}},\ }\bibfield
  {title} {\enquote {\bibinfo {title} {{Adaptive aggregation based domain
  decomposition Multigrid for the lattice Wilson Dirac operator}},}\ }\href
  {\doibase 10.1137/130919507} {\bibfield  {journal} {\bibinfo  {journal} {SIAM
  J. Sci. Comput.}\ }\textbf {\bibinfo {volume} {36}},\ \bibinfo {pages}
  {A1581} (\bibinfo {year} {2014})},\ \Eprint {http://arxiv.org/abs/1303.1377}
  {arXiv:1303.1377 [hep-lat]} \BibitemShut {NoStop}%
\end{thebibliography}%
\end{document}